\documentclass[journal]{IEEEtran}
\usepackage{epsfig}
\usepackage{amssymb}
\usepackage[cmex10]{amsmath, nccmath}
\usepackage{array}
\usepackage{upgreek}
\usepackage{graphicx}
\usepackage{mathtools}
\usepackage{enumerate}
\usepackage{bm}
\usepackage{cite}
\newtheorem{theorem}{Theorem}
\newtheorem{remark}{Remark}

\newtheorem{lemma}{Lemma}
\newtheorem{definition}{Definition}
\newtheorem{corollary}{Corollary}
\newcommand{\norm}[1]{\left\lVert#1\right\rVert}
\newcommand{\cB}[1]{\left\{#1\right\}}
\newcommand{\nB}[1]{\left(#1\right)}
\newcommand{\sB}[1]{\left[#1\right]}

\newcommand{\Ip}{H}
\newcommand{\ip}{h}
\newcommand{\vX}{{\mathbf X}}
\newcommand{\vY}{{\mathbf Y}}
\newcommand{\vx}{{\mathbf x}}
\newcommand{\vy}{{\mathbf y}}
\newcommand{\vZ}{{\mathbf Z}}
\newcommand{\vXltwo}{{\mathbf{X}_1 + \mathbf{X}_2}}

\newcommand{\bbP}{\mathbb{P}}
\newcommand{\bbE}{\mathbb{E}}

\newcommand{\Ponetwo}{P_1 + P_2}
\newcommand{\PS}{P_{\left\langle \mc{S} \right \rangle}}
\newcommand{\bi}{\boldsymbol{\tilde{\imath}}_2}

\newcommand{\E}[1]{\mathbb E\left[#1\right]}
\newcommand{\Prob}[1]{\mathbb P\left[#1\right]}

\newcommand{\bs}[1]{\boldsymbol{#1}}
\newcommand{\mc}[1]{\mathcal{#1}}
\newcommand{\mb}[1]{\mathbf{#1}}
\newcommand{\ms}[1]{\mathsf{#1}}
\newcommand{\bigo}[1]{O\left(#1\right)}

\newcommand{\Var}[1]{\mathrm{Var}\left[#1\right]}
\newcommand{\Cov}[1]{\mathrm{Cov}\left[#1\right]}
\newcommand{\iK}{\tilde{\boldsymbol{\imath}}_K}
\newcommand{\thmref}[1]{Theorem~\ref{#1}}

\newcommand{\secref}[1]{Section~\ref{#1}}
\newcommand{\lemref}[1]{Lemma~\ref{#1}}

\newcommand{\corref}[1]{Corollary~\ref{#1}}
\newcommand{\appref}[1]{Appendix~\ref{#1}}

\newcommand{\figref}[1]{Fig.~\ref{#1}}
\newcommand{\subemp}[1]{\mathcal{P}(#1)}
\usepackage{url}
\usepackage{cases}
\usepackage{ifthen}
\newif\ifshowtodo
\showtodotrue

\newcommand{\VersionLength}{long}
\providecommand{\ver}{\ifthenelse{\equal{\VersionLength}{long}}}

\allowdisplaybreaks
\begin{document}
\title{Gaussian Multiple and Random Access Channels: Finite-Blocklength Analysis
}
\author{Recep Can Yavas, Victoria Kostina, and Michelle Effros
\thanks{R. C.~Yavas, V.~Kostina, and M.~Effros are with the Department of Electrical Engineering, California Institute of Technology, Pasadena, CA~91125, USA (e-mail:  ryavas, vkostina, effros@caltech.edu). This work was supported in part by the National Science Foundation (NSF) under grant CCF-1817241. A part of this work was presented at the 2020 International Symposium on Information Theory (ISIT’20) \cite{yavas2020Gaussian}.}}
\date{\today}
\IEEEoverridecommandlockouts

	\maketitle 
	
\begin{abstract}
This paper presents finite-blocklength achievability bounds 
for the Gaussian multiple access channel (MAC) 
and random access channel (RAC) 
under average-error and maximal-power constraints.  
Using random codewords uniformly distributed on a sphere 
and a maximum likelihood decoder, 
the derived MAC bound on each transmitter's rate 
matches the MolavianJazi-Laneman bound (2015) 
in its first- and second-order terms, 
improving the remaining terms to 
$\frac12\frac{\log n}{n}+O \left(\frac 1 n \right)$
bits per channel use. 
The result then extends to a RAC model
in which neither the encoders nor the decoder 
knows which of $K$ possible transmitters are active. 
In the proposed rateless coding strategy, 
decoding occurs at a time $n_t$ 
that depends on the decoder's estimate $t$ 
of the number of active transmitters $k$. 
Single-bit feedback from the decoder to all encoders 
at each potential decoding time $n_i$, $i \leq t$, 
informs the encoders when to stop transmitting. 
For this RAC model, the proposed code achieves 
the same first-, second-, and third-order performance 
as the best known result for the Gaussian MAC in operation.
\end{abstract}


	\section{Introduction}
	Emerging communication systems such as the Internet of Things, wireless cellular networks, and machine-to-machine communication systems impose two significant requirements on the code design: low latency constraints and random activity in a large number of communicating devices. These constraints lead us to study random access channels in the finite blocklength regime, where an unknown number of transmitters is active, and communication delay is finite. Current random access strategies mostly use either orthogonalization (TDMA, FDMA, and CDMA) or collision avoidance (e.g., slotted ALOHA). Orthogonalization methods divide up resources (e.g., time, frequency, or signal space) among the transmitters. In slotted ALOHA, each transmitter randomly chooses a time slot to transmit its message, and the decoder declares an error if two or more transmitters are active in a time slot. Performance of these methods is inferior to the information-theoretic bounds achieved through simultaneous resource use. For example, slotted ALOHA achieves only 37\% of the single-transmitter capacity \cite{roberts1975aloha}. 
	
We consider a communication scenario where $K$ transmitters are communicating with a single receiver through a Gaussian channel. We study two problems in this network: multiple access and random access communication. In the multiple access problem, the identity of active transmitters is known to all transmitters and to the receiver. In the random access problem, the set of active transmitters is unknown to the transmitters and to the receiver.

For $K=1$, 
Shannon's 1948 paper \cite{shannon1948A} 
derives the capacity 
\begin{equation}
	C(P) = \frac 1 2 \log (1 + P) \label{eq:capacity} 
\end{equation}
using codewords with symbols drawn independently and identically distributed (i.i.d.) 
according to the Gaussian distribution with variance $P - \delta$ 
for a very small value $\delta$; 
here $P$ is the maximal (per-codeword) power constraint, 
and the noise variance is 1.
In~\cite{shannon1959Probability}, 
Shannon shows the performance improvement in the achievable reliability function 
using codewords drawn uniformly at random
on an $n$-dimensional sphere of radius $\sqrt{nP}$ and a maximum likelihood decoder.
Tan and Tomamichel~\cite{tan2015Third} 
use the same distribution and decoder
to prove the achievability of a maximal rate of 
\begin{equation}
C(P) - \sqrt{\frac{V(P)}{n}} Q^{-1}(\epsilon)  	+ \frac 1 2 \frac{\log n}{n} + \bigo{\frac 1 {n}} \label{eq:tantomamichel3}
\end{equation}
at blocklength $n$, 
average error probability $\epsilon$, and maximal power $P$,
where 
\begin{equation}
	V(P) =  \frac{P(P + 2)}{2(1 + P)^2} \label{eq:dispersion}
\end{equation}
is the \emph{dispersion} of the point-to-point Gaussian channel; 
Polyanskiy \emph{et al.} prove a matching converse 
in~\cite{polyanskiy2010Channel}. Under a maximal-error constraint, the first- and second-order terms in \eqref{eq:tantomamichel3} remain the same under both maximal- and average-power constraints across codewords; under an average-error constraint, average- and maximal-power constraints yield different first- and second-order performance limits \cite[Ch.~4]{polyanskiy2010thesis}. In this paper, we only consider average-error and maximal-power constraints.
	
	MolavianJazi and Laneman~\cite{molavianjazi2015second} 
    and Scarlett \emph{et al.}~\cite{scarlett2015constantcompMAC} 
    generalize the asymptotic expansion in \eqref{eq:tantomamichel3} to the two-transmitter Gaussian MAC, 
    bounding the achievable rate 
    as a function of the $3 \times 3$ dispersion matrix $\ms{V}(P_1, P_2)$, 
    an analogue of $V(P)$ assuming transmitters 
    with per-codeword power constraints $P_{1}$ and $P_{2}$.
    The bound in \cite{molavianjazi2015second} 
    uses codewords uniformly distributed on the power sphere 
    and threshold decoding 
    based on the \emph{information density}; 
    the bound in \cite{scarlett2015constantcompMAC} 
    uses constant composition codes 
    and a quantization argument for the Gaussian channel. 
    This paper improves those bounds using 
    codewords uniformly distributed on the power sphere 
    and a maximum likelihood decoding rule.
		
    The literature on RAC communications 
    includes works like~\cite{dyachkov1981mac, mathys1990class,bassalygo1983restricted}, 
    where the number of active transmitters is known, 
    and~\cite{ordentlich2017mac}, where 
    neither the transmitters nor the receiver knows 
    the number of active transmitters. In~\cite{ordentlich2017mac}, 
    Ordentlich and Polyanskiy propose a concatenated code with a linear inner code that detects the number of active users and an outer code that decodes their messages.
    A two-layer code for joint erasure correction and collision resolution appears in~\cite{ebrahimi2017coded}. 

Recently, RACs with massive numbers of users
have attracted significant attention. 
The Gaussian ``many access'' channel, 
with a total number of users, $K$, 
that grows with the blocklength, $n$, as $K=O(n)$, 
is considered in~\cite{ordentlich2017mac, chen2014many, chen2017capacity}. 
Chen and Guo \cite{chen2014many} find the capacity 
of the Gaussian many access channel,
and Chen \emph{et al.} \cite{chen2017capacity} 
derive the capacity of the Gaussian many access channel 
in a random access scenario where the number of users, $K$, is unknown. 
For the criterion of average per-user error probability, Polyanskiy \cite{polyanskiy2017perspective} and Zadik \emph{et al.} \cite{zadik2019improved} derive non-asymptotic random coding achievability bounds when $K$ transmitters are active. 
Extensions of these ideas to 
quasi-static fading MACs and RACs
appear in \cite{kowshik2021fundamental} and \cite{kowshik2019energy}, respectively. 
Unlike \cite{ordentlich2017mac, chen2014many, chen2017capacity, polyanskiy2017perspective, zadik2019improved, kowshik2019energy, kowshik2021fundamental}, we here assume that $K$ does not grow with $n$.
    
    In \cite{yavas2020Random}, 
    we develop a communication strategy for a general RAC 
    where neither the transmitters 
    nor the receiver knows the set of active transmitters. 
    A central result of that work shows that for permutation-invariant RACs, 
    under mild conditions it is possible to achieve performance 
    identical in the first- and second-order terms 
    to the best performance known to be achievable 
    for the underlying MAC. 
    These results are obtained using a rateless coding scheme 
    where decoding occurs at one of a fixed collection of possible decoding times 
    $n_0, \dots, n_K$, and $K$ is the maximal number of transmitters. The chosen decoding time $n_t$ depends on the receiver's estimate $t$ 
    of the number of active transmitters. 
    At each decoding time, the receiver makes an attempt to decode 
    by applying a single threshold rule; 
    the receiver sends a single bit of feedback to all transmitters
    in order to specify when communication is completed. 
    In~\cite{liu2020finiteblocklength}, 
    Liu and Effros achieve improved third-order bounds
    using a maximum likelihood decoder under both i.i.d. random code and random LDPC code constructions. 
    Although the coding strategies proposed in~\cite{yavas2020Random, liu2020finiteblocklength}
    apply to the Gaussian RAC, 
    the random encoder design in \cite{yavas2020Random} uses an i.i.d. input distribution. 
    As shown in~\cite{molavianjaziThesis}, this codeword distribution 
    guarantees performance strictly inferior to that
    obtained when blocklength-$n$ codewords are uniformly distributed
    on the $n$-dimensional sphere of radius $\sqrt{nP}$. 

    Motivated by the desire to build superior RAC codes for Gaussian channels, we here propose a new code design for the Gaussian RAC. In the proposed code design, random codewords are designed by concatenating $K$ partial codewords
    of blocklengths $n_1, n_2-n_1, \dots, n_K - n_{K-1}$, 
    each drawn from a uniform distribution 
    on a sphere of radius $\sqrt{(n_i-n_{i-1})P}$.  
    When $k$ transmitters are active, 
    the resulting codewords are uniformly distributed 
    on a restricted subset 
    of the sphere of radius $\sqrt{n_{k}P}$. 
    The receiver uses output typicality to determine the number of transmitters 
    and then applies a maximum likelihood decoding rule. 
    Despite the restricted subset of codewords that result from our design, 
    we achieve the same first-, second-, and third-order performance 
    as the MAC code. While this paper focuses on Gaussian channels with maximal-power and average-error constraints, we note that the ideas developed here may be useful beyond this example channel and communication scenario.
    
    The use of codes with multiple possible decoding times arises in a variety of communication scenarios. 
    Codes that allow arbitrary decoding times are called rateless or variable-length codes. In \cite{burnashev1976data}, Burnashev finds the optimal error exponent of variable-length feedback (VLF) codes for discrete memoryless channels. Extending \cite{burnashev1976data} to the finite blocklength regime, Polyanskiy \emph{et al.} \cite{polyanskiy2011feedback} formalize VLF and variable-length feedback with termination (VLFT) codes for discrete memoryless channels. They show that the second-order term in the asymptotic expansion of the maximum achievable message size for VLF and VLFT codes is $O(\log{n})$, which means that the dispersion is zero.
    Truong and Tan \cite{truong2016gaussian} extend the zero-dispersion result for VLF and VLFT codes to the Gaussian point-to-point channel under an average-power constraint. In \cite{truong2018Journal}, Truong and Tan study VLF and VLFT codes for the Gaussian MAC. Building upon the results in the present paper, in \cite{yavas2021VLF} we derive an achievability bound for the Gaussian point-to-point channel with a maximal-power constraint, an average decoding delay constraint, and at most $K$ decoding times. The coding schemes in \cite{polyanskiy2011feedback, truong2016gaussian, truong2018Journal, yavas2021VLF} utilize so-called `stop-feedback' \cite{polyanskiy2011feedback}, i.e., feedback that is only used to direct the encoder to stop transmitting. While VLF and VLFT codes in \cite{burnashev1976data, polyanskiy2011feedback, truong2016gaussian, truong2018Journal, yavas2021VLF} employ multiple decoding times for a fixed number of transmitters, the Gaussian RAC codes studied in the present paper allow a single decoding time $n_k$ to decode messages from $k$ active transmitters. 

    The proposed coding problem is also related to the sparse recovery problem, which is a RAC problem in which each transmitter has only a single message to transmit; this message conveys the encoder's identity, and the decoder is tasked only with determining the identities of active transmitters. The group testing problem is a special case of the sparse recovery problem where the MAC output applies the logical OR operation to all inputs. For example, \cite{malyutov1978, atia2012, scarlettPhase2016, scarlett2017limits} study a group testing problem where an unknown subset of $k$ out of $K$ items is defective, and the decoder uses item signatures to identify which items are defective with an average probability of error approaching to zero. In the scenario where $k = O(1)$, Atia and Saligrama \cite{atia2012} show that the number of measurements (i.e., the blocklength $n$) required to identify $k$ out of $K$ items behaves as $O(k \log {\frac{K}{k}})$. In \cite{scarlettPhase2016}, Scarlett and Cevher extend this result to the scenario where $k = O(K^{\theta})$ and $\theta \in (0, 1)$; in \cite{scarlett2017limits}, they study a general channel model, which also covers the Gaussian MAC. In \cite{scarlettPhase2016, scarlett2017limits}, $2^k - 1$ information density threshold tests are used at the decoder, and a fixed number of defective items is considered. In the present paper, we combine a maximum likelihood decoding rule with a single threshold test based on the received power to decode messages from an unknown number of active transmitters.

	The organization of the paper is as follows. In \secref{sec:notation}, we define notation. The system model, main result, and discussions for the Gaussian MAC and Gaussian RAC appear in Sections \ref{sec:GaussianMAC} and \ref{sec:GaussianRACnotproof}. The proofs of the achievability bounds for the two-transmitter Gaussian MAC, the $K$-transmitter Gaussian MAC, and the Gaussian RAC appear in Sections \ref{sec:MACproof}, \ref{sec:proofMACK}, and \ref{sec:proof:nonasymRAC}--\ref{sec:proofRAC}. \secref{sec:conclusion} concludes the paper.

	\section{Notation} \label{sec:notation}
	
	We use bold symbols to denote vectors (e.g., $\vx$). For any integer $k \geq 1$, 
    we define $[k]\triangleq\{1,\ldots,k\}$. For any set $\mc{A}$, we denote by $\subemp{\mc{A}} \triangleq \{\mc{S} \subseteq \mc{A}, \mc{S} \neq \emptyset\}$ the set of non-empty subsets of $\mc{A}$. For any $\vx = (x_1, \dots, x_n) \in \mathbb{R}^n$ and $\mc{N} \subseteq [n]$, 
    $\vx^{\mc{N}} = (x_i \colon i \in \mc{N})$ denotes the sub-vector of $\vx$ 
    with components in $\mc{N}$. 
    For vectors $\vx_{1},\ldots,\vx_{K}$ of the same dimension 
    and index set $\mathcal{S} \in \subemp{[K]}$, 
    $\vx_{\mathcal{S}} = (\vx_s\colon s \in \mathcal{S})$. 
    and $\vx_{\langle \mc{S} \rangle} \triangleq \sum_{s \in \mc{S}} \vx_s$.
	Our notation for vectors and their collections is summarized in Table \ref{tab:notation}, below.
	
	For vectors $\mathbf{x} = (x_1, \dots, x_n)$ and $\mathbf{y} = (y_1, \dots, y_n)$, we write $
	\mathbf{x} \stackrel{\pi}{=} \mathbf{y} $ if there exists a permutation $\pi$ of the elements of $\mathbf{y}$ such that $\mathbf{x} = \pi\left(\mathbf{y}\right)$, and
	$\mathbf{x} \stackrel{\pi}{\neq} \mathbf{y} $ if $\mathbf{x} \neq \pi\left(\mathbf{y}\right)$ for all permutations $\pi$ of the elements of $\mathbf{y}$. We denote the inner product of $\mathbf{x}$ and $\mathbf{y}$ by 
    $\langle \mathbf{x}, \mathbf{y} \rangle = \sum_{i = 1}^n x_i y_i$ 
    and the Euclidean norm of $\mathbf{x}$ 
    by $\norm{\mathbf{x}} \triangleq \sqrt{\langle \mathbf{x}, \mathbf{x} \rangle}$. Vector inequalities are understood element-wise, i.e., $\mathbf{x} \leq \mathbf{y}$ if and only if $x_i \leq y_i$ for all $i \in [n]$. 
	All-zero and all-one vectors are denoted by $\bm{0}$ and $\bm{1}$, respectively.

	Matrices are denoted by sans serif font (e.g., $\mathsf{A}$). 
	The $n \times n$ identity matrix is denoted by $\mathsf{I}_n$. 
	Logarithms and exponents are base $e$.  The indicator function is denoted by $1\cB{\cdot}$. For any scalar function $f(\cdot)$ and any vector $\mathbf{x} \in \mathbb{R}^n$, we form the vector of function values $f(\mathbf{x}) = (f(x_i)\colon i \in [n])$. For a set $\mathcal{D} \subseteq \mathbb{R}^n$, a vector $\mathbf{c} \in \mathbb{R}^n$, and a scalar $a$, $a\mathcal{D} + c \triangleq \{a\mathbf{x} + \mathbf{c}: \mathbf{x} \in \mathcal{D}\}$. The sphere with dimension $n$, radius $r$, and center at the origin is denoted by $\mathbb{S}^{n}(r)  \triangleq \{ \vx \in \mathbb{R}^n: \norm{\vx} = r\}$.
	
	The distribution of a random variable $X$ is denoted by $P_X$. We write $P_X \to P_{Y|X} \to P_Y$ to indicate that $P_Y$ is the marginal distribution of $P_X P_{Y|X}$. To indicate that the random variables (or vectors) $X$ and $Y$ are identically distributed, we write $X\sim Y$. The multivariate Gaussian distribution with mean $\bm{\mu}$ and covariance matrix $\mathsf{\Sigma}$ is denoted by $\mathcal{N}(\bm{\mu}, \mathsf{\Sigma})$. We employ the complementary Gaussian cumulative distribution function $Q(x) = \frac{1}{\sqrt{2 \pi}} \int_{x}^{\infty} \exp\cB{-\frac{t^2}{2}} dt$. The functional inverse of $Q(\cdot)$ is denoted by $Q^{-1}(\cdot)$. 
	
	We use big-O notation $f(n) = O(g(n))$ if and only if there exist constants $c$ and $n_0$ such that $|f(n)| \leq c |g(n)|$ for all $n > n_0$; we use little-o notation $f(n) = o(g(n))$ if and only if for every $\epsilon > 0$, there exists a constant $n_0$ such that $|f(n)| \leq \epsilon |g(n)|$ for all $n > n_0$. 
	
    \begin{table}[!htbp]
    \centering
    \caption{Vector Notation Example}
    \begin{tabular}{ | m{3.7cm}| m{4.2cm} |} 
    \hline
    Notation & Description\\ 
    \hline
    $\vx_s = (x_{s, 1}, \dots, x_{s, n})$ & The length-$n$ vector that is a member of a collection indexed by {$s \in \mc{S}$}\\
    \hline
    $\vx_{\mc{S}} = (\vx_s \colon s \in \mc{S}) $ & The size-$|\mc{S}|$ ordered collection of length-$n$ vectors \\
    \hline
    $\vx_{\mc{S}}^{\mc{N}} = ((x_{s,t}\colon t \in \mc{N}) \colon s \in \mc{S})$ & The size-$|\mc{S}|$ ordered collection of length-$|\mc{N}|$ vectors with time indices in $\mc{N} \subseteq [n]$  \\
    \hline
    $\vx_{\langle \mc{S} \rangle} = \sum_{s \in \mc{S}} \vx_s$ & Summation of length-$n$ vectors from the collection $\mc{S}$\\
    \hline
    \end{tabular}
     \label{tab:notation}
    \end{table}
	
	\section{An RCU Bound and its Analysis for the Gaussian MAC} \label{sec:GaussianMAC}
	\subsection{An RCU Bound for General MACs}
	We begin by defining a two-transmitter MAC code.
		\begin{definition}\label{def:MACcodegen}
		An $(M_1, M_2, \epsilon)$-MAC code
		 for the channel with transition law $P_{Y_2|X_1 X_2}$ 
		 consists of two encoding functions
		 $\mathsf{f}_1\colon [M_1] \to \mathcal{X}_1$ and 
		 $\mathsf{f}_2\colon [M_2] \to \mathcal{X}_2$ 
		 and a decoding function 
		 $\mathsf{g}\colon \mathcal{Y}_2 \to [M_1] \times [M_2]$ 
		 such that 
		\begin{align}
		\frac{1}{M_1 M_2} \sum_{m_1 = 1}^{M_1} \sum_{m_2 = 1}^{M_2} &\bbP \bigl[ \mathsf{g}(Y_2) \neq (m_1, m_2) \mid \notag\\
		& (X_1, X_2) =  (\mathsf{f}_1(m_1), \mathsf{f}_2(m_2)) \bigr] \leq \epsilon,
		\end{align}
		where $Y_2$ is the channel output 
		under inputs $X_1$ and $X_2$, 
		and $\epsilon$ is the average-error constraint.  
	\end{definition}
	
	We define the information densities for a MAC with channel transition law $P_{Y_2|X_1 X_2}$ as
	\begin{IEEEeqnarray}{rCl}
	\IEEEyesnumber
	\IEEEyessubnumber*
	\imath_1(x_1; y | x_2) &\triangleq& \log \frac{P_{Y_2|X_1 X_2}(y|x_1, x_2)}{P_{Y_2|X_2}(y|x_2)} \label{eq:imath1}\\
	\imath_2(x_2; y | x_1) &\triangleq& \log \frac{P_{Y_2|X_1 X_2}(y|x_1, x_2)}{P_{Y_2|X_1}(y|x_1)} \label{eq:imath2} \\
	\imath_{1, 2}(x_1, x_2; y) &\triangleq& \log \frac{P_{Y_2|X_1 X_2}(y|x_1, x_2)}{P_{Y_2}(y)}, \label{eq:imath12}
	\end{IEEEeqnarray}
	where $P_{X_1}$ and $P_{X_2}$ are the channel input distributions, and $P_{X_1}P_{X_2} \to P_{Y_2|X_1 X_2} \to P_{Y_2}$. The information density random vector is defined as
		\begin{align}
		\boldsymbol{\imath}_2 \triangleq  \begin{bmatrix} \imath_1(X_1; Y_2 | X_2) \\ \imath_2(X_2; Y_2 | X_1) \\ \imath_{1, 2}(X_1, X_2; Y_2) \end{bmatrix}, \label{eq:i2vector}
		\end{align}
		where $(X_1, X_2, Y_2)$ is distributed according to $P_{X_1} P_{X_2} P_{Y_2|X_1 X_2}$.
	
Theorem~\ref{thm:RCUMac}, below, generalizes Polyanskiy \emph{et al.}'s
random-coding union (RCU) achievability bound ~\cite[Th. 16]{polyanskiy2010Channel} to the MAC.  
Theorem~\ref{thm:RCUMac} is derived by Liu and Effros \cite{liu2020finiteblocklength}
in their work on LDPC codes and is inspired by a new RCU bound for the 
Slepian-Wolf setting \cite[Th.~2]{chen2019lossless}. Its proof
combines random code design and a maximum likelihood decoder, which decodes to the message pair with the maximum information density $\imath_{1, 2}(X_1, X_2; Y_2)$. 
Our main result on the Gaussian MAC, 
Theorem~\ref{thm:MAC}, below, 
analyzes the RCU bound 
with $P_{X_1}$ and $P_{X_2}$ uniform on the power spheres. 
		\begin{theorem}[RCU bound for the MAC {\cite[Th.~6]{liu2020finiteblocklength}}]\label{thm:RCUMac}
		Fix input distributions $P_{X_1}$ and $P_{X_2}$.  
		Let 
		\begin{IEEEeqnarray}{rCl}
		    \IEEEeqnarraymulticol{3}{l}{P_{X_1, \bar{X}_1, X_2, \bar{X}_2, Y_2}(x_1, \bar{x}_1, x_2, \bar{x}_2, y)} \notag \\
		    &&= P_{X_1}(x_1)P_{X_1}(\bar{x}_1)P_{X_2}(x_2) P_{X_2}(\bar{x}_2)P_{Y_2|X_1 X_2}(y|x_1, x_2). \IEEEeqnarraynumspace
		\end{IEEEeqnarray}
		There exists an $(M_1, M_2, \epsilon)$-MAC code for $P_{Y_2 | X_1 X_2}$ such that
		\begin{align}
		\epsilon &\leq \mathbb{E}\Bigl[\min\Bigl\{1, \notag \\
		& (M_1-1) \bbP \bigl[\imath_1(\bar{X}_1; Y_2 | X_2) \geq \imath_1(X_1; Y_2 | X_2) \mid X_1, X_2, Y_2 \bigr] \notag \\
		+& (M_2-1)\bbP \bigl[\imath_2(\bar{X}_2; Y_2 | X_1) \geq \imath_{2}(X_2; Y_2 | X_1) \mid  X_1, X_2, Y_2 \bigr] \notag \\ +&(M_1-1)(M_2-1) \bbP \bigl[\imath_{1,2}(\bar{X}_1, \bar{X}_2; Y_2) \geq \imath_{1, 2}(X_1, X_2; Y_2) \notag \\
		&  \mid X_1, X_2, Y_2\bigr]\Bigr\}\Bigr]. \label{eq:RCUmac}
		\end{align}
	\end{theorem}
	\begin{remark}\label{rem:RCU}
	As noted in~\cite{liu2020Arxiv},
	\thmref{thm:RCUMac} generalizes to the $K$-transmitter MAC. Define the conditional information densities for the $K$-transmitter MAC as
	\begin{align}
	\imath_{\mc{S}}(x_{\mc{S}}; y| x_{\mc{S}^c}) \triangleq \log \frac{P_{Y_K|X_{[K]}}(y | x_{[K]})}{P_{Y_K|X_{\mc{S}^c}}(y | x_{\mc{S}^c})},
	\end{align}
    where $\mc{S} \subset [K]$, $\mc{S} \neq \emptyset$, and $\mc{S}^c = [K] \setminus \mc{S}$, and the unconditional information density as
    \begin{align}
    \imath_{[K]}(x_{[K]}; y) \triangleq \log \frac{P_{Y_K|X_{[K]}}(y | x_{[K]})}{P_{Y_K}(y)}.
    \end{align}
    Following arguments identical to those in the proof of \thmref{thm:RCUMac}, inequality \eqref{eq:RCUmac} extends to the $K$-transmitter MAC as
    \begin{IEEEeqnarray}{rCl}
    \epsilon &\leq & \mathbb{E} \biggl[\min \bigg\{1, \sum_{\substack{\mc{S} \in \subemp{[K]}}} \bigg(\prod_{s \in \mc{S}} (M_s - 1) \bigg) \bbP\bigl[\imath_{\mc{S}}(\bar{X}_{\mc{S}}; Y_{K}| X_{\mc{S}^c}) \notag\\
    && \geq \imath_{\mc{S}}(X_{\mc{S}}; Y_K| X_{\mc{S}^c}) \mid X_{[K]}, Y_K \bigr] \bigg\} \biggr]. \label{eq:RCUKnonasymp}
\end{IEEEeqnarray}
	\end{remark}	
		
	\subsection{A Third-order Achievability Bound for the Gaussian MAC}
	
	We begin by modifying our code definition to incorporate 
	maximal-power constraints $(P_1,P_2)$ on the channel inputs.  
	Let $(\vX_1,\vX_2)$ and $\vY_2$ be the MAC inputs and output, respectively. 
	\begin{definition} \label{def:MAC}
	An $(n, M_1, M_2, \epsilon, P_1, P_2)$-MAC code 
	for a two-transmitter MAC 
	comprises encoding functions 
	$\mathsf{f}_1\colon [M_1] \to \mathbb{R}^{n}$ and 
	$\mathsf{f}_2\colon [M_2] \to \mathbb{R}^{n}$,
	and a decoding function
	$\mathsf{g}\colon \mathbb{R}^{n} \to [M_1] \times [M_2]$
	such that 
	\begin{IEEEeqnarray*}{rCl}
	\norm{\mathsf{f_i}(m_i)}^2 \leq n P_i && \forall i\in\{1,2\},\ m_i \in [M_i] \IEEEeqnarraynumspace \\
	\frac1{M_1 M_2} \sum_{m_1 = 1}^{M_1} \sum_{m_2 = 1}^{M_2} 
	\bbP 	\left[\mathsf{g}(\vY_2)\right. &  \neq & (m_1, m_2) \mid \\
	\left. \, (\vX_1,\vX_2) \right.& = & \left.(\mathsf{f}_1(m_1),\mathsf{f}_2(m_2)) \right] 
	\leq \epsilon.
	\end{IEEEeqnarray*}
	\end{definition}	
		
    The following notation 
    is used in presenting our achievability result for the Gaussian MAC with $k \geq 1$ transmitters. 
    Over $n$ channel uses, the channel has inputs $\vX_1, \dots, \vX_k \in \mathbb{R}^n$, 
    additive noise $\vZ \sim \mathcal{N}(\boldsymbol{0}, \mathsf{I}_n)$, 
    and output 
    \begin{align}
    \vY_k = \vX_{\langle[k]\rangle} + \vZ. \label{eq:ysum}
    \end{align}
    The channel transition law induced by \eqref{eq:ysum} can be written as
	\begin{align}
	P_{\vY_k|\vX_{[k]}}(\vy|\vx_{[k]}) =  \prod_{i = 1}^n P_{Y_k|X_{[k]}}(y_i|x_{1i}, \dots, x_{ki}), \label{eq:noisedensityblock}
	\end{align}
	where
	\begin{align}
	P_{Y_k|X_{[k]}}(y|x_{[k]}) = \frac{1}{\sqrt{2\pi}} \exp\cB{-\frac{\nB{y-x_{\langle [k] \rangle}}^2}{2}}. \label{eq:noisedensity}
	\end{align}
    When $\vZ \sim \mc{N}(\bs{0}, \ms{V})$,
    and $\ms{V}$ is a $d \times d$ positive semi-definite matrix, 
    the multidimensional analogue 
    of the inverse $Q^{-1}(\cdot)$ of the complementary Gaussian cumulative distribution is 
	\begin{align}
	\mc{Q}_{\mathrm{inv}}(\mathsf{V}, \epsilon) \triangleq \cB{ \mathbf{z} \in \mathbb{R}^d: \Prob{\mathbf{Z} \leq \mathbf{z}} \geq 1-\epsilon}. \label{eq:defqinv}
	\end{align}
	For $d = 1$, we have $Q^{-1}(\epsilon) = \min\{z \colon z \in Q_{\mathrm{inv}}(1, \epsilon)\}$. 
	
	Recall that $C(P)$ is the capacity function \eqref{eq:capacity}. 
	The capacity vector for the two-transmitter Gaussian MAC is defined as
	\begin{align}
	\mathbf{C}(P_1, P_2) &\triangleq \begin{bmatrix} C(P_1) \\ C(P_2) \\ C(\Ponetwo) \end{bmatrix}. \label{eq:capacityvector}
	\end{align}
    The dispersion matrix \cite[eq.~(25)]{molavianjazi2015second} for the two-transmitter Gaussian MAC is defined as
	\begin{align}
	&\mathsf{V}(P_1, P_2)  \notag \\
	& \triangleq \setlength\arraycolsep{2.5pt} \begin{bmatrix} V(P_1)& 			V_{1, 2}(P_1, P_2)  & V_{1, 12}(P_1, P_2)\\ 		V_{1, 2}(P_1, P_2) & V(P_2) & 			V_{2, 12}(P_1, P_2) \\ 
	V_{1, 12}(P_1, P_2) & V_{2, 12}(P_1, P_2)& V_{12}(P_1, P_2) \end{bmatrix}, \label{eq:dispersionmatrix}
	\end{align}
	where $V(P)$ is the dispersion function \eqref{eq:dispersion}, and 
	\begin{align}
	V_{1, 2}(P_1, P_2) &= \frac 1 2 \frac{P_1 P_2}{(1 + P_1)(1+ P_2)} \label{eq:v1and2} \\
	V_{i, 12}(P_1, P_2) &= \frac 1 2 \frac {P_i (2 + \Ponetwo)}{(1 + P_i)(1 + \Ponetwo)}, \quad i \in \{1, 2\} \label{eq:vi12} \\
	V_{12}(P_1, P_2) &= V(P_1 + P_2) + \frac{P_1 P_2}{(1 + \Ponetwo)^2}. \label{eq:v12}
	\end{align}
	The following theorem is the main result of this section.
	
	\begin{theorem}\label{thm:MAC}
		For any $\epsilon \in (0, 1)$ and any $P_1, P_2 > 0$, an $(n, M_1, M_2, \epsilon, P_1, P_2)$-MAC code for the two-transmitter Gaussian MAC exists provided that
		\begin{IEEEeqnarray}{rCl}
		\begin{bmatrix}\log M_1 \\ \log M_2 \\ \log M_1 M_2 \end{bmatrix} &\in& n\mathbf{C}(P_1, P_2) - \sqrt{n}  Q_{\mathrm{inv}}(\mathsf{V}(P_1, P_2), \epsilon) \notag \\
		&&+ \frac 1 2 \log n \boldsymbol{1} + O(1) \boldsymbol{1}. \label{eq:MACmainresult}
		\end{IEEEeqnarray}
	\end{theorem}
	\begin{IEEEproof}
    See \secref{sec:MACproof}.
	\end{IEEEproof}
	
		\thmref{thm:MAC} extends to the general $K$-transmitter Gaussian MAC. The definition of an $(n,M_{[K]}, \epsilon, P_{[K]})$-MAC code for the $K$-transmitter Gaussian MAC with message set sizes $M_1, \dots, M_K$ and power constraints $P_1, \dots, P_K$ is a natural extension of Definition~\ref{def:MAC}, which defines the two-transmitter MAC code. The following theorem bounds the achievable region for the $K$-transmitter Gaussian MAC.
	\begin{theorem}\label{thm:KMAC}
	For any $\epsilon \in (0, 1)$, and $P_i > 0$, $i \in [K]$, an $(n,M_{[K]}, \epsilon, P_{[K]})$-MAC code for the $K$-transmitter Gaussian MAC exists provided that
	\begin{IEEEeqnarray}{rCl}
	 \IEEEeqnarraymulticol{3}{l}{\nB{\sum_{s \in \mc{S}} \log M_s \colon \mathcal{S} \in \subemp{[K]}} \in n \mathbf{C}(P_{[K]})} \notag \\
	 \IEEEeqnarraymulticol{3}{r}{\quad -  {\sqrt{n}}  Q_{\mathrm{inv}}(\mathsf{V}(P_{[K]}), \epsilon) + \frac 1 {2} \log n \boldsymbol{1} + \bigo{1} \boldsymbol{1}}, \label{eq:MACmainresultK}
	\end{IEEEeqnarray}
	where $\mathbf{C}(P_{[K]})$ is the capacity vector 
	\begin{align}
	\mathbf{C}(P_{[K]}) &\triangleq \nB{C(\PS): {\mathcal{S}}\in \subemp{[K]}} \in \mathbb{R}^{2^{K} - 1} \label{eq:capacityvectorK},
	\end{align}
	and $\mathsf{V}(P_{[K]})$ is the $\nB{2^{K} - 1} \times \nB{2^{K} - 1}$ dispersion matrix with the elements $\mathsf{V}_{\mathcal{S}_1, \mathcal{S}_2}(P_{[K]})$, $\mathcal{S}_1, \mathcal{S}_2 \in \subemp{[K]}$, 
	given by 
	\begin{IEEEeqnarray}{rCl}
	\IEEEeqnarraymulticol{3}{l}{\mathsf{V}_{\mathcal{S}_1, \mathcal{S}_2}(P_{[K]})} \notag\\
	&\triangleq& \frac{P_{\langle{\mathcal{S}_1}\rangle} P_{\langle{\mathcal{S}_2}\rangle} + 2 P_{\langle {\mathcal{S}_1 \cap \mathcal{S}_2}\rangle} + \nB{P_{\langle{\mathcal{S}_1 \cap \mathcal{S}_2}\rangle}}^2 -   P_{\langle \mathcal{S}_1 \cap \mathcal{S}_2 \rangle }^2}{2 (1 + P_{\langle \mathcal{S}_1 \rangle}) (1 + P_{\langle \mathcal{S}_2 \rangle}) }. \label{eq:VPvec} \IEEEeqnarraynumspace
	\end{IEEEeqnarray}
	\end{theorem}
\begin{IEEEproof}
See \secref{sec:proofMACK}.
\end{IEEEproof}	
	
	Before concluding this section, we make several remarks about Theorems~\ref{thm:MAC} and \ref{thm:KMAC} above.
	\begin{enumerate}
	\item Theorems \ref{thm:MAC} and \ref{thm:KMAC} 
    apply the RCU bound (\thmref{thm:RCUMac}) 
    with independent inputs uniformly distributed 
    on the $n$-dimensional origin-centered spheres with radii $\sqrt{nP_i}$, $i \in [K]$.
    \thmref{thm:MAC} matches 
    the first- and second-order terms 
    of MolavianJazi and Laneman \cite{molavianjazi2015second} and Scarlett \emph{et al.} \cite{scarlett2015constantcompMAC}, 
    and improves the third-order term 
    from $O\left(n^{1/4}\right)\bs{1}$ in \cite{molavianjazi2015second} and $O\left(n^{1/4} \log n\right) \bs{1}$ in \cite{scarlett2015constantcompMAC} 
    to $\frac 1 2 \log n \boldsymbol{1} + O(1) \bs{1}$.
	
	\item Our proof technique in \thmref{thm:MAC} differs from the technique in \cite{molavianjazi2015second} in two key ways. First, we use a maximum likelihood decoder 
    in place of the set of simultaneous threshold rules 
    based on unconditional and conditional information densities from \cite{molavianjazi2015second}; 
    the change of the decoding rule is essential for obtaining 
    the third-order term $\frac 1 2 \log n \bs{1} + O(1) \bs{1}$ in \thmref{thm:MAC}.
    Second, we refine the analysis 
    bounding the probability 
    that the information density random vector $\bs{\imath}_2$ belongs to a set $\mc{D} \subseteq \mathbb{R}^3$. Our non-i.i.d. input distribution prevents direct application of 
    the Berry-Esseen theorem. However, given that the inner product of the inputs $\langle \vX_1, \vX_2 \rangle$ equals a constant, the information density random vector $\bs{\imath}_2$ can be written as a sum of independent random vectors. Therefore, we apply the Berry-Esseen theorem after conditioning on the inner product $\langle \vX_1, \vX_2 \rangle$ and then integrate the resulting probabilities over the range of the inner product. In order to approximate the resulting probability by the probability that a Gaussian vector belongs to the same set, we use a result (\lemref{lem:gotze} in \secref{sec:tools}, below) that approximates the normalized inner product $\frac{1}{\sqrt{n P_1 P_2}} \langle \vX_1, \vX_2 \rangle $ by a standard Gaussian random variable. We then derive a bound  (\lemref{lem:gaussiancdf} in \secref{sec:tools}, below) on the total variation distance between two Gaussian vectors. This analysis appears in \secref{sec:be}.
	
This approach contrasts with \cite{molavianjazi2015second}, which bounds the probability that the information density random vector $\bs{\imath}_2$ belongs to a set $\mc{D}$. Writing $\bs{\imath}_2$ as a vector-valued function of an average of i.i.d. Gaussian vectors, \cite[Prop.~1]{molavianjazi2015second} applies a central limit theorem for functions of sums to prove $\bigo{\frac{1}{n^{1/4}}}$ convergence to normality. Our technique, described above, improves the rate of convergence to normality to $\bigo{\frac{1}{\sqrt{n}}}$, which is the rate of convergence for i.i.d. sums. This improvement implies that the threshold-based decoding rule in \cite{molavianjazi2015second} achieves a third-order term $O(1)\bs{1}$.
	
	\item Our technique for proving Theorems~\ref{thm:MAC} and \ref{thm:KMAC} parallels those used for non-singular discrete memoryless channels \cite[Th. 53]{polyanskiy2010thesis} and for the point-to-point Gaussian channel \cite{tan2015Third}. In \cite[Th. 53]{polyanskiy2010thesis}, Polyanskiy applies the RCU bound using a refined large deviations result \cite[Lemma 47]{polyanskiy2010Channel}; the use of a non-i.i.d. input distribution for the Gaussian channel prevents the direct application of \cite[Lemma 47]{polyanskiy2010Channel}. In \cite[eq. (52)]{tan2015Third}, Tan and Tomamichel derive an alternative to \cite[Lemma 47]{polyanskiy2010Channel} for the point-to-point Gaussian channel in order to accommodate codewords drawn uniformly on an $n$-dimensional sphere. While evaluating the RCU bound in this paper, we extend the bound in \cite[eq. (52)]{tan2015Third} to the Gaussian MAC. 
	
	\item For the symmetric setting, where $P_i = P$ and $M_i = M$ for all $i \in [K]$, \thmref{thm:KMAC} reduces to the scalar inequality, below. This result refines the result in \cite[Th. 2]{molavianjazi2015second} to the third-order
    term and generalizes it to the $K$-transmitter MAC.
	\begin{corollary}\label{cor:MACsym}
		For any $\epsilon \in (0, 1)$ and $P > 0$, an $(n, M \bs{1},  \epsilon, P \bs{1})$-MAC code for the $K$-transmitter Gaussian MAC exists provided that
		\begin{align}
		&K \log M \leq  n C(K P) \notag \\
		& - \sqrt{n (V(K P) + V_{\textnormal{cr}}(K, P))} Q^{-1}(\epsilon) + \frac 1 2 \log n + O(1). \label{eq:corollaryMAC}
		\end{align}
		Again, $C(\cdot)$ and $V(\cdot)$ are the capacity \eqref{eq:capacity} and dispersion \eqref{eq:dispersion} functions, respectively, and $V_{\textnormal{cr}}(K, P)$ is the cross dispersion term
		\begin{align}
		V_{\textnormal{cr}}(K, P) \triangleq \frac{K (K-1)P^2}{2 (1 + K P)^2}. \label{eq:crossdispersion}
		\end{align}
	\end{corollary}
	\begin{IEEEproof}
	See \appref{app:proofcorMAC}.
	\end{IEEEproof}
	 
	 \item
	 In~\cite{fong2016gaussianstrong}, Fong and Tan
	derive a converse for the Gaussian MAC with a second-order term $O(\sqrt{n \log n}) \bs{1}$. Kosut \cite{kosut2020converse} improves the second-order term in the converse to $\bigo{\sqrt{n}} \bs{1}$.
	The coefficients of the second-order term in \cite{kosut2020converse} do not match the second-order term 
	in the achievability bounds proven in \thmref{thm:MAC}.
	As discussed in~\cite{yavas2020Random}, closing the gap between the second-order terms of the MAC achievability and converse results
	is a challenging open problem.
	
	\end{enumerate}
	
	\section{A Nonasymptotic Bound and its Analysis for the Gaussian Random Access Channel} \label{sec:GaussianRACnotproof}
	\subsection{System Model}
	\emph{Channel model}:	
    Given an unknown set of active  transmitters~$\mc{A}$, 
	the Gaussian channel \eqref{eq:noisedensity} depends on $\mc{A}$ only through the number of active transmitters, $|\mc{A}| = k$, that is, $P_{Y_{k} | X_{\mc{A}}} = P_{Y_k | X_{[k]}}$. Therefore, in order to capture the scenario 
 	of a memoryless Gaussian channel 
 	with $K$ possible transmitters, 
 	a single receiver, 
 	and an unknown activity pattern $\mc{A}\subseteq[K]$,
	we describe the Gaussian RAC 
	by a family of Gaussian MACs $\{P_{Y_k|X_{[k]}}\}_{k = 0}^K$ \eqref{eq:noisedensity}, 
	each indexed by the number of active transmitters $k \in \{0, \dots, K\}$. This RAC model is introduced in \cite{yavas2020Random} for general channels satisfying permutation-invariance and reducibility assumptions; the Gaussian RAC satisfies these assumptions.
	As in \cite{yavas2020Random}, we choose a \emph{compound channel} model 
	in order to avoid the need to assign probabilities to each activity pattern $\mathcal{A}$.
	
	\emph{Communication strategy}:
    We apply the epoch-based \emph{rateless} communication strategy that
    we proposed in \cite{yavas2020Random}.
    Each transmitter is either active or silent during a whole epoch. 
    At each of times $n_0, n_1, \dots$, the decoder broadcasts to all transmitters a single bit --- sending value 1 if it can decode and 0 otherwise.
    The transmission of 1 at time $n_t$ ends the current epoch and starts the next, indicating that the
    decoder's estimate of the number of transmitters is $t$.
    As in \cite{polyanskiy2017perspective, yavas2020Random}, 
    we employ identical encoding, 
    with each active transmitter $i$ using the same encoding function 
    to describe its message $W_i\in[M]$. 
    Identical encoding here requires $P_i=P$ and $M_i=M$ for all $i$.
    The task of the decoder is to decode a list of messages 
    sent by the active transmitters $\mc{A}$ but not the identities of those transmitters. 
    The messages in $W_{\mc{A}}$ are independent and uniformly distributed on alphabet $[M]$. 
    
    Since encoding is identical and the channel is invariant to permutation of its inputs, 
    we assume without loss of generality that $|\mc{A}|=k$ implies $\mc{A} = [k]$.
    Intuitively, given identical encoding and our Gaussian channel,
    one would expect that interference increases with the number of active transmitters $k$, 
    and therefore that the decoding time $n_k$ increases with $k$. Since the capacity per transmitter for the $k$-transmitter Gaussian MAC, $\frac{1}{k}{C(kP)}$, decreases with $k$, we can choose $n_0 < \dots < n_K$ for $M$ large enough. 
    (See~\cite[Lemma~1]{yavas2020Random} for more general sufficient conditions under which $n_0<\dots<n_K$ is optimal.) As a notational convenience, we use $n_K$ to represent the largest decoding time.  
    Unless it stops the encoders' transmissions earlier, at time $n_K$, the decoder sees 
	\begin{align}
	\vY_k = \vX_{\langle [k] \rangle} + \vZ \in \mathbb{R}^{n_K} \quad \mbox{ for } k \in [K], \label{eq:yRAC}
	\end{align}
	where $\vX_1, \dots, \vX_k$ are $n_K$-dimensional channel inputs, 
	$\vZ \sim \mathcal{N}(\boldsymbol{0}, \mathsf{I}_{n_K})$ is the Gaussian noise, 
	and $\vY_k$ is the $n_K$-dimensional output when $k$ transmitters are active. 
	When no transmitters are active, $\vX_{\langle[0] \rangle} = \bs{0}$ and $\vY_0 = \vZ$.  
	At each time $n_t < n_K$, the decoder has access to the first $n_t$ components of vector $\vY_k$, which is denoted by $\vY_k^{[n_t]}$.
	
	As in \cite{yavas2020Random}, we assume an \emph{agnostic} random access model, 
	where the transmitters know nothing about the set $\mathcal{A}$ of active transmitters 
	except their own membership and the feedback from the receiver. The receiver knows nothing about $\mathcal{A}$ except what it can learn from the channel output $\vY_k$. 
	
	\emph{Code definition}: The following definition formalizes the rateless Gaussian RAC code described above.
	\begin{definition}\label{def:RAC}
	    An $\left( \{n_j, \epsilon_j\}_{j = 0}^K, M, P\right)$-RAC code 
		for the Gaussian RAC with $K$ transmitters 
		consists of a single encoding function 
		$\mathsf{f}\colon \mathcal{U} \times [M] \to \mathbb{R}^{n_K}$ 
		and decoding functions
		$\mathsf{g}_k \colon \mathcal{U} \times \mathbb{R}^{n_k} \to [M]^k \cup \{ \mathsf{e} \}$ for $k = 0, \dots, K$, where the input $u \in \mc{U}$ to the encoder and decoders is common randomness shared by all transmitters and the receiver.\footnote{The
		realization $u$ of the common randomness random variable $U$ 
		initializes the encoders and the decoder. At the start of each communication epoch, 
		$u$ is shared by all transmitters and the receiver. 
		We show in \cite[Appendix D]{yavas2020Random} that the alphabet size of $U$ 
		need never exceed $K + 1$.} If it cannot decode at time $n_k$, the decoder outputs the erasure symbol ``$\mathsf{e}$" and broadcasts value 0 to the transmitters, informing them that they should keep transmitting. If it can decode at time $n_k$, the decoder broadcasts value 1 to the transmitters, informing them that they should stop transmitting.
		The codewords satisfy the maximal-power constraints
		\begin{align}
		&\norm{\mathsf{f}(u, m)^{[n_j]}}^2 \leq n_j P \mbox{ for } m \in [M], u \in \mathcal{U}, j \in [K]. \label{eq:powerRAC}
		\end{align}
		If $k$ transmitters are active, then the average error probability 
		in decoding $k$ messages at time $n_k$ is bounded as 
		\begin{align}
		&\frac{1}{M^k} \sum_{m_{[k]} \in [M]^k} \bbP \biggl[ \bigg\{\bigcup_{t \colon n_t \leq n_k, t \neq k} \cB{\mathsf{g}_t(U, \vY_k^{[n_t]}) \neq \mathsf{e}}\bigg\} {{\bigcup}} \notag\\
		&\bigg\{ \mathsf{g}_k(U, \vY_k^{[n_k]}) \stackrel{\pi}{\neq} m_{[k]} \bigg\} \bigg| \vX_{[k]}^{[n_k]} = \mathsf{f}(U, m_{[k]})^{[n_k]} \biggr] \leq \epsilon_k, \label{eq:errorprobcons}
		\end{align}
		where $\mathsf{f}(U, m_i)$ is the codeword for the message $m_i \in [M]$, 
		$U$ is the common randomness random variable, 
		and the output  $\vY_k$ is generated according to \eqref{eq:yRAC}. 
		If no transmitters are active, then the decoder decodes to the unique message $[M]^0 \triangleq \{0\}$ 
		with probability of error bounded as 
		\begin{align}
		\Prob{\mathsf{g}_0(U, \vY_0^{[n_0]}) \neq 0} \leq \epsilon_0.
		\end{align}
	\end{definition}

\subsection{A Third-order Achievability Result for the Gaussian RAC}
	The following theorem is the main result of this section. 
	
	\begin{theorem}\label{thm:GRAC}
	Fix $K < \infty$, $\epsilon_k \in (0, 1)$ for $k \in \{0\} \cup [K]$, and $M$. An $\left( \{n_j, \epsilon_j\}_{j = 0}^K, M, P\right)$-RAC code exists for the Gaussian RAC with $K$ possible transmitters provided that
		\begin{IEEEeqnarray}{rCl}
		k \log M &\leq& n_k C(kP) -  \sqrt{n_k (V(kP) + V_{\textnormal{cr}}(k, P))}  Q^{-1}(\epsilon_k) \notag\\
		&& + \frac 1 2 \log n_k + O(1) \label{eq:RACmainresult} 
		\end{IEEEeqnarray}
for $k \in [K]$, and
\begin{align}
		n_0 &\geq \frac{4(1 + P^2)}{P^2} \log n_1 + o(\log n_1), \label{eq:n0C}
		\end{align}
	where $C(\cdot)$, $V(\cdot)$, and $V_{\textnormal{cr}}(\cdot, \cdot)$ are the capacity \eqref{eq:capacity}, dispersion \eqref{eq:dispersion}, and cross dispersion functions  \eqref{eq:crossdispersion}, respectively. All uses of $O(\cdot)$ and $o(\cdot)$ are taken with respect to $n_1$.
	\begin{remark} 
	From \eqref{eq:RACmainresult}, $n_1 \to \infty$ implies that $n_2, \dots, n_K$ also grow without bound. Since all target error values $\epsilon_k$ are assumed to be constants with respect to $n_1$, choosing decoding times $n_0, \dots, n_K$ so that \eqref{eq:RACmainresult} and \eqref{eq:n0C} hold with equality results in $n_k = O(n_1)$ for $k \geq 2$, and $n_0 = O(\log n_1)$ (see \eqref{eq:nkn1}, below).
	\end{remark}
	\end{theorem}
	
	\begin{IEEEproof}
	\thmref{thm:GRAC} follows from the non-asymptotic achievability bound in \thmref{thm:nonasymRAC}, below, which bounds the average error probability of the proposed Gaussian RAC code. See \secref{sec:proofRAC} for details.
\end{IEEEproof}

	\begin{theorem} \label{thm:nonasymRAC}
Fix constants $\lambda_k >0$ for $k \in \{0\} \cup [K]$ and distribution $P_{\vX}$ on $\mathbb{R}^{n_K}$. Then, there exists an $\left( \{n_j, \epsilon_j\}_{j = 0}^K, M, P\right)$-RAC code with
\begin{IEEEeqnarray}{rCl}
\epsilon_0 &\leq& \bbP\Bigl[\Big \lvert \Big \lVert  \vY_0^{[n_0]} \Big \rVert^2 - n_0 \Big\rvert > n_0 \lambda_0 \Bigr]\\
\epsilon_k &\leq& \frac{k (k-1)}{2 M} + \Prob{\bigcup_{i = 1}^k \bigcup_{\substack{j:n_j \leq n_k \\ j \geq 1}} \Big\{ \norm{\vX_i^{[n_j]}}^2 > n_j P\Big\}}  \IEEEyesnumber  \label{eq:ekbound37}
 \IEEEyessubnumber* \label{eq:repetitionpower} \\
&+&\bbP \Bigg[ \bigcup_{\substack{t: n_t \leq n_k \\ t \neq k}} \Big\{\Big\lvert  \norm{\vY_k^{[n_t]}}^2 - n_t(1 + tP) \Big \rvert \leq n_t\lambda_t \Big\}   \notag \\
&&\quad  \bigcup \Big\{ \Big\lvert \norm{\vY_k^{[n_k]}}^2 - n_k(1 + kP) \Big \rvert > n_k\lambda_k \Big\} \Bigg] \label{eq:decodingtimeerror}  \\
&+&\bbE \Bigl[\min \Big\{1, \sum_{s = 1}^k \binom{k}{s}  \binom{M-k}{s} \notag \\
&& \bbP\bigl[\imath_{[s]}(\bar{\vX}^{[n_k]}_{[s]}; \vY_{k}^{[n_k]} | \vX_{[s+1:k]}^{[n_k]})  \notag \\
&& \geq \imath_{[s]}(\vX_{[s]}^{[n_k]}; \vY_k^{[n_k]}| \vX_{[s+1:k]}^{[n_k]}) \mid \vX_{[k]}^{[n_k]}, \vY_k^{[n_k]} \bigr] \Big\} \Bigr]\label{eq:ekbound} \IEEEeqnarraynumspace
\end{IEEEeqnarray}
for all $k \in [K]$, 
where $\vX_{[K]},\bar{\vX}_{[K]},\vY_k\in\mathbb{R}^{n_K}$ 
are distributed according to 
${P_{\vX_{[K]}, \bar{\vX}_{[K]}, \vY_k}(\vx_{[K]}, \bar{\vx}_{[K]}, \vy_k) }= \Big(\prod_{j \in [K]} P_{\vX}(\vx_j) P_{\vX}(\bar{\vx}_j) \Big)   P_{\vY_k|\vX_{[k]}}(\vy_k | \vx_{[k]})$,
and $P_{\vY_k|\vX_{[k]}}$ is given in \eqref{eq:yRAC}.
\end{theorem}
\begin{IEEEproof}
The terms in \eqref{eq:repetitionpower} capture the probability that at least two transmitters send the same message and the probability of a power constraint violation, respectively. We treat the event that at least two transmitters send the same message as an error because the analysis relies on the codeword independence across transmitters. The probability in \eqref{eq:decodingtimeerror} captures the probability that the decoder decodes at a wrong decoding time, and the expectation in \eqref{eq:ekbound} captures the probability that the decoder decodes an incorrect message list at the correct decoding time $n_k$ for $k$ active transmitters. See \secref{sec:proof:nonasymRAC} for details.
\end{IEEEproof}

We conclude this section with some remarks concerning Theorems~\ref{thm:GRAC} and \ref{thm:nonasymRAC}.
\begin{enumerate}

\item \thmref{thm:GRAC} shows that for the Gaussian RAC, our proposed rateless code performs as well in the first-, second-, and third-order terms as the best known MAC communication scheme without feedback (Corollary \ref{cor:MACsym}). In other words, the first three terms on the right-hand side of \eqref{eq:RACmainresult} for $k$ active transmitters match the first three terms of the largest achievable sum-rate in our achievability bound in \eqref{eq:corollaryMAC}
for the $k$-transmitter MAC. 

\item To prove \thmref{thm:GRAC}, we particularize the distribution of the random codewords, $P_{\vX}$, in \thmref{thm:nonasymRAC} as follows. The first $n_1$ symbols are drawn uniformly from $\mathbb{S}^{n_1}(\sqrt{n_1P})$. The sub-vector of symbols indexed from $n_{j-1} + 1$ to $n_j$ is drawn uniformly from $\mathbb{S}^{n_j - n_{j-1}}(\sqrt{(n_j - n_{j-1})P})$ for $j = 2, \dots, K$. These $K$ sub-codewords, each uniformly distributed on an incremental power sphere, are independent. Under this $P_{\vX}$, the maximal-power constraint in \eqref{eq:powerRAC} is satisfied with equality for each number of active transmitters. Rather than using an encoding function that depends on the feedback from the receiver to the transmitters, we use an encoding function that is suitable for all possible transmitter activity patterns and does not depend on the receiver's feedback. Given that a decision is made at time $n_k$, the
 active transmitters have transmitted only the first $n_k$ symbols of the codewords representing their messages during that epoch, and the remaining $n_K - n_k$ symbols of the codewords are not transmitted. 

\item As noted in \cite{ordentlich2017mac}, our achievability proofs leverage the fact that the number of active transmitters can be reliably estimated from the total received power. This is possible because when $k$ active transmitters send $k$ distinct messages, the average received power $\frac{1}{n_k}\E{\norm{\vY_{k}^{[n_k]}}^2}$ at time $n_k$, 
concentrates around its mean value, $1 + kP$, and this mean is distinct for each $k \in \{0\} \cup [K]$. The decoding function used at time $n_k$ combines the maximum likelihood decoding rule for the $k$-transmitter MAC with a typicality rule based on the power of the output. The typicality rule decides to decode at time $n_k$ if the average received power at time $n_k$ lies in the interval $\left[ 1 + kP - \frac{P}{2}, 1 + kP + \frac{P}{2}\right]$ for $k \geq 1$ and $\left[ 1 - O({n_0}^{-\frac{1}{2}}), 1 + O({n_0}^{-\frac{1}{2}})\right]$ for $k = 0$. In this case, the decoder decodes $k$ messages at time $n_k$ by using the maximum likelihood decoding rule. When at least two transmitters send the same message (e.g., $\vX_1^{[n_k]} = \vX_2^{[n_k]}$), $\frac{1}{n_k}\E{\norm{\vY_{k}^{[n_k]}}^2} \geq 1 + (k+2)P$. In our decoder design, we choose not to handle this scenario because the probability that at least two transmitters send the same message is negligible as shown in \eqref{eq:erep} in \secref{sec:RACerroranalysis}, below.

\item \thmref{thm:nonasymRAC} applies without change to non-Gaussian RACs with power constraints satisfying the conditions in \cite[Th.~1]{yavas2020Random};  the tightness of the bound depends on how well $k$ can be estimated from the received power.

\item The proof of \thmref{thm:GRAC} indicates that the constant term $O(1)$ in \eqref{eq:RACmainresult} depends on the number of active transmitters $k$, but not on the total number of transmitters $K$. Not requiring the decoder to determine transmitter identity is crucial for this $O(1)$ bound to hold.

\item By choosing $n_1, \dots, n_K$ such that the inequalities in \eqref{eq:RACmainresult} are satisfied with equality for each $k$, we can express each $n_k$ as a function of $n_1, \epsilon_1$, $\epsilon_k$, $k$, and $P$, given by
    \begin{align}
        &n_k =  n_1 \frac{k C_1}{C_k} + \sqrt{n_1} \biggl(\frac{1}{C_k} \sqrt{\frac{k C_1 V_k}{C_k}} Q^{-1}(\epsilon_k) \notag \\
        &- \frac{k}{C_k} \sqrt{ V_1} Q^{-1}(\epsilon_1) \biggr) + \frac{k-1}{2 C_k} \log n_1 + O(1), \label{eq:nkn1}
    \end{align}
    where $C_k = C(kP)$ and $V_k = V(kP) + V_{\textrm{cr}}(k, P)$. We derive \eqref{eq:nkn1} by computing the Taylor series expansion of the equation for $n_k$ \eqref{eq:RACmainresult} in terms of $k, P, \epsilon_k,$ and $\log M$; we then replace $\log M$ by \eqref{eq:RACmainresult} for $k = 1$. \figref{fig:nk} shows the approximate decoding times $\{n_k\}_{k = 1}^{6}$ (neglecting the $O(1)$ term in \eqref{eq:nkn1}), where $P = 1$, $\epsilon_k = 10^{-3}$ for all $k$, and the smallest decoding time $n_1 \in [100, 1000]$.
    
    \begin{figure}
\center
\includegraphics[width=1\linewidth]{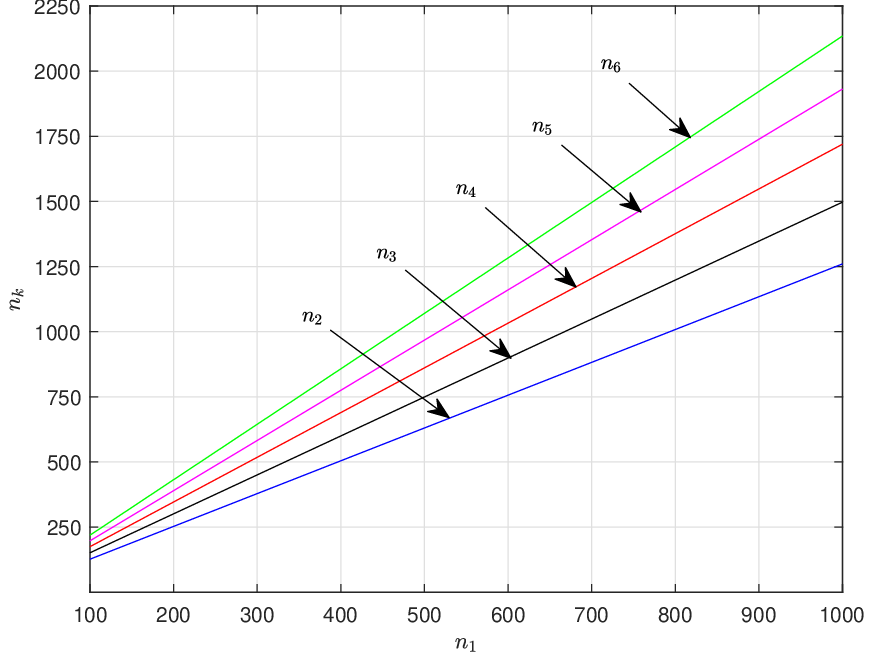}
\caption{Let $P = 1$, $\epsilon_k = 10^{-3}$ for all $k$. The decoding times $\{n_k\}_{k = 1}^{6}$ are given according to \eqref{eq:nkn1}, where the $O(1)$ term in \eqref{eq:nkn1} is ignored,} and $n_1 \in [100, 1000]$.
\label{fig:nk}
\end{figure}

\item \thmref{thm:GRAC} implies that the input distribution used for the Gaussian RAC also achieves the performance in \thmref{thm:KMAC} for the $K$-transmitter Gaussian MAC. As long as $n_j - n_{j-1} \geq c n_K$ holds for some constant $c > 0$ for all $j \in [K]$, requiring separate power constraints on each sub-block of the codewords as
\begin{align}
\norm{\mathsf{f}_i(m_i)^{[n_j]}}^2 \leq n_j P_i  \text{ for } m_i \in [M_i], i \in [K], j \in [K]
\end{align}
does not degrade our performance bound, which matches the first three terms in the expansion in \thmref{thm:KMAC}. The support of the distribution from which the codewords are drawn for the Gaussian MAC and RAC is illustrated in \figref{fig:sphere}. 

\begin{figure}
\center
\includegraphics[width=1\linewidth]{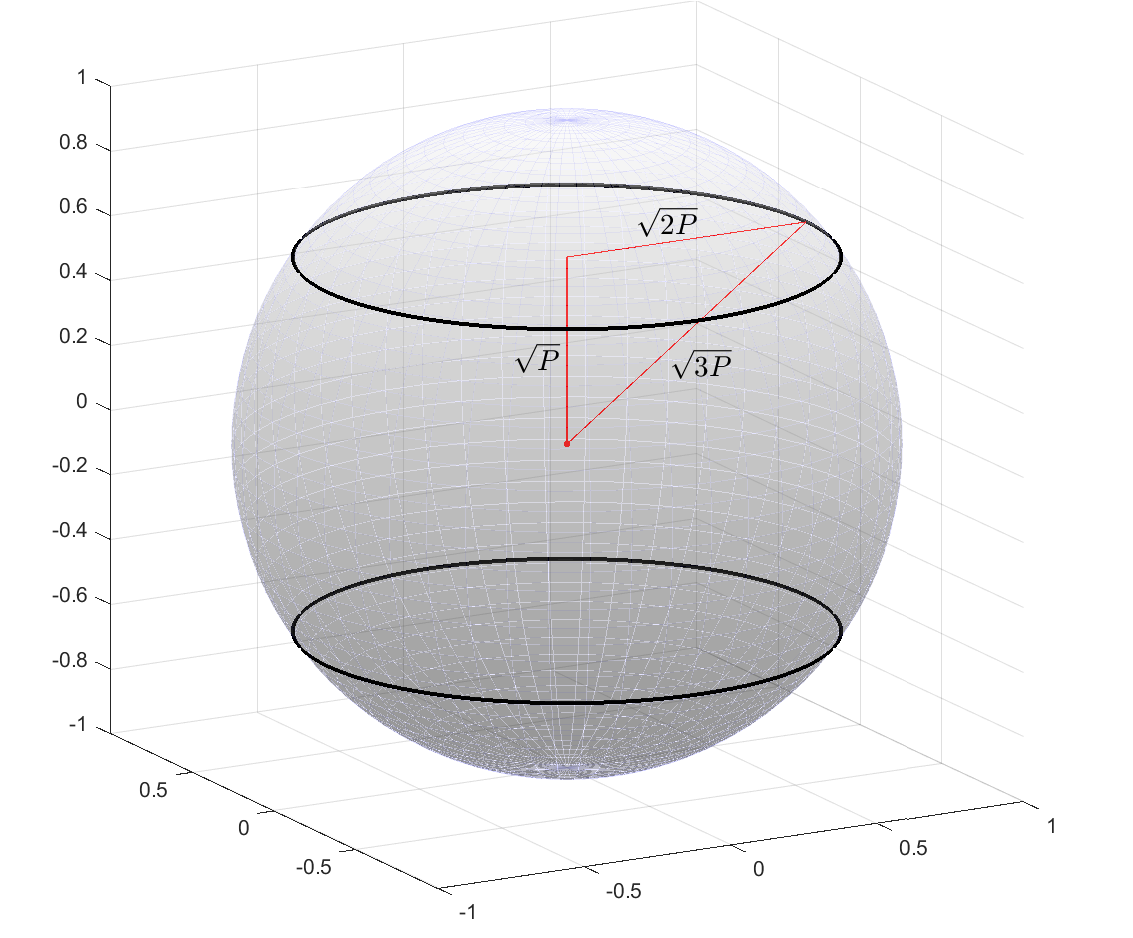}
\caption{Let $K = 2$, $n_1 = 2, n_2 = 3$, and $P_1 = P_2 = P = \frac 1 3$. 
The support of the input distribution for the Gaussian RAC
is the Cartesian product of 
$\mathbb{S}^{n_1}(\sqrt{n_1P})$ (here a circle with radius $\sqrt{2P}$) and 
$\mathbb{S}^{n_2 - n_1}(\sqrt{(n_2 - n_1)P})$
(here the set $\{-\sqrt{P}, \sqrt{P}\}$.) 
This set, shown above as a pair of circles, is a subset of $\mathbb{S}^{n_2}(\sqrt{n_2 P})$; the set $\mathbb{S}^{n_2}(\sqrt{n_2 P})$
is the support of the input distribution used in \thmref{thm:KMAC} for the Gaussian MAC.}
\label{fig:sphere}
\end{figure}

\item The coding strategy we propose in \cite[Th.~1]{yavas2020Random} requires an i.i.d. input distribution. One can also employ the coding strategy in \cite[Th.~1]{yavas2020Random} to the Gaussian MAC, drawing codewords i.i.d. from $\mc{N}(0, P')$ for some $P' = P - \delta_n$ and $\delta_n$ decaying to 0 sufficiently quickly as blocklength $n$ grows without bound, provided that we discard codewords that violate the maximal-power $P$ constraint. However, \cite[eq. (5.113)]{molavianjaziThesis} shows that the resulting achievable second-order term is inferior to that achieved by the uniform distribution on the sphere.

\item As described above, the number of active transmitters in an epoch is estimated via a sequence of decodability tests. An alternative strategy is to estimate the number of active transmitters in one shot from the received power at time $n_0$, and to inform the transmitters about the estimate, $t$, of the number of active transmitters via a $\lceil \log(K + 1) \rceil$-bit feedback at time $n_0$. Given this knowledge, active transmitters can employ an encoding function matched to $t$. We show in \appref{app:mod1} that this modified coding strategy affects our bound in \eqref{eq:RACmainresult} only in the $O(1)$ term.

\item By using distinct codebooks for each transmitter, the decoder can associate transmitter identities with the decoded messages. We show that the first three terms of the expansion in \eqref{eq:RACmainresult} are still achievable in this setting. This scenario is discussed in \appref{app:mod2}.

\end{enumerate}
	
	\section{Proof of \thmref{thm:MAC}} \label{sec:MACproof}
	\subsection{Supporting Lemmas} \label{sec:tools}
	We begin by presenting the lemmas that play a key role in the proof of \thmref{thm:MAC}. The first two lemmas are used to bound the probability that the squared norm of the output of the channel, $\vY_2 = \vXltwo + \vZ$, does not belong to its typical interval around $1 + \Ponetwo$. 
	
	\lemref{lem:kappa} from \cite[Prop.~2]{molavianjazi2015second} uniformly bounds the Radon-Nikodym derivative of the conditional and unconditional output distributions of the Gaussian MAC \eqref{eq:noisedensityblock} in response to the spherical inputs with respect to the output distributions that result under i.i.d. Gaussian inputs. The squared norm of the output in response to the i.i.d. Gaussian inputs has a chi-squared distribution.
	\begin{lemma}[MolavianJazi and Laneman {\cite[Prop.~2]{molavianjazi2015second}}] \label{lem:kappa}
	\begin{enumerate}
		\item 2-Transmitter MAC: Let $\vX_1$ and $\vX_2$ be independent, distributed uniformly on $\mathbb{S}^{n}(\sqrt{nP_1})$ and $\mathbb{S}^{n}(\sqrt{nP_2})$, respectively. Let $\tilde{\vX}_i \sim\mathcal{N}(\bs{0}, P_i \mathsf{I}_n)$, $i \in [2]$, be independent of each other. Let $P_{\vX_1 \vX_2} \to P_{\vY_2|\vX_1 \vX_2} \to P_{\vY_2}$, and $P_{\tilde{\vX}_1 \tilde{\vX}_2} \to P_{\vY_2|\vX_1 \vX_2} \to P_{\tilde{\vY}_2}$, where $P_{\vY_2|\vX_1 \vX_2}$ is the Gaussian MAC \eqref{eq:noisedensityblock} with $k = 2$ transmitters. Then there exists $n_0 \in \mathbb{N}$ such that for all $n \geq n_0$, $\forall \, (\vx_1, \vx_2, \vy) \in \mathbb{R}^{n \otimes 3}$, it holds that
		\begin{align}
		\frac{P_{\vY_2|\vX_2}(\vy | \vx_2)}{P_{\tilde{\vY}_2|\tilde{\vX}_2}(\vy | \vx_2)} &\leq \kappa_1(P_1) = {27}{\sqrt{\frac{\pi}{8}}}\frac{1 + P_1}{\sqrt{1 + 2 P_1}} \label{eq:molavianjazi1} \\
		\frac{P_{\vY_2}(\vy)}{P_{\tilde{\vY}_2}(\vy)} &\leq \kappa_2(P_1, P_2) = \frac{9}{2 \pi \sqrt{2}} \frac{\Ponetwo}{\sqrt{P_1 P_2}}.  \label{eq:kappa2}
		\end{align}
		If there is no additive noise $\vZ$ in \eqref{eq:ysum}, \eqref{eq:kappa2} continues to hold.
	Inequalities \eqref{eq:molavianjazi1}--\eqref{eq:kappa2} are generalized to the $K$-transmitter Gaussian MAC as follows. 
	
	\item $K$-Transmitter MAC: Let $\vX_1, \dots, \vX_K$ be independent, and for each $i \in [K]$, let $\vX_i$ be distributed uniformly on $\mathbb{S}^{n}(\sqrt{nP_i})$. Let $\tilde{\vX}_i \sim \mathcal{N}(\bs{0}, P_i \mathsf{I}_n)$ for $i \in [K]$, where $\vX_i$ are independent of each other. Let $P_{\vX_{[K]}} \to P_{\vY_K|\vX_{[K]}} \to P_{\vY_K}$, and $P_{\tilde{\vX}_{[K]}} \to P_{\vY_K|\vX_{[K]}} \to P_{\tilde{\vY}_K}$, where $P_{\vY_K|\vX_{[K]}}$ is the Gaussian MAC in \eqref{eq:noisedensityblock} with $K$ transmitters.  
	Then there exists $n_K \in \mathbb{N}$ such that for all $n \geq n_K$, for any $\vx_{[K]} \in \mathbb{R}^{n \otimes K}$,  $\vy \in \mathbb{R}^n$, and non-empty $\mc{S} \in \subemp{[K]}$, it holds that 
		\begin{align}
		\frac{P_{\vY_K|\vX_{\mc{S}^c}}(\vy | \vx_{\mc{S}^c})}{P_{\tilde{\vY}_K|\tilde{\vX}_{\mc{S}^c}}(\vy |\vx_{\mc{S}^c})} &\leq \kappa_{|\mc{S}|}(P_s: s \in \mc{S}), \label{eq:kappaS}
		\end{align}
	where $\kappa_{|\mc{S}|}(P_s: s \in \mc{S})$ is a constant depending only on the power values $(P_s: s \in \mc{S})$. 
	\end{enumerate}
	\end{lemma}
	The proof of \eqref{eq:kappaS}, which is given in \cite[eq. (5.138)]{molavianjaziThesis}, relies on a recursive formula for the distribution of $\vY_K$. 
	
     \lemref{lem:chi2}, stated next, bounds the tail probabilities of the chi-squared distribution from above.
	\begin{lemma}[Laurent and Massart {\cite[Lemma 1]{laurent2000Chi}}]\label{lem:chi2}
	Let $\chi_n^2$ be a random variable with a chi-squared distribution and $n$ degrees of freedom. Then for $t > 0$,
	\begin{align}
	\Prob{\chi_n^2 - n \geq 2 \sqrt{nt} + 2t} &\leq \exp\{-t\} \label{eq:chiuppermas2} \\
	\Prob{\chi_n^2 - n \leq -2 \sqrt{nt}} &\leq \exp\{-t\}. \label{eq:chilowermas2}
	\end{align}
\end{lemma}

 \lemref{lem:tanKP2}, stated next, is used as the main tool to obtain large deviation bounds on the information density random variables that arise when we apply the RCU bound. 
		\begin{lemma}[Tan and Tomamichel {\cite[eq.~(52)]{tan2015Third}}]\label{lem:tanKP2}
			Let $\vZ = (Z_1, \dots, Z_n) \sim \mathcal{N}(\boldsymbol{0}, \mathsf{I}_n)$, $\vx = (\sqrt{nP}, 0, \dots, 0)$, and let $s > 0$ and $P > 0$ be constants. Then for any $a \in \mathbb{R}$, $\mu > 0$, and $n$ large enough,
			\begin{align}
			\mathbb{P}\left[ Z_1 \in \left. \left[\frac{a}{\sqrt{n P}}, \frac{a + \mu}{\sqrt{n P}} \right] \middle| \norm{\vx + \vZ}^2 = n s \right. \right] \leq \frac{L(P, s) \mu}{\sqrt{n}},
			\end{align}
			where
			\begin{align}
			L(P, s) &\triangleq \frac{8(Ps)^{3/2}}{\sqrt{2 \pi}} \, \sqrt{\frac{1 + 4 P s - \sqrt{1 +  4Ps}}{(\sqrt{1 + 4Ps} - 1)^5}}. \label{eq:Lfunction}
			\end{align}
		\end{lemma}
		
			We state the multidimensional Berry-Esseen theorem for sums of independent but not necessarily identical random vectors. The theorem is used as the main tool to bound the probability that the information density random vector belongs to a given set.
		\begin{theorem}[Bentkus \cite{bentkus2005be}] \label{thm:mbe}
			Let $\mathbf{U}_1, \dots, \mathbf{U}_n$ be zero mean, independent random vectors in $\mathbb{R}^d$, and let $\mathbf{Z} \sim \mathcal{N}(\boldsymbol{0}, \mathsf{I}_d)$. Denote $\mathbf{S} = \sum_{i = 1}^n \mathbf{U}_i$, and $T = \sum_{i = 1}^n \E{ \Vert{ \mathbf{U}_i }\Vert^3 }$. Assume that $\Cov{\mathbf{S}} = \mathsf{I}_d$. Then, there exists a constant $c > 0$ such that
			\begin{align}
			\sup_{\mathcal{A} \in \mathfrak{C}_d} \left \lvert \Prob{\mathbf{S} \in \mathcal{A}} - \mathbb{P}[\mathbf{Z} \in \mathcal{A}] \right \rvert \leq {c d^{1/4} T}, \label{eq:multbe}
			\end{align}
			where $\mathfrak{C}_d$ is the set of all convex, Borel measurable subsets of $\mathbb{R}^d$.
		\end{theorem}
		Rai{\v{c}} \cite[Th. 1.1]{raic2019} establishes that the constant $c d^{1/4}$ in \eqref{eq:multbe} can be replaced by $42 d^{1/4} + 16$.
		Tan and Kosut \cite{tan2014dispersions} provide the following corollary to \thmref{thm:mbe} for the case of a general nonsingular $\Cov{\mathbf{S}}$.
		\begin{corollary} [Tan and Kosut {\cite[Corollary 8]{tan2014dispersions}}] \label{cor:be}  For the setup in \thmref{thm:mbe}, assume that $\Cov{\mathbf{S}} = n \mathsf{V}$, where $\lambda_{\min}(\mathsf{V}) > 0$ denotes the minimum eigenvalue of $\mathsf{V}$, and $T = \frac 1 n \sum_{i = 1}^n \E{ \Vert{ \mathbf{U}_i }\Vert^3 }$. Let $\mathbf{Z} \sim \mathcal{N}(\boldsymbol{0}, \mathsf{V})$. Then, there exists a constant $c > 0$ such that
			\begin{align}
			\sup_{\mathcal{A} \in \mathfrak{C}_d} \left \lvert \Prob{\frac 1 {\sqrt{n}} \mathbf{S} \in \mathcal{A}} - \mathbb{P}[\mathbf{Z} \in \mathcal{A}] \right \rvert \leq \frac{c d^{1/4} T}{\sqrt{n} \lambda_{\min}(\mathsf{V})^{3/2}}. \label{eq:multbecor}
			\end{align}
		\end{corollary}
		
	Lemmas \ref{lem:gaussiancdf} and \ref{lem:gotze}, below, are used to bound the probability that the information density random vector belongs to a set. The total variation distance between the measures $P_X$ and $P_Y$ on $\mathbb{R}^d$ is defined as 
		\begin{align}
		\mathrm{TV}(P_X, P_Y) &\triangleq \sup_{\mathcal{D} \in \mathbb{R}^d} \left \lvert \Prob{X \in \mathcal{D}} - \Prob{Y \in \mathcal{D}} \right \rvert \notag \\
		&= \frac{1}{2} \int_{x \in \mathbb{R}^d} \left \lvert dP_X(x) - dP_Y(x) \right \rvert.
		\end{align}
	\lemref{lem:gaussiancdf}, stated next, bounds the total variation distance between two Gaussian vectors. 
	\begin{lemma} \label{lem:gaussiancdf} Let $\mathsf{\Sigma}_1$ and $\mathsf{\Sigma}_2$ be two positive definite $d \times d$ matrices, and let $\bm{\mu}_1, \bm{\mu}_2 \in \mathbb{R}^d$ be two constant vectors. Then, 
		\begin{IEEEeqnarray}{lCr}
		\IEEEeqnarraymulticol{3}{l}{\mathrm{TV}(\mathcal{N}(\bm{\mu}_1, \mathsf{\Sigma}_1), \mathcal{N}(\bm{\mu}_2, \mathsf{\Sigma}_2))} \notag \\
		&\leq& \frac{2 + \sqrt{6}}{4} \norm{\mathsf{\Sigma}_1^{-1/2}  \mathsf{\Sigma}_2  \mathsf{\Sigma}_1^{-1/2} - \mathsf{I}_d }_F \notag \\
		&& + \frac 1 2 \sqrt{(\bm{\mu}_1 - \bm{\mu}_2)^T \mathsf{\Sigma}_1^{-1} (\bm{\mu}_1 - \bm{\mu}_2)}  \label{eq:C2const},
		\end{IEEEeqnarray}
		where $\norm{\cdot}_F$ denotes the Frobenius norm.
	\end{lemma}
	\begin{IEEEproof}
		See \appref{app:gaussian}.
	\end{IEEEproof}	
	A weaker version of the bound  in \lemref{lem:gaussiancdf} by Devroye \emph{et al.} appears in \cite[Th.~1.1]{devroye2018total}. Like our proof, the proof of \cite[Th.~1.1]{devroye2018total} relies on Pinsker's inequality. We improve the factor in front of the Frobenius norm from 1.5 in \cite[Th~1.1]{devroye2018total} to $\frac{2 + \sqrt{6}}{4} \approx 1.113$ by using the result in \cite[Th.~1.1]{rump2018estimates} to lower bound the logdeterminant of the matrix $\mathsf{\Sigma}_1^{-1/2}  \mathsf{\Sigma}_2  \mathsf{\Sigma}_1^{-1/2} - \mathsf{I}_d$ in \eqref{eq:C2const}.
	
	\lemref{lem:gotze}, stated next, gives an upper bound on the total variation distance between the marginal distribution of the first $k$ dimensions of a random variable distributed uniformly on $\mathbb{S}^{n}(\sqrt{n})$ and the $k$-dimensional standard Gaussian random vector. 
		\begin{lemma}[Stam {\cite[Th. 2]{stam1982}}]\label{lem:gotze}
		Let $\vX = (X_1, \dots, X_n)$ be distributed uniformly on $\mathbb{S}^{n}({\sqrt{n}})$. Let $\vX^{[k]} = (X_1, \dots, X_k)$ contain the first $k$ coordinates of $\vX$. Then,
		\begin{align}
		\mathrm{TV}(P_{\vX^{[k]}},  \mathcal{N}(\boldsymbol{0}, \mathsf{I}_k)) \leq n^{\frac 1 2 k} (n - k - 2)^{-\frac 1 2 k } - 1  \label{eq:C1const}
		\end{align}
		for $n > k + 2$.
	\end{lemma}
	
	We use \lemref{lem:gotze} with $k = 1$ to approximate the inner product $\langle \vX_1, \vX_2 \rangle$ by a Gaussian random variable, which facilitates an application of the Berry-Esseen theorem in \secref{sec:be}.
	
	The proof of \thmref{thm:MAC} relies on a random coding argument and \thmref{thm:RCUMac}. The asymptotic analysis of the RCU bound (\thmref{thm:RCUMac}) borrows some techniques from the point-to-point case \cite{tan2015Third}.
	\subsection{Encoding and Decoding for the MAC}
	We select the distributions of the independent inputs $\vX_1$ and $\vX_2$ as the uniform distributions on $\mathbb{S}^n(\sqrt{n P_1})$ and $\mathbb{S}^n(\sqrt{n P_2})$, which are the $n$-dimensional spheres centered at the origin with radii $\sqrt{nP_1}$ and $\sqrt{nP_2}$, respectively. The resulting distribution is
	\begin{align}
	P_{\vX_1}(\vx_1)P_{\vX_2}(\vx_2) = \frac{\delta(\norm{\vx_1}^2 - n P_1)}{S_{n}(\sqrt{nP_1})} \frac{\delta(\norm{\vx_2}^2 - n P_2)}{S_{n}(\sqrt{nP_2})}, \label{eq:inputMAC}
	\end{align}
	where $\delta(\cdot)$ is the Dirac delta function, and
	\begin{align}
	S_n(r) \triangleq \frac{2 \pi^{n/2}}{\Gamma(n/2)}r^{n-1}
	\end{align}
	is the surface area of an $n$-dimensional sphere $\mathbb{S}^n(r)$ with radius $r$. We draw $M_1$ codewords i.i.d. from $P_{\vX_1}$ and $M_2$ codewords i.i.d. from $P_{\vX_2}$, respectively. We denote these by $\mathsf{f}_i(m_i)$ for $m_i \in [M_i]$, $i \in \{1, 2\}$.

	In order to use \thmref{thm:RCUMac}, the channel $P_{Y_2|X_1 X_2}$ is particularized to the two-transmitter Gaussian MAC in \eqref{eq:noisedensityblock}.
	Upon receiving the output sequence $\vy$, the decoder employs a maximum likelihood decoding rule, given by
	\begin{align}
	\mathsf{g}(\vy) = \begin{cases} 
	(m_1, m_2) &\mbox {if } 
	\imath_{1, 2}(\mathsf{f}_1(m_1), \mathsf{f}_{2}(m_2); \vy) \\
	&> \imath_{1, 2}(\mathsf{f}_1(m_1'), \mathsf{f}_2(m_2'); \vy)  \\
	&\textnormal{for all } (m_1', m_2') \neq (m_1, m_2),\\
	&(m_1', m_2') \in [M_1] \times [M_2] \\
	\mbox{error} &\mbox{otherwise.} \label{eq:ML}
	\end{cases}
	\end{align}
	We treat all ties in \eqref{eq:ML} as errors because the probability that two codewords result in exactly the same information density is negligible due to the continuity of the noise.
	Substituting the transition law of the Gaussian MAC \eqref{eq:noisedensityblock} and the uniform input distributions on the power spheres \eqref{eq:inputMAC} into \eqref{eq:imath1}--\eqref{eq:imath12}, we compute for any $(\vx_1, \vx_2, \vy) \in {\mathbb{R}^{n \otimes 3}}$
	\begin{IEEEeqnarray}{rCl}
	\imath_1(\vx_1; \vy|\vx_2) &=& \frac{n}{2} \log \frac{1}{2\pi} + \langle \vy-\vx_2, \vx_1 \rangle - \frac{\norm{\vy-\vx_2}^2}{2} \notag \\
	&&- \frac{nP_1}{2} - \log P_{\vY_2|\vX_2}(\vy|\vx_2) \label{eq:mutualinfo2cond}\\
	\imath_2(\vx_2; \vy|\vx_1) &=& \frac{n}{2} \log \frac{1}{2\pi} + \langle \vy-\vx_1, \vx_2 \rangle - \frac{\norm{\vy-\vx_1}^2}{2} \notag \\
	&&- \frac{nP_2}{2} - \log P_{\vY_2|\vX_1}(\vy|\vx_1)	\\
	\imath_{1, 2}(\vx_1, \vx_2; \vy) &=& \frac{n}{2} \log \frac{1}{2\pi} + \langle \vy, \vx_1 + \vx_2 \rangle - \frac{\norm{\vy}^2}{2}  \notag \\
	&&-\frac{\norm{\vx_1 + \vx_2}^2}{2}  - \log P_{\vY_2}(\vy). \label{eq:mutualinfo2uncond}
	\end{IEEEeqnarray}
	Observe that for each $\vx_2$ and $\vy$, $\imath_1(\vx_1; \vy|\vx_2)$ depends on $\vx_1$ only through the inner product $\langle \vy-\vx_2, \vx_1 \rangle$, and for each $\vy$, $\imath_{1, 2}(\vx_1, \vx_2; \vy)$ depends on $(\vx_1, \vx_2)$ only through $\langle \vy, \vx_1 + \vx_2 \rangle - \langle \vx_1, \vx_2 \rangle$. By the input-output relation in \eqref{eq:ysum}, the conditional information density for two transmitters, $\imath_1(\vx_1; \vy|\vx_2)$, can be reduced to the unconditional information density for a single transmitter as
	\begin{align}
	\imath_1(\vx_1; \vy|\vx_2) = \imath_1(\vx_1; \vy - \vx_2) \triangleq \log \frac{P_{\vY_1| \vX_1}( \vy - \vx_2 | \vx_1)}{P_{\vY_1}(\vy - \vx_2)} \label{eq:imathreduction},
	\end{align}
	where $\vY_1 = \vX_1 + \vZ$ is the output of the channel with a single transmitter. 
	\subsection{Typical Set for the MAC}
	For the rest of the proof, $\vZ\sim \mathcal{N}(\boldsymbol{0}, \mathsf{I}_n)$ denotes the Gaussian noise, which is independent of the channel inputs $\vX_1$ and $\vX_2$. Note that the expectations of the squared norms of $\vX_1 + \vZ, \vX_2 + \vZ$ and $\vY_2$ are $n(1 + P_1), n(1 + P_2)$, and $n(1 + \Ponetwo)$, respectively. We define a typical set for vector $(\vX_1 + \vZ, \vX_2 + \vZ, \vY_2)$ by
	\begin{align}
	\mc{F} &\triangleq \bigtimes_{\substack{\mc{S} \in \subemp{[2]}}} \mc{F}(\mc{S}) \subseteq \mathbb{R}^{n \otimes 3}, \label{eq:F2}
	\end{align}
	where
	\begin{align}
		\mc{F}(\mc{S}) &\triangleq \cB{\vx_{\langle \mc{S} \rangle} + \mathbf{z} \in \mathbb{R}^n \colon  \frac{1}{n} \norm{\vx_{\langle \mc{S} \rangle} + \mathbf{z}}^2 \in \mc{I}(\mc{S})} \label{eq:FS} \\
		\mc{I}(\mc{S}) &\triangleq [1 + \PS - n^{-1/3}, 1 + \PS + n^{-1/3}]. \label{eq:IS}
	\end{align}
	We next show that for $n$ large enough,
	\begin{align}
	\Prob{(\vX_1 + \vZ, \vX_2 + \vZ, \vY_2) \notin \mc{F}} \leq \exp\{-c_2 n^{1/3}\} \label{eq:consf2},
	\end{align}
	where $c_2 > 0$ is a constant.
	
	To bound the probability that the triplet $(\vX_1 + \vZ, \vX_2 + \vZ, \vY_2)$ does not belong to the typical set $\mc{F}$, we use \lemref{lem:kappa} to approximate 
	the squared norms $\norm{\vX_1 + \vZ}^2$,  $\norm{\vX_2 + \vZ}^2$, and $\norm{\vY_2}^2$ by multiples of chi-squared distributed random variables with $n$ degrees of freedom. We then use \lemref{lem:chi2} to bound the two-sided tail probability of these chi-squared distributed random variables. Weakening the upper bound \eqref{eq:chiuppermas2} in \lemref{lem:chi2} using $2 \sqrt{2nt} \geq 2 \sqrt{nt} + 2t$ for $0 < t \leq \frac{n}{8} \leq (3 - 2\sqrt 2)n$, we get the following concentration inequalities for the squared norms of the random vectors $\vX_1 +\vZ$ and $\vY_2$
	\begin{IEEEeqnarray}{rCl}
	&&\Prob{\left\lvert \norm{\vX_1 + \vZ}^2 - n(1 + P_1) \right\rvert > n t_1} \notag \\
	&& \quad \quad \quad \quad  \leq 2 \kappa_1(P_1) \exp\left\{-\frac{nt_1^2}{8(1 + P_1)^2}\right\} \label{eq:probchi1}\\
	&&\Prob{\left\lvert \norm{\vY_2}^2 - n(1 + \Ponetwo) \right\rvert > n t_2} \notag \\
	&& \quad \quad \quad \quad \leq 2 \kappa_2(P_1, P_2) \exp\left\{-\frac{nt_2^2}{8(1 + \Ponetwo)^2}\right\} \label{eq:probchi2}
	\end{IEEEeqnarray}
	for $t_1 \in (0, 1 + P_1)$, and $t_2 \in (0, 1 + \Ponetwo)$, 
	where $\kappa_1(P_1)$ and $\kappa_2(P_1, P_2)$ are constants defined in \lemref{lem:kappa}.
	We deduce \eqref{eq:consf2} by the union bound and setting $t_1 = t_2 = n^{-1/3}$ in \eqref{eq:probchi1}--\eqref{eq:probchi2}.
	\subsection{A Large Deviation Bound on the Mutual Information Random Variables} \label{sec:boundingmutualinfo}
	We introduce the following functions that are analogous to the one used in the point-to-point channel in \cite[eq.~(27)]{tan2015Third}
	\begin{align}
	g_1(t; \vy, \vx_2) &\triangleq  \Prob{\imath_1(\bar{\vX}_1; \vY_2|\vX_2) \geq t \mid \vX_2 = \vx_2, \vY_2 = \vy} \label{eq:g1def}\\
	g_2(t; \vy, \vx_1) &\triangleq  \Prob{\imath_2(\bar{\vX}_2; \vY_2|\vX_1) \geq t \mid \vX_1 = \vx_1, \vY_2 = \vy} \\
	g_{1,2}(t; \vy) &\triangleq \Prob{\imath_{1,2}(\bar{\vX}_1, \bar{\vX}_2; \vY_2) \geq t \mid \vY_2 = \vy} \label{eq:g2def},
	\end{align}
where 
\begin{IEEEeqnarray*}{rCl}
\IEEEeqnarraymulticol{3}{l}{P_{\vX_1 \vX_2 \bar{\vX}_1 \bar{\vX}_2 \vY_2}(\vx_1, \vx_2, \bar{\vx}_1, \bar{\vx}_2, \vy)}  \\ 
&=& P_{\vX_1}(\vx_1) P_{\vX_2}(\vx_2) P_{\vX_1}(\bar{\vx}_1) P_{\vX_2}(\bar{\vx}_2) P_{\vY_2|\vX_1 \vX_2}(\vy|\vx_1, \vx_2). 
\end{IEEEeqnarray*} 

The following lemma, which generalizes \cite[eq. (53)]{tan2015Third} to the Gaussian MAC, gives upper bounds on these functions. We use \lemref{lem:gbound} in the evaluation of the RCU bound.

\begin{lemma}{\label{lem:gbound}}
Let $(\vy - \vx_2, \vy - \vx_1, \vy) \in \mc{F}$, where the set $\mc{F}$ is defined in \eqref{eq:F2}. Then, for $n$ large enough, 
\begin{IEEEeqnarray}{rCl}
 \IEEEyesnumber \label{eq:g2global}
 \IEEEyessubnumber*
		g_1(t; \vy, \vx_2) &\leq& \frac{G_1 \exp\cB{-t}}{\sqrt{n}}  \label{eq:g1lemma}\\
		g_2(t; \vy, \vx_1) &\leq& \frac{G_2 \exp\cB{-t}}{\sqrt{n}} \label{eq:g1lemma2}\\
		g_{1,2}(t; \vy) &\leq& \frac{G_{1,2} \exp\cB{-t}}{\sqrt{n}}, \label{eq:g2lemma}
\end{IEEEeqnarray}
where $G_1, G_2$, and $G_{1,2}$ are positive constants depending only on $P_1, P_2$, and $(P_1, P_2)$, respectively.
\end{lemma}

\begin{IEEEproof} The bounds in \eqref{eq:g1lemma} and \eqref{eq:g1lemma2} follow from the equivalence (stated in \eqref{eq:imathreduction}) between the  conditional information density for two transmitters and the unconditional information density for a single transmitter combined with the analysis in \cite[Sec. IV-E]{tan2015Third}. The constants in \eqref{eq:g1lemma} and \eqref{eq:g1lemma2} are
\begin{align} 
G_i = (3 \log 2) L(P_i, 1 + P_i), \quad i \in \{1, 2\},
\end{align}
where $L(\cdot, \cdot)$ is the function defined in \eqref{eq:Lfunction}.

Bounding the function $g_{1,2}(t; \vy)$ is more challenging. While $\norm{\vX_1}^2$ is a constant under the uniform distribution on a power sphere, $\norm{\vXltwo}^2$ is not. The proof of \eqref{eq:g2lemma} follows steps similar to \cite[Sec. IV-E]{tan2015Third}. First, we change the measure from $P_{\vX_1}P_{\vX_2} P_{\vY_2}$ to $P_{\vX_1}P_{\vX_2} P_{\vY_2 | \vX_1 \vX_2}$ to get
\begin{IEEEeqnarray}{rCl}
g_{1,2}(t; \vy) &=& \mathbb{E}[\exp\{-\imath_{1, 2}(\vX_1, \vX_2; \vY_2) \} \notag \\
&& \quad  1\{\imath_{1, 2}(\vX_1, \vX_2; \vY_2) \geq t\} \mid \vY_2 = \vy]. \label{eq:g12}
\end{IEEEeqnarray}
To bound \eqref{eq:g12}, we define function $h_{1, 2}(\vy; a, \mu)$ for constants $a \in \mathbb{R}$ and $\mu > 0$ as
\begin{IEEEeqnarray}{rCl}
\IEEEeqnarraymulticol{3}{l}{h_{1, 2}(\vy; a, \mu)} \notag \\
&\triangleq& \bbP \Bigl[\imath_{1, 2}(\vX_1, \vX_2; \vY_2) \in [a, a + \mu] \Big| \vY_2 = \vy \Bigr] \IEEEeqnarraynumspace \\
&=& \mathbb{P} \bigg[ \langle \vXltwo, \vY_2 \rangle - \frac{\norm{\vXltwo}^2}{2} \notag \\
&&\quad \in [a', a' + \mu] \,\bigg| \vY_2 = \vy \bigg], \label{eq:h2yn}
\end{IEEEeqnarray}
where $a'$ is shifted from $a$ by some amount depending on $\vy$, and \eqref{eq:h2yn} follows from~\eqref{eq:mutualinfo2uncond}. 
By spherical symmetry of the distribution of $\vY_2$, \eqref{eq:h2yn} depends on $\vy$ only through its norm $\norm{\vy}$. We have 
		\begin{align}
		h_{1,2}(s; a, \mu) &\triangleq h_{1,2}(\vy; a, \mu) \notag \\
		&= \mathbb{P} \bigg[\langle \vXltwo, \vY_2 \rangle - \frac{\norm{\vXltwo}^2}{2} \notag \\
		&\quad \in [a', a' + \mu] \biggm | \norm{\vY_2}^2 = ns \bigg], \label{eq:h12s}
		\end{align}
		where $\norm{\vy}^2 = ns$, and $s \in \mathcal{I}([2])$, and  $\mathcal{I}(\mathcal{S})$ is defined in \eqref{eq:IS}. Recall that the support of the norm $\norm{\vXltwo}^2$ is $[n(\sqrt{P_1} - \sqrt{P_2})^2, n(\sqrt{P_1} + \sqrt{P_2})^2]$. To avoid the cases where $\norm{\vXltwo}^2$ is too small, we separate the probability term \eqref{eq:h12s} according to whether or not the event
		\begin{align}
		\mc{B} = \left\{\norm{\vXltwo}^2 < n(\Ponetwo - \sqrt{P_1 P_2})\right\}
		\end{align} 
		occurs under the condition that $\norm{\vY_2}^2 = ns$. Here, the choice $\sqrt{P_1 P_2}$ is arbitrary and can be replaced by any constant in $(0, 2\sqrt{P_1 P_2})$.
		
		In \eqref{eq:h12s}, conditioning on the event $\mc{B}$ and then bounding the corresponding probability terms by 1 gives 
		\begin{align}
		&h_{1,2}(s; a, \mu) \leq \Prob{\mc{B} \,\middle| \,  \norm{\vY_2}^2 = ns} +  \bbP \biggl[\langle \vXltwo, \vY_2 \rangle \notag \\
		& \quad - \frac{\norm{\vXltwo}^2}{2}\in [a', a' + \mu] \biggm| \norm{\vY_2}^2 = ns, \mc{B}^c \biggr]. \label{eq:h2split}
		\end{align}
 For $n$ large enough, we bound the first term on the right-hand side of \eqref{eq:h2split} by
		\begin{align}
		\Prob{\mc{B} \,\middle| \,  \norm{\vXltwo + \vZ}^2 = ns} \leq \exp\{-nC\}, \label{eq:concentU}
		\end{align}
		where $C > 0$ is a constant. The proof of \eqref{eq:concentU} appears in \appref{app:Ubound}.

By spherical symmetry, the distribution of $\langle \vX_1 + \vX_2, \vX_1 + \vX_2 + \vZ \rangle$ depends on $\vXltwo$ only through the norm $\norm{\vXltwo}$. Therefore, fixing $\vXltwo$ to $\vx = (\sqrt{n u}, 0, \dots, 0)$, we find that for any $u \in [\Ponetwo - \sqrt{P_1 P_2}, (\sqrt{P_1} + \sqrt{P_2})^2]$, $s \in \mc{I}([2])$, and $n$ large enough,
		\begin{align}
		&\bbP \Bigl[\langle \vXltwo, \vXltwo + \vZ \rangle - \frac{nu}{2}\in [a', a' + \mu] \notag \\
		&\quad \Bigm| \norm{\vXltwo + \vZ}^2 = ns, \norm{\vXltwo}^2 = nu \Bigr]  \notag \\
		&= \Prob{Z_1 + \frac{\sqrt{nu}}{2} \in \left. \left[ \frac{a'}{\sqrt{nu}}, \frac{a' + \mu}{\sqrt{nu}} \right] \, \middle| \right. \norm{\vx+ \vZ}^2 = ns} \label{eq:hy2tanlemma0} \\
		&\leq \frac{L(u, s) \mu}{\sqrt{n}} \label{eq:hy2tanlemma} \\
		&\leq \frac 3 2 \frac{ L(u, 1 + \Ponetwo) \mu}{\sqrt{n}},  \label{eq:hy2tanlemma2}
		\end{align}
		where \eqref{eq:hy2tanlemma} follows by \lemref{lem:tanKP2}, and \eqref{eq:hy2tanlemma2} holds for $n$ large enough by the continuity of the map $s \mapsto L(u, s)$ since $s \in \mathcal{I}([2])$. Using \eqref{eq:hy2tanlemma2}, we bound the second term in \eqref{eq:h2split} as
		\begin{IEEEeqnarray}{rCl}
		\IEEEeqnarraymulticol{3}{l}{\bbP \Bigl[\langle \vXltwo, \vY_2 \rangle - \frac{\norm{\vXltwo}^2}{2}\in [a', a' + \mu]} \notag \\
		&&\Bigm| \norm{\vXltwo + \vZ}^2 = ns, \mc{B}^c \Bigr] \notag \\
		&\leq& \max_{u \in [\Ponetwo - \sqrt{P_1 P_2}, (\sqrt{P_1} + \sqrt{P_2})^2]} \frac {3 \mu L(u, 1 + \Ponetwo)}{2 \sqrt{n}}. \IEEEeqnarraynumspace \label{eq:K2bound}
		\end{IEEEeqnarray}
		By \eqref{eq:h2split}, \eqref{eq:concentU}, \eqref{eq:K2bound}, and because $L(u, 1 + \Ponetwo)$ is bounded above for $u \in [\Ponetwo - \sqrt{P_1 P_2}, (\sqrt{P_1} + \sqrt{P_2})^2]$, there exists a constant $K_2(P_1, P_2) > 0$ such that
		\begin{align}
		h_{1,2}(s; a, \mu) \leq K_2(P_1, P_2) \frac{\mu}{\sqrt{n}}
		\end{align}
		for $n$ large enough.
		By following the same steps as \cite[eq. (55)-(57)]{tan2015Third}, we conclude that
		\begin{align}
		g_{1,2}(t; \vy) \leq \frac{G_{1,2} \exp\cB{-t}}{\sqrt{n}},
		\end{align}
		where $G_{1,2} = (2 \log 2) K_2(P_1, P_2)$.

\end{IEEEproof}
		
		\subsection{Evaluating the RCU Bound for the MAC}\label{sec:evalRCU}
		We here bound the right-hand side of \eqref{eq:RCUmac} in \thmref{thm:RCUMac}.
		The information density random vector is defined as
	\begin{align}
	    \bs{\imath}_2 \triangleq \begin{bmatrix}
	        \imath_1(\vX_1; \vY_2 |\vX_2) \\
	        \imath_2(\vX_2; \vY_2|\vX_1) \\
	        \imath_{1, 2}(\vX_1, X_2; \vY_2)
	    \end{bmatrix},
	\end{align}
	where $\vX_1$ and $\vX_2$ are distributed according to \eqref{eq:inputMAC}, and $P_{\vX_1} P_{\vX_2} \to P_{\vY_2 | \vXltwo} \to P_{\vY_2}$.
		
		Define the typical events
		\begin{align}
		\mc{E}(\mc{S}) &\triangleq \cB{\vX_{\langle \mc{S} \rangle} + \vZ \in \mc{F}(\mc{S})} \label{eq:setE1}\\
		\mc{E} &\triangleq \bigcap_{\substack{\mc{S} \in \subemp{[2]}}}\mc{E}(\mc{S}) \label{eq:setE2} \\
		\mc{A} &\triangleq \cB{\boldsymbol{\imath}_2 \geq \log \begin{bmatrix} {M_1 (G_1)^2 \alpha_1} \\ {M_2 (G_2)^2 \alpha_1} \\ {M_1 M_2 (G_{1,2})^2 \alpha_2} \end{bmatrix} - \frac 1 2 \log n \bs{1}}, \label{eq:setA}
		\end{align}
		where $G_1$, $G_2$ and $G_{1,2}$ are the constants given in \eqref{eq:g2global}, $\mc{F}(\mc{S})$ is defined in \eqref{eq:FS}, and
		\begin{align}
		    \alpha_s \triangleq 2 \binom{2}{s}, \quad s = 1, 2.
		\end{align}
		Denote for brevity
		\begin{IEEEeqnarray}{rCl}
		g_1 &\triangleq& g_1(\imath_1(\vX_1; \vY_2 | \vX_2); \vY_2, \vX_2)  \IEEEyesnumber \IEEEyessubnumber* \\
		g_2 &\triangleq& g_2(\imath_2(\vX_2; \vY_2 | \vX_1); \vY_2, \vX_1) \\
		g_{1,2} &\triangleq& g_{1,2}(\imath_{1, 2}(\vX_1, \vX_2; \vY_2); \vY_2),
		\end{IEEEeqnarray}
		where $g_1(\cdot), g_2(\cdot)$, and $g_{1, 2}(\cdot)$, are defined in \eqref{eq:g1def}--\eqref{eq:g2def}.
		
The right-hand side of \eqref{eq:RCUmac} is bounded in \eqref{eq:expsplit}--\eqref{eq:MACprobgeneralE2} at the top of the next page.
\begin{table*}[ht!]
\newcounter{mytempeqncnt}
\normalsize
\setcounter{mytempeqncnt}{\value{equation}}
\setcounter{equation}{85}
\vspace*{4pt}
\begin{IEEEeqnarray}{rCl}
\IEEEeqnarraymulticol{3}{l}{\mathbb{E}\Bigl[\min\Big\{1, (M_1-1) \Prob{\imath_1(\bar{\vX}_1; \vY_2 | \vX_2) \geq \imath_1(\vX_1; \vY_2 | \vX_2) \mid \vX_1, \vX_2, \vY_2}} \notag \\
&&+ (M_2-1) \Prob{\imath_2(\bar{\vX}_2; \vY_2 | \vX_1) \geq \imath_2(\vX_2; \vY_2 | \vX_1) \mid \vX_1, \vX_2, \vY_2} \notag \\\
&&+ (M_1-1)(M_2-1) \Prob{\imath_{1, 2}(\bar{\vX}_1, \bar{\vX}_2; \vY_2) \geq \imath_{1, 2}(\vX_1, \vX_2; \vY_2) \mid \vX_1, \vX_2, \vY_2}\Big\}\Bigr] \notag \\
&=&  \mathbb{E}\Bigl[\min\Big\{1, (M_1-1)g_1 +  (M_2-1)g_2 + (M_1-1)(M_2-1) g_{1,2} \Big\} 1 \cB{\mc{A}^c \cup \mc{E}^c}\Bigr] \notag\\
&&+  \mathbb{E}\Bigl[\min\Big\{1, (M_1-1)g_1 +  (M_2-1)g_2 + (M_1-1)(M_2-1) g_{1,2} \Big\} 1 \cB{\mc{A} \cap \mc{E}} \Bigr]  \label{eq:expsplit}\\
& \leq& \Prob{\mc{A}^c \cup \mc{E}^c} + \Prob{\mc{E}(\{1\})} \, M_1 \, \E{ g_1 1\cB{\imath_1(\vX_1; \vY_2 | \vX_2) \geq \log \frac{M_1 (G_1)^2 \alpha_1}{\sqrt{n}}} \biggm | \mc{E}(\{1\})}   \notag\\
&&+ \Prob{\mc{E}(\{2\})} \, M_2 \, \E{g_2 1 \cB{\imath_2(\vX_2; \vY_2 | \vX_1) \geq \log \frac{M_2 (G_2)^2 \alpha_1}{\sqrt{n}}} \biggm | \mc{E}(\{2\})}  \notag\\
&&+ \Prob{\mc{E}(\{1, 2\})} \, M_1 M_2 \,\E{g_{1,2} 1 \cB{\imath_{1, 2}(\vX_1, \vX_2; \vY_2) \geq \log \frac{M_1 M_2 (G_{1,2})^2 \alpha_2}{\sqrt{n}}} \biggm | \mc{E}(\{1, 2\})} \label{eq:indicatordist} \\
&\leq& \Prob{\mc{A}^c \cup \mc{E}^c} + \frac{M_1 G_1}{\sqrt{n}}  \E{ \exp\{-\imath_1(\vX_1; \vY_2 | \vX_2)\} 1 \cB{\imath_1(\vX_1; \vY_2 | \vX_2) \geq \log \frac{M_1 (G_1)^2 \alpha_1}{\sqrt{n}}} \biggm | \mc{E}(\{1\})}  \notag\\
&&+ \frac{ M_2 G_2}{\sqrt{n}}\E{\exp\{-\imath_2(\vX_2; \vY_2 | \vX_1)\} 1 \cB{\imath_2(\vX_2; \vY_2 | \vX_1) \geq \log \frac{M_2 (G_2)^2 \alpha_1}{\sqrt{n}}} \biggm | \mc{E}(\{2\})}  \notag\\
&&+ \frac{M_1 M_2 G_{1,2}}{\sqrt{n}}  \E{\exp\{-\imath_{1, 2}(\vX_1, \vX_2; \vY_2)\} 1 \cB{\imath_{1, 2}(\vX_1, \vX_2; \vY_2) \geq \log \frac{M_1 M_2 (G_{1,2})^2 \alpha_2}{\sqrt{n}}} \biggm | \mc{E}(\{1, 2\})} \label{eq:Glemmaused}\\
&\leq&  \Prob{\mc{A}^c} + \Prob{ \mc{E}^c}  + \frac{\frac{2}{ \alpha_1} + \frac{1}{\alpha_2}}{\sqrt{n}}  \label{eq:MACprobgeneral} \\
& \leq&  \Prob{\mc{A}^c} + \exp \cB{-c_2 n^{1/3}} + \frac{1}{\sqrt{n}} \label{eq:MACprobgeneralE2}
	\end{IEEEeqnarray}
\hrulefill
\vspace{-10pt}
\setcounter{equation}{\value{mytempeqncnt}}
\end{table*}
\setcounter{equation}{90}
Here, $c_2$ is the positive constant defined in \eqref{eq:consf2}. Equality \eqref{eq:expsplit} follows from the definitions of the functions $g_{1}(t; \vy, \vx_2)$ and $g_{1, 2}(t; \vy)$ and splitting the expectation into two cases according to whether the event $\{\mc{A}^c \cup \mc{E}^c\}$ occurs or not. Inequality \eqref{eq:indicatordist} follows by bounding the minimum inside the first expectation in \eqref{eq:expsplit} by 1; bounding the minimum inside the second expectation in \eqref{eq:expsplit} by its second argument; writing the indicator function $1\cB{\mc{A} \cap \mc{E}}$ as a multiplication of 3 indicator functions using the definitions in \eqref{eq:setE2} and \eqref{eq:setA} and distributing that multiplication over the summation. Inequality \eqref{eq:Glemmaused} follows from \lemref{lem:gbound} and by bounding the probability terms by 1. Inequality \eqref{eq:MACprobgeneral} is obtained by applying the union bound to $\Prob{\mc{A}^c \cup \mc{E}^c}$ and by using \lemref{lem:gbound} with $t = \log \frac{M_1 (G_1)^2 \alpha_1}{\sqrt{n}}$, $t = \log \frac{M_2 (G_2)^2 \alpha_1}{\sqrt{n}}$, and $t = \log \frac{M_1 M_2 (G_{1,2})^2 \alpha_2}{\sqrt{n}}$ to bound the three remaining terms, respectively. Inequality \eqref{eq:MACprobgeneralE2} follows from \eqref{eq:consf2}. 
		
		\subsection{A Multidimensional Berry-Esseen Type Inequality} \label{sec:be}
		
	    To complete the proof of \thmref{thm:MAC}, it remains only to evaluate the probability $\Prob{\mc{A}^c}$ in \eqref{eq:MACprobgeneralE2}. If the operational rate pair $\nB{\frac{\log M_1}{n}, \frac{\log M_2}{n}}$ does not lie at a corner point of the achievable capacity region, applying the union bound to $\Prob{\mc{A}^c}$ gives a tight achievability bound since two of the three probability terms that appear after applying the union bound to $\Prob{\mc{A}^c}$ are $O\nB{\frac{1}{\sqrt{n}}}$. However, for the corner points, $\Prob{\mc{A}^c}$ needs to be bounded without using the union bound in order to obtain a tighter achievability bound (see \cite[Sec.~5.1.1]{molavianjaziThesis}). In this section, we bound $\Prob{\mc{A}^c}$ jointly by deriving a multidimensional Berry-Esseen type inequality. 
		
		Due to the non-i.i.d. input distribution, the random vector $\bs{\imath}_2$ cannot be separated into a sum of $n$ random vectors. Therefore, to approximate $\bs{\imath}_2$, we define the modified conditional and unconditional information densities whose denominators have Gaussian distributions corresponding to
		\begin{IEEEeqnarray}{rCl}
		\IEEEyesnumber
		\IEEEyessubnumber*
		\tilde{\imath}_1(\vx_1; \vy | \vx_2) &\triangleq& \sum_{i = 1}^n \log \frac{P_{Y_2|X_1 X_2}(y_i | x_{1i}, x_{2i})}{P_{\tilde{Y}_2|\tilde{X}_2}(y_i | x_{2i})} \label{eq:i2tilde1} \\
		\tilde{\imath}_2(\vx_2; \vy | \vx_1) &\triangleq& \sum_{i = 1}^n \log \frac{P_{Y_2|X_1 X_2}(y_i | x_{1i}, x_{2i})}{P_{\tilde{Y}_2|\tilde{X}_1}(y_i | x_{1i})} \label{eq:i2tilde12} \\	
		\tilde{\imath}_{1,2}(\vx_1, \vx_2; \vy) &\triangleq& \sum_{i = 1}^n \log \frac{P_{Y_2|X_1 X_2}(y_i | x_{1i}, x_{2i})}{P_{\tilde{Y}_2}(y_i)}, \label{eq:i2tilde}
		\end{IEEEeqnarray}
		where $\tilde{X}_i \sim \mathcal{N}(0, P_i)$, $i \in [2]$, and $P_{\tilde{X}_1} P_{\tilde{X}_2} \to P_{Y_2|X_1 X_2} \to P_{\tilde{Y}_2} = \mathcal{N}(0, 1 + \Ponetwo)$. Denote the modified and centered information density random vector by
		\begin{align}
		\boldsymbol{\tilde{\imath}}_2 \triangleq  \frac{1}{\sqrt{n}} \nB{ \begin{bmatrix} \tilde{\imath}_1(\vX_1; \vY_2 | \vX_2) \\ \tilde{\imath}_2(\vX_2; \vY_2 | \vX_1) \\ \tilde{\imath}_{1,2}(\vX_1, \vX_2; \vY_2) \end{bmatrix} - n \mathbf{C}(P_1, P_2) }, \label{eq:i2modified}
		\end{align}
		where $ \mathbf{C}(P_1, P_2) = \frac 1 n \E{\boldsymbol{\imath}_2}$ is the capacity vector defined in \eqref{eq:capacityvector}.
Define the threshold vector
		\begin{align}
		\bm{\tau} &\triangleq  \log \begin{bmatrix} {M_1 (G_1)^2 \kappa_1(P_1)} \alpha_1 \\ {M_2 (G_2)^2 \kappa_1(P_2)} \alpha_1 \\ {M_1 M_2 (G_{1,2})^2 \kappa_2(P_1, P_2) \alpha_2} \end{bmatrix} \notag \\
		&\quad - \frac 1 2 \log n \bs{1} - n \mathbf{C}(P_1, P_2).
\label{eq:tauvec}
		\end{align}
		
		Our method to bound the probability $\Prob{\mc{A}^c}$ involves 5 steps. 
	
		\textbf{Step 1:} We first replace $\boldsymbol{\imath}_2$ by $\boldsymbol{\tilde{\imath}}_2$. Unlike $\bs{\imath}_2$, $\bs{\tilde{\imath}}_2$ can be written as a sum of $n$ dependent random vectors. Prior uses of this approach include \cite[eq. (65)]{tan2015Third} for the point-to-point channel and \cite[eq. (2)]{molavianjazi2015second} for the MAC. We then bound $\Prob{\mc{A}^c}$ in terms of the modified information density random vector~$\boldsymbol{\tilde{\imath}}_2$. By \eqref{eq:setA} and \lemref{lem:kappa}, 
		\begin{align}
		\Prob{\mc{A}^c} &= 1 - \Prob{\boldsymbol{\imath}_2 - \E{\boldsymbol{\imath}_2}  \geq   \nB{\boldsymbol{\tau} - \log \begin{bmatrix} \kappa_1(P_1) \\\kappa_1(P_2) \\  \kappa_2(P_1, P_2) \end{bmatrix} }} \\
		&\leq 1 - \Prob{\boldsymbol{\tilde{\imath}}_2 \geq \frac 1 {\sqrt{n}} \bm{\tau}}. \label{eq:itildeprob}
		\end{align}
		From \eqref{eq:i2tilde1}--\eqref{eq:i2tilde}, we see that
		\begin{align}
		\boldsymbol{\tilde{\imath}}_2  \sim \frac{1}{\sqrt{n}}  \begin{bmatrix}  \frac{(n - \norm{\vZ}^2) P_1 + 2 \langle \vX_1, \vZ \rangle}{2(1 + P_1)} \\  \frac{(n - \norm{\vZ}^2) P_2 + 2 \langle \vX_2, \vZ \rangle}{2(1 + P_2)} \\   \frac{(n - \norm{\vZ}^2) (\Ponetwo) +  2 \langle \vX_1, \vX_2 \rangle + 2 \langle \vZ, \vXltwo \rangle}{2(1 + \Ponetwo)}\end{bmatrix}. \label{eq:itildedist}
		\end{align}
		
		Although the right-hand side of \eqref{eq:itildedist} is not a sum of $n$ independent random vectors, the conditional distribution of $\boldsymbol{\tilde{\imath}}_2 $ given $(\vX_1, \vX_2)$ is such a sum. Therefore, the multidimensional Berry-Esseen theorem is applicable to the corresponding conditional probability. In the remainder of Step 1, we detail the distribution of $\bs{\tilde{\imath}}_2$.
		
		By spherical symmetry, the conditional distribution of $\boldsymbol{\tilde{\imath}}_2$ given $(\vX_1, \vX_2) = (\vx_1, \vx_2)$ depends on $(\vx_1, \vx_2)$ only through the inner product $\langle \vx_1, \vx_2 \rangle$ given that each squared norm satisfies $\norm{\vx_i}^2 = nP_i$, $i \in [2]$. Define the normalized inner product random variable
		\begin{align}
		\Ip \triangleq \frac{\langle \vX_1, \vX_2 \rangle}{\sqrt{n P_1 P_2}},
		\end{align}
		and set
		\begin{align}
		\vx_1 &= (\sqrt{nP_1}, 0, \dots, 0) \label{eq:x1n}\\
		\vx_2 &= (\ip \sqrt{P_2}, \sqrt{(n- \ip^2)P_2}, 0, \dots, 0) \label{eq:x2n}
		\end{align}
		for some $\ip \in [-\sqrt{n}, \sqrt{n}]$,
		which satisfy
		\begin{align}
		\frac{\langle \vx_1, \vx_2\rangle}{\sqrt{n P_1 P_2}} = \ip.
		\end{align}
		Putting \eqref{eq:x1n}--\eqref{eq:x2n} into \eqref{eq:itildedist} gives that the conditional distribution of $\bi$ given $\Ip = \ip$ equals the conditional distribution of $\bi$ given $(\vX_1, \vX_2) = (\vx_1, \vx_2)$, which equals the conditional distribution of the random variable
		\begin{align}
		\boldsymbol{\mu}(\ip) +  \frac{1}{\sqrt{n}} \sum_{i = 1}^n {\mathbf{J}_i(\ip)}, \label{eq:itildegivenQ}
		\end{align}
		where
		\begin{align}
		\boldsymbol{\mu}(\ip) &\triangleq \E{\bi \middle | \Ip = \ip} = \ip \begin{bmatrix} 0 \\ 0 \\ \frac{\sqrt{P_1 P_2}}{1 + \Ponetwo} \end{bmatrix} \label{eq:muq}\\
		\mathbf{J}_i(\ip) &\triangleq \begin{bmatrix}  \frac{(1 - Z_i^2) P_1 + 2 x_{1i} Z_i}{2(1 + P_1)} \\  \frac{(1 - Z_i^2) P_2 + 2 x_{2i} Z_i}{2(1 + P_2)} \\  \frac{(1 - Z_i^2) (\Ponetwo) + 2 (x_{1i}+x_{2i})Z_i}{2(1 + \Ponetwo)}\end{bmatrix}, \quad  i \in [n]. \label{eq:Ji}
		\end{align}
		Here, $\mathbf{J}_i(\ip)$ depends on $\ip$ through the vectors $\vx_1$ and $\vx_2$ given in \eqref{eq:x1n}--\eqref{eq:x2n}.
		Conditioned on the event that $\Ip = \ip$, the modified information density random vector $\bi$ behaves as a sum of conditionally independent but not identical random vectors $\frac{1}{n}\bm{\mu}(h) + \frac{1}{\sqrt{n}} \mathbf{J}_i(h)$ in \eqref{eq:itildegivenQ}.
		
		We next find the distribution of $\Ip$. By spherical symmetry, the distribution of $\Ip$ does not depend on $\vX_1$. Therefore, we can set $\vX_1 = \vx_1$ and get
		\begin{align}
		\Ip \sim \frac{X_{21}}{\sqrt{P_2}}, \label{eq:innerprod}
		\end{align} 
		where $X_{21}$ denotes the first coordinate of $\vX_2$. Therefore, $\Ip$ is distributed according to the marginal distribution of the first coordinate of a random vector distributed uniformly on $\mathbb{S}^{n}(\sqrt{n})$.
		The distribution of $\Ip$ is computed as (e.g.,~\cite[Th.~1]{stam1982})
		\begin{align}
		P_\Ip(\ip) = \frac{\Gamma(\frac n 2)}{\sqrt{\pi n} \Gamma(\frac{n-1}{2})} \nB{1 - \frac{\ip^2}{n}}_{+}^{\frac{n-3}{2}}, \label{eq:qdist}
		\end{align}
		where $\Gamma(\cdot)$ denotes the Gamma function, and $x_+ \triangleq \max\cB{0, x}$ for all $x \in \mathbb{R}$. The support of $\Ip$ is $[-\sqrt{n}, \sqrt{n}]$. From \eqref{eq:qdist}, we compute
		\begin{align}
		\E{\Ip} = 0,  \quad\Var{\Ip} = 1. \label{eq:qexpvar}
		\end{align}
		By Stirling's approximation, $\Ip \to \mathcal{N}(0, 1)$ in distribution as $n \to \infty$ (e.g., \cite[Th. 1]{stam1982}). Recall that an upper bound on the total variation distance between $P_{\Ip}$ and $\mathcal{N}(0, 1)$ is given in \lemref{lem:gotze}.
		
		From \eqref{eq:itildegivenQ}, we find the conditional covariance matrix of the modified information density random vector as
		\begin{align}
		\mathsf{\Sigma}(\ip) &\triangleq \textnormal{Cov}\left[\bi \middle | \Ip = \ip \right] \label{eq:SigmahCov} \\
		&= \textnormal{Cov}\left[\frac{1}{\sqrt{n}} \sum_{i = 1}^n {\mathbf{J}_i(\ip)}\right] \\
		&= \mathsf{\Sigma} + \frac{\ip}{\sqrt{n}} \mathsf{B} \label{eq:sigmaq},
		\end{align}
		where 
		\begin{align}
		\mathsf{\Sigma} &\triangleq \begin{bmatrix} V(P_1)& 			V_{1, 2}(P_1, P_2)  & V_{1, 12}(P_1, P_2)\\ 		V_{1, 2}(P_1, P_2) & V(P_2) & 			V_{2, 12}(P_1, P_2), \label{eq:Sigma}\\ 
		V_{1, 12}(P_1, P_2) & V_{2, 12}(P_1, P_2)& V(\Ponetwo)  \end{bmatrix} \\
		\mathsf{B} &\triangleq  \frac{\sqrt{P_1 P_2}}{(1 + P_1)(1 + P_2)(1 + \Ponetwo)} \notag \\
		&\quad \cdot \begin{bmatrix} 0 & 	1 + \Ponetwo  & 	1 + P_2 \\	1 + \Ponetwo & 0 & 		1 + P_1  \\ 
		1 + P_2 & 1 + P_1 & 	\frac{(1 + P_1)(1 + P_2)}{ (1 + \Ponetwo)}  \end{bmatrix}, \label{eq:B}
		\end{align}
		and $V(P), V_{1, 2}(P_1, P_2)$, and $	V_{i, 12}(P_1, P_2)$, $i \in [2]$, are given in \eqref{eq:dispersion}, \eqref{eq:v1and2}, and \eqref{eq:vi12}, respectively.
		Note that $\mathsf{\Sigma}$ and $\mathsf{B}$ depend only on $P_1$ and $P_2$.
		Using \eqref{eq:muq}, \eqref{eq:qexpvar}, \eqref{eq:sigmaq}, by the law of total expectation and variance, we compute
		\begin{align}
		\E{\bi} &=0 \\
		\Cov{\bi } &= \mathsf{V}(P_1, P_2),
		\end{align}
		where $\mathsf{V}(P_1, P_2)$ is the dispersion matrix defined in \eqref{eq:dispersionmatrix}.
		
		\textbf{Step 2:} We next approximate the distribution of $\boldsymbol{\tilde{\imath}}_2$ by a Gaussian. Toward that end, we consider some auxiliary random variables. Based on our observation in \eqref{eq:itildegivenQ}, we express the probability on the right-hand side of \eqref{eq:itildeprob} by conditioning on $\Ip$ and taking the expectation with respect to $P_{\Ip}$. 
		 Define the probability measure $P_{\tilde{\Ip}}$, and the transition probability kernels $P_{\mathbf{V}|\Ip}$ and $P_{\mathbf{W}|\Ip}$ as
		 \begin{align}
		 P_{\tilde{\Ip}} &\triangleq \mathcal{N}(0, 1)\\
		 P_{\mathbf{V}|\Ip = \ip} &\triangleq \begin{cases} \mathcal{N} \nB{\bm{\mu}(\ip) , \mathsf{\Sigma}(\ip)} \quad &\textnormal{ if } |\ip| \leq \sqrt{n} \\ \mathcal{N} \nB{\bm{\mu}(\ip) , \mathsf{\Sigma}} \quad &\textnormal{ if } |\ip| > \sqrt{n}
		 \end{cases} \\
		  P_{\mathbf{W}|\Ip = \ip} &\triangleq \mathcal{N} \nB{\bm{\mu}(\ip) , \mathsf{\Sigma}} \quad \quad \textnormal{ for } \ip \in (-\infty, \infty). \label{eq:WgivenQ}
		  \end{align}
		 As with $P_{\mathbf{V} | \Ip}$, we extend the definition of the kernel $P_{\bi | \Ip}$ given in \eqref{eq:itildegivenQ} for $|\Ip| > \sqrt{n}$ by choosing $P_{\bi |\Ip = \ip} = \mathcal{N}(\boldsymbol{\mu}(\ip), \mathsf{\Sigma})$ for $|\ip| > \sqrt{n}$ in order for the joint distribution $P_{\tilde{\Ip}} P_{\bi | \Ip}$ to be valid. Recall that $\tilde{\Ip}$ is a Gaussian random variable with the same mean and variance as $\Ip$, and the mean and covariance matrix according to $P_{\mathbf{V}| \Ip = \ip}$ are the same as those for $P_{\bi | \Ip = \ip}$. The Gaussian kernel $P_{\mathbf{W}| \Ip}$ is obtained from $P_{\mathbf{V}| \Ip}$ by replacing its covariance matrix $\mathsf{\Sigma}(\Ip)$ by the mean value of $\mathsf{\Sigma}(\Ip)$, $\mathsf{\Sigma}$.
		 
		 We define the joint distributions $P_{\Ip \, \bi}$, $P_{\tilde{\Ip} \, \bs{\imath}_2^*}$, $P_{\tilde{\Ip} \, \mathbf{V}}$ and $P_{\tilde{\Ip} \, \mathbf{W}}$ as
		 \begin{IEEEeqnarray}{rCl}
		 \IEEEyesnumber
		 \IEEEyessubnumber*
		 P_{\Ip \,\bi} &=& P_{\Ip} P_{\bi | \Ip} \\
		 P_{\tilde{\Ip} \, \bs{\imath}_2^*} &=& P_{\tilde{\Ip}} P_{\bi | \Ip} \\
		 P_{\tilde{\Ip} \, \mathbf{V}} &=& P_{\tilde{\Ip}} P_{\mathbf{V} | \Ip} \\
		  P_{\tilde{\Ip} \, \mathbf{W}} &=& P_{\tilde{\Ip}} P_{\mathbf{W} | \Ip}, \label{eq:WQ}
		 \end{IEEEeqnarray}
where
		 \begin{align}
		{\mathbf{W}} \sim \mathcal{N}(\boldsymbol{0}, \mathsf{V}(P_1, P_2)), \label{eq:WQtilde}
		 \end{align}
which has the desired Gaussian distribution in our Berry-Esseen type bound. 

Let $\mathcal{D}$ be any convex, Borel-measurable subset of $ \mathbb{R}^3$. Then,
\begin{IEEEeqnarray}{rCl}
\IEEEyesnumber
\IEEEyessubnumber*
&& \hspace{-2em}\left \lvert \Prob{\bi \in \mathcal{D}} - {\mathbb{P}}\left[ \mathbf{W} \in \mathcal{D} \right] \right \rvert \label{eq:ineqtriang} \\
&\leq& \left \lvert \Prob{\bi \in \mathcal{D}} - \mathbb{P}\left[ \boldsymbol{\imath}_2^* \in \mathcal{D} \right] \right \rvert \label{eq:dif1}\\
&&+ \left \lvert \mathbb{P}\left[ \boldsymbol{\imath}_2^* \in \mathcal{D} \right] - \mathbb{P} \left[ \mathbf{V} \in \mathcal{D} \right] \right \rvert \label{eq:dif2} \\
&&+ \left \lvert \mathbb{P} \left[ \mathbf{V} \in \mathcal{D} \right]- \mathbb{P} \left[ \mathbf{W} \in \mathcal{D} \right] \right \rvert, \label{eq:dif3}
\end{IEEEeqnarray}
where the inequality in \eqref{eq:dif1} follows from the triangle inequality. The absolute differences in \eqref{eq:dif1}, \eqref{eq:dif2}, and \eqref{eq:dif3} reflect the change of the input measure from $P_{\Ip}$ to $P_{\tilde{\Ip}}$, the change of the transition probability kernel from $P_{\bi |\Ip}$ to $P_{\mathbf{V}|\Ip}$, and the change of the transition probability kernel from $P_{\mathbf{V} | \Ip}$ to $P_{\mathbf{W}|\Ip}$, respectively. We next bound \eqref{eq:ineqtriang} by showing that the absolute difference in each of \eqref{eq:dif1}--\eqref{eq:dif3} is $\bigo{\frac 1 {\sqrt{n}}}$. In the next three steps, we bound each of these absolute differences in turn.
	
		\textbf{Step 3: } We bound the absolute difference in the right-hand side of \eqref{eq:dif1} as
		\begin{IEEEeqnarray}{rCl}
	\IEEEeqnarraymulticol{3}{l}{\left \lvert \Prob{\bi \in \mathcal{D}} - \mathbb{P}\left[ \boldsymbol{\imath}_2^* \in \mathcal{D} \right] \right \rvert \notag} \\
	    &=& \left \lvert \int_{-\infty}^{\infty} \Prob{\bi \in \mc{D} | \Ip = \ip} \left(P_{\Ip}(\ip) - P_{\tilde{\Ip}}(\ip)  \right) d\ip \right \rvert \label{eq:qchange}\\		
		&\leq& \int_{-\infty}^{\infty}  \left \lvert P_{\Ip}(\ip) - P_{\tilde{\Ip}}(\ip) \right \rvert d\ip \label{eq:intdefd1} \\
		&=& 2 \, \mathrm{TV}(P_{\Ip}, P_{\tilde{\Ip}}) \\
		&\leq& 2 \frac{\sqrt{n}}{\sqrt{n - 3}} - 2 \label{eq:gotzeused} \\
		&\leq& \frac {C_{\mathrm{\Ip}}} n, \label{eq:Gaussianswitch}
		\end{IEEEeqnarray}
		where $C_{\mathrm{\Ip}} = 8$. Inequality \eqref{eq:intdefd1} follows by moving the absolute value to the inside of the integral and bounding the conditional probability by 1 for all $\ip$, and \eqref{eq:gotzeused} holds for any $n \geq 4$ by \lemref{lem:gotze}. Inequality \eqref{eq:Gaussianswitch} holds for $n \geq 4$. We conclude that \eqref{eq:Gaussianswitch} holds for any $n$ since \eqref{eq:qchange} is trivially bounded by 1. 
		
		\textbf{Step 4: } We bound the absolute difference due to changing the transition probability kernel from $P_{\bi | \Ip}$ to the Gaussian kernel $P_{\mathbf{V}|\Ip}$ as
		\begin{IEEEeqnarray}{rCl}
	 \IEEEeqnarraymulticol{3}{l}{\left \lvert \mathbb{P}\left[ \boldsymbol{\imath}_2^* \in \mathcal{D} \right] - \mathbb{P} \left[ \mathbf{V} \in \mathcal{D} \right] \right \rvert} \notag \\
	    &=&  \left\lvert \E{\Prob{ \bs{\imath}_2^* \in \mc{D} \middle| \tilde{\Ip}} - \Prob{ \mathbf{V} \in \mc{D} \middle| \tilde{\Ip}} } \right \rvert \label{eq:step2cond}  \\
		&\leq&  \E{ \left \lvert \Prob{ \bs{\imath}_2^* \in \mc{D} \middle| \tilde{\Ip}} - \Prob{ \mathbf{V} \in \mc{D} \middle| \tilde{\Ip}} \right \rvert 1 \cB{ \left \lvert \tilde{\Ip} \right \rvert \leq \frac{\sqrt{n}}{2} }}   \notag \\
		&& + \mathbb{P} \left[\left \lvert \tilde{\Ip} \right \rvert > \frac{\sqrt{n}}{2}\right] \label{eq:separate} \\
		&\leq& \max_{\ip \in \sB{-\frac{\sqrt{n}}{2}, \frac{\sqrt{n}}{2}}} \frac{C(\ip)}{\sqrt{n}}  + \mathbb{P} \left[\left \lvert \tilde{\Ip} \right \rvert > \frac{\sqrt{n}}{2}\right] \label{eq:beused} \\
		&\leq& \frac{C_{\textnormal{BE}}}{\sqrt{n}} + 2 \exp \cB{-\frac n 8} \label{eq:Cher} \\
		&\leq& \frac{C_{\textnormal{BE}} + C_{\textnormal{Ch}}}{\sqrt{n}}, \label{eq:step3}
		\end{IEEEeqnarray}
		where 
		\begin{align}
		T(\ip) &\triangleq \frac 1 n \sum_{i = 1}^n \E{\norm{\mathbf{J}_i(\ip)}^3} \\
		C(\ip) &\triangleq \frac{c \, 3^{1/4} T(\ip)}{\lambda_{\min}(\mathsf{\Sigma}(\ip))^{3/2}} \\
		C_{\textnormal{BE}} &\triangleq \max_{\ip \in \sB{-\frac{\sqrt{n}}{2}, \frac{\sqrt{n}}{2}}} C(\ip) \\
		C_{\textnormal{Ch}} &\triangleq  4 \exp\cB{-\frac 1 2},
		\end{align}
		each $\mathbf{J}_i(\ip)$ is defined in \eqref{eq:Ji}, and $c$ is the Berry-Esseen constant given in \thmref{thm:mbe}. Here, \eqref{eq:separate} moves the absolute value in \eqref{eq:step2cond} to the inside of the expectation. We then separate the expectation into two cases in order to guarantee that we apply the Berry-Esseen theorem for values of $\ip$ such that $\mathsf{\Sigma}(\ip)$ is positive-definite. Inequality \eqref{eq:beused} follows from Corollary \ref{cor:be}, and \eqref{eq:Cher} follows from the Chernoff bound applied to a Gaussian random variable. Inequality \eqref{eq:step3} holds for any $n$. For every $\ip \in \sB{-\frac{\sqrt{n}}{2}, \frac{\sqrt{n}}{2}}$, $\mathsf{\Sigma}(\ip)$ is a non-degenerate covariance matrix, and $T(\ip) < \infty$. Therefore, we conclude that $C_{\textnormal{BE}} < \infty$.

		\textbf{Step 5:} We next bound the probability in \eqref{eq:dif3}, which is the absolute difference due to changing the covariance matrix of the Gaussian kernel from $\mathsf{\Sigma}(\ip)$ to $\mathsf{\Sigma}$, using \lemref{lem:gaussiancdf}, which bounds the total variation distance between two Gaussian vectors. Denote the spectral radius of a $d \times d$ symmetric matrix $\ms{M}$ by
		\begin{align}
		    \rho(\ms{M}) \triangleq \max_{i \in [d]} \left \lvert \lambda_i(\ms{M}) \right \rvert,
		\end{align}
		where $\lambda_i(\cdot)$ is the $i$-th largest eigenvalue of its matrix argument. Let
		\begin{align}
		\mathsf{A} \triangleq \mathsf{ \mathsf{\Sigma}^{-1/2} \mathsf{B} \mathsf{\Sigma}^{-1/2}},
		\end{align}
		where matrices $\mathsf{\Sigma}$ and $\mathsf{B}$ are defined in \eqref{eq:Sigma}--\eqref{eq:B}.
		Then 
		\begin{IEEEeqnarray}{rCl}
	\IEEEeqnarraymulticol{3}{l}{\left \lvert \mathbb{P} \left[ \mathbf{V} \in \mathcal{D} \right]- \mathbb{P} \left[ \mathbf{W} \in \mathcal{D} \right] \right \rvert} \notag \\
	&=&\left\lvert \E{\Prob{ \mathbf{V} \in \mc{D} \middle| \tilde{\Ip}} - \Prob{ \mathbf{W} \in \mc{D} \middle| \tilde{\Ip}} } \right \rvert  \label{eq:s5bef} \\
		 &\leq& \E{ \left \vert \Prob{ \mathbf{V} \in \mc{D} \middle| \tilde{\Ip}} - \Prob{ \mathbf{W} \in \mc{D} \middle| \tilde{\Ip}} \right \rvert } \label{eq:s5move} \\		
		 &\leq& \mathbb{E}\sB{\mathrm{TV}(\mathcal{N}(\boldsymbol{\mu}(\tilde{\Ip}), \mathsf{\Sigma}), \mathcal{N}(\boldsymbol{\mu}(\tilde{\Ip}), \mathsf{\Sigma}(\tilde{\Ip})))} \\
		&\leq& \frac{2 + \sqrt{6}}{4}  \norm{\mathsf{A} }_F \frac{\E{\left \lvert \tilde{\Ip} \right \rvert}}{\sqrt{n}},  \label{eq:tv}
		\end{IEEEeqnarray}
		where \eqref{eq:s5move} follows by moving the absolute value inside the expectation in \eqref{eq:s5bef}, $\mu(\cdot)$ and $\ms{\Sigma}(\cdot)$ are defined in \eqref{eq:muq} and \eqref{eq:SigmahCov}, respectively, and \eqref{eq:tv} follows from \lemref{lem:gaussiancdf}.
		
The matrices $\mathsf{\Sigma}$, $\mathsf{\Sigma} + \mathsf{B}$, and $\mathsf{\Sigma} - \mathsf{B}$ are all positive semidefinite as they are special cases of $\textnormal{Cov}\left[\bi \middle | \Ip = \ip \right]$ in \eqref{eq:SigmahCov} with $h$ equal to $0, \sqrt{n}$, and $-\sqrt{n}$, respectively. Hence $\ms{\Sigma}^{-1/2} (\mathsf{\Sigma} + \mathsf{B}) \ms{\Sigma}^{-1/2}$ and $\ms{\Sigma}^{-1/2} (\mathsf{\Sigma} - \mathsf{B}) \ms{\Sigma}^{-1/2}$ are also positive semidefinite. Since their eigenvalues are respectively given by $1 + \lambda_i(\mathsf{A})$ and $1-\lambda_i(\mathsf{A})$, it follows then that $-1 \leq \lambda_i(\mathsf{A}) \leq 1$ for $i \in [d]$, giving $\rho(\mathsf{A}) \leq 1$.\footnote{Actually, $\rho(\ms{A}) = 1$. Indeed, for $\ip = \sqrt{n}$, the random variables in the first and the second index of the vectors in \eqref{eq:Ji} are identical. Therefore, both $\mathsf{\Sigma}(\sqrt{n}) = \mathsf{\Sigma} + \mathsf{B}$ and $\ms{\Sigma}^{-1/2} (\mathsf{\Sigma} + \mathsf{B}) \ms{\Sigma}^{-1/2}$ have an eigenvalue 0, and $\mathsf{A}$ has an eigenvalue $-1$.} Using the fact that $\norm{\mathsf{M}}_F \leq \sqrt{d} \rho(\ms{M})$ for any $d \times d$ symmetric matrix $\mathsf{M}$, and employing the value of the expectation in \eqref{eq:tv}, we conclude that
		\begin{align}
			\left \lvert \mathbb{P} \left[ \mathbf{V} \in \mathcal{D} \right]- \mathbb{P} \left[ \mathbf{W} \in \mathcal{D} \right] \right \rvert  &\leq \frac{C_{\textnormal{G}}}{\sqrt{n}}, \label{eq:Gaussiantilt}
		\end{align}
		where $C_{\textnormal{G}} = \frac{2 \sqrt{6} + 6}{4 \sqrt{\pi}}$.

		Combining the bounds in \eqref{eq:Gaussianswitch}, \eqref{eq:step3}, and \eqref{eq:Gaussiantilt}, we have the following Berry-Esseen-type inequality
		\begin{align}
		\left \lvert \Prob{\bi  \in \mathcal{D}} - \mathbb{P}\left[ \mathbf{W} \in \mathcal{D} \right] \right \rvert 
		& \leq \frac{C_{\textnormal{\Ip}} +  C_{\textnormal{BE}} + C_{\textnormal{Ch}} + C_{\textnormal{G}} }{\sqrt{n}}  \label{eq:BEfinal}
		\end{align}
		for the modified information density random vector.
		
		\subsection{Completion of the Proof of \thmref{thm:MAC}}\label{sec:completion}
		We employ the set $\mathcal{D} = \cB{\mathbf{x} \in \mathbb{R}^3: \mathbf{x} \geq \frac 1 {\sqrt{n}} \bm{\tau}}$ in \eqref{eq:BEfinal}, where $\bm{\tau}$ is given in \eqref{eq:tauvec}.
		Combining \eqref{eq:itildeprob} and \eqref{eq:BEfinal}, we conclude that the probability $\Prob{\mc{A}^c}$ in \eqref{eq:MACprobgeneralE2} satisfies
		\begin{align}
		\Prob{\mc{A}^c}& \leq 1 - \mathbb{P}\sB{\mathbf{W} \geq  \frac 1 {\sqrt{n}}  \bm{\tau}} + \frac{C_{\textnormal{\Ip}} + C_{\textnormal{BE}} +  C_{\textnormal{Ch}} + C_{\textnormal{G}}}{\sqrt{n}} \\
		&=  1 - \mathbb{P}\sB{\mathbf{W} \leq -  \frac 1 {\sqrt{n}} \bm{\tau}} + \frac{C_{\textnormal{Out}}}{\sqrt{n}}, \label{eq:Zdirection}
		\end{align}
		where $\mathbf{W} \sim \mathcal{N}(\boldsymbol{0}, \mathsf{V}(P_1, P_2))$ and 
		\begin{align}
		C_{\textnormal{Out}} \triangleq C_{\textnormal{\Ip}} + C_{\textnormal{BE}} +  C_{\textnormal{Ch}} + C_{\textnormal{G}}.
		\end{align} 
		Equality \eqref{eq:Zdirection} follows since $\mathbf{W} \sim -\mathbf{W}$. Suppose that $\bm{\tau}$ satisfies
		\begin{IEEEeqnarray}{rCl}
		-\frac{1}{\sqrt{n}} \bm{\tau} &\in& Q_{\mathrm{inv}}\nB{\mathsf{V}(P_1, P_2), \epsilon  - \gamma_n} \label{eq:tau} \\
		\gamma_n &\triangleq& \exp\cB{-c_2 n^{1/3}} + \frac{1 + C_{\textnormal{Out}} }{\sqrt{n}}, \IEEEeqnarraynumspace
		\end{IEEEeqnarray}
		where the constant $c_2$ is as in \eqref{eq:MACprobgeneralE2}. Then, the right-hand side of \eqref{eq:MACprobgeneralE2} is bounded by $\epsilon$. From the Taylor series expansion of $Q_{\mathrm{inv}}(\mathsf{V}, \cdot)$ (e.g., \cite[Lemma~13]{chen2019lossless}), we conclude that \eqref{eq:tau} is equivalent to the inequality in \eqref{eq:MACmainresult},
		which completes the proof.
		
				\section{Proof of \thmref{thm:KMAC}}\label{sec:proofMACK}
		In this section, we sketch the proof of \thmref{thm:KMAC} by detailing the modifications to generalize the proof of \thmref{thm:MAC} from 2 to $K \geq 2$ transmitters. Assume that $\mc{S} \in \subemp{[K]}$. Define the information densities as
		\begin{align}
		\imath_{\mathcal{S}}(\vx_{\mathcal{S}}; \vy | \vx_{\mathcal{S}^c}) \triangleq \log \frac{P_{\vY_K|\vX_{[K]}}(\vy | \vx_{[K]})}{P_{\vY_K|\vX_{\mathcal{S}^c}}(\vy | \vx_{\mathcal{S}^c})}, \label{eq:imathS}
		\end{align}
		where $\mc{S}^c = [K] \setminus \mc{S}$. The information density random vector for $K$ transmitters is
		\begin{align}
		\bs{\imath}_K \triangleq (\imath_{\mathcal{S}}(\vX_{\mc{S}}; \vY_K | \vX_{\mc{S}^c}) \colon \mc{S} \in \subemp{[K]}) \in \mathbb{R}^{2^K - 1},
		\end{align}
		where $\vX_k$ is distributed uniformly on $\mathbb{S}^{n}(\sqrt{nP_k})$ for $k \in [K]$, $\vZ \sim \mc{N}(\bs{0}, \ms{I}_n)$, $\vX_1, \dots, \vX_K$ and $\vZ$ are independent, and $\vY_K = \vX_{\langle [K] \rangle} + \vZ$.

		Below, we use Lemma \ref{lem:kappa} and the generalization of Lemma \ref{lem:gbound} given in \eqref{eq:gS}. The following lemma, which generalizes \lemref{lem:gotze} to $K$ transmitters, is the critical part of the proof of \thmref{thm:KMAC}.

		\begin{lemma}\label{lem:QK}
		Let $\vX_i = (X_{i1}, \dots, X_{in})$, $i = 1, \dots, K$, be $K$ independent random vectors, distributed uniformly on $\mathbb{S}^{n}(1)$. Let $\Ip_{ij} = \sqrt{n} \langle \vX_i, \vX_j \rangle $ for $1 \leq i < j \leq K$, and $\mathbf{\Ip} = (\Ip_{ij} \colon 1 \leq i < j \leq K)$. Then
		\begin{align}
		\mathrm{TV}\left(P_{\mathbf{\Ip}}, \mathcal{N}\left(\boldsymbol{0}, \mathsf{I}_{\frac{K(K-1)}{2} }\right)\right) \leq \frac{C_K}{\sqrt{n}} \label{eq:tvK}
		\end{align} 
for some constant $C_K$ depending only on $K$.
		
		\end{lemma}
		
		\begin{IEEEproof}
		See Appendix \ref{app:proofQK}.
		\end{IEEEproof}
The modifications in \secref{sec:MACproof} are as follows.
\begin{enumerate}
\item The two-transmitter maximum likelihood decoder given in \eqref{eq:ML} is replaced by a $K$-transmitter maximum likelihood decoder, which chooses the message vector $m_{[K]} = (m_1, \dots, m_K)$ corresponding to the maximal information density $\imath_{[K]}(\mathsf{f}_{[K]}(m_{[K]}); \vy)$.
\item The typical set $\mc{F}$ defined in \eqref{eq:F2} is replaced by
\begin{align}
	\mathcal{F}_K &\triangleq \bigtimes_{\mc{S} \in \subemp{[K]}} \mc{F}(\mc{S}) \subseteq \mathbb{R}^{n \otimes (2^K - 1)}, 
\end{align}
where $\mathcal{F}(\mc{S})$ is defined in \eqref{eq:FS}.
Inequality \eqref{eq:consf2} extends to $\mc{F}_K$ by \lemref{lem:kappa}.
\item The functions given in \eqref{eq:g1def}--\eqref{eq:g2def} are extended as
\begin{IEEEeqnarray}{rCl}
	\IEEEeqnarraymulticol{3}{l}{g_{\mc{S}}(t; \vy, \vx_{\mc{S}^c})} \notag \\
	&&\triangleq  \bbP \bigl[\imath_{\mc{S}}(\bar{\vX}_{\mc{S}}; \vY_K|\vX_{\mc{S}^c}) \geq t  \mid \vX_{\mc{S}^c} = \vx_{\mc{S}^c}, \vY_K = \vy \bigr]. \IEEEeqnarraynumspace\label{eq:gKdef}
\end{IEEEeqnarray}
In the proof of \lemref{lem:gbound}, we replace $\Ponetwo$ by $\PS$, and $P_1 P_2$ by $\sum_{\substack{i, j \in [K] \\ i < j}} P_i P_j$. Inequality \eqref{eq:concentU} generalizes to the $K$-transmitter MAC by applying its proof from \appref{app:Ubound} with \lemref{lem:kappa} from \secref{sec:tools}.
Hence, \lemref{lem:gbound} generalizes as
\begin{align}
g_{\mc{S}}(t; \vy, \vx_{\mc{S}^c}) \leq \frac{G(\mc{S})  \exp\cB{-t}}{\sqrt{n}}, \label{eq:gS}
\end{align}
where $G(\mc{S})$ is a constant depending only on the powers $(P_s: s\in \mc{S})$.
\item The high probability events given in \eqref{eq:setE2} and \eqref{eq:setA} are replaced by 
\begin{IEEEeqnarray}{rCl}
\mc{E}_K &\triangleq& \bigcap_{\mc{S} \in \subemp{[K]}}\mc{E}(\mc{S}) \label{setEK},  \\
\mathcal{A}_K &\triangleq& \bigg\{\boldsymbol{\imath}_K \geq \bigg( \log \bigg( \bigg(\prod_{s \in \mc{S}} M_s \bigg) (G(\mc{S})^2) \alpha_{|\mc{S}|, K} \bigg) \colon \notag \\ && \mc{S} \in \subemp{[K]} \bigg) - \frac 1 2 \log n \bs{1} \bigg\}, \label{eq:setAK}
\end{IEEEeqnarray}
where 
\begin{align}
    \alpha_{s, K} \triangleq K \binom{K}{s}, \quad s = 1, \dots, K. \label{eq:alphas}
\end{align}
Using the extension of the RCU bound for $K$ transmitters given in Remark \ref{rem:RCU} and following the same steps as \secref{sec:evalRCU}, we replace the right-hand side of the inequality in \eqref{eq:MACprobgeneralE2} by
\begin{align}
\Prob{\mathcal{A}_K^c} + \exp \cB{-c_K n^{1/3}} + \frac{1}{\sqrt{n}}, \label{eq:RCUboundK}
\end{align}
where $c_K$ is a constant.

\item To understand the differences between bounding $\Prob{\mc{A}_K^c}$ and $\Prob{\mc{A}^c}$, we first extend the definition of the modified and centered information density random vector to $K$ transmitters by defining
	\begin{align}
	&\tilde{\imath}_{\mathcal{S}}(\vx_{\mc{S}}; \vy_K | \vx_{\mc{S}^c}) \triangleq  \sum_{i = 1}^n \log \frac{P_{Y_K|{X}_{[K]}}(y_i | x_{[K] i})}{P_{\tilde{Y}_K| \tilde{X}_{\mathcal{S}^c}}(y_i | x_{\mathcal{S}^c i})}   \\
	&\tilde{\boldsymbol{\imath}}_K \triangleq \frac 1 {\sqrt{n}} \bigl[\nB{\tilde{\imath}_{\mathcal{S}}(\vX_{\mc{S}}; \vY_K | \vX_{\mathcal{S}^c}) \colon \mathcal{S} \in \subemp{[K]}} \notag \\
	&\quad \quad - n \mathbf{C}(P_{[K]})\bigr],
	\end{align}
	where $\mathbf{C}(P_{[K]})$ is the capacity vector defined in \eqref{eq:capacityvectorK}, $\tilde{X}_k \sim \mathcal{N}(0, P_k)$ for $k \in [K]$, and $\prod_{k = 1}^K P_{\tilde{X}_k} \to P_{Y_K | X_{[K]}} \to P_{\tilde{Y}_K} = \mathcal{N}(0, 1 + P_{[K]})$. 
	
	We replace the threshold value in \eqref{eq:tauvec} by
	\begin{align}
	\bm{\tau} &\triangleq \log \bigg(\frac{\nB{\prod_{s \in \mc{S}} M_s} (G({\mc{S}}))^2  \kappa_{|\mc{S}|}(P_{\mc{S}}) \alpha_{|\mc{S}|, K}}{\sqrt{n}} \colon \notag \\
	&\quad \mathcal{S} \in \subemp{[K]} \bigg)  - n \mathbf{C}(P_{[K]}),
	\end{align}
	where $\kappa_{|\mc{S}|}(P_{\mc{S}})$ is the constant (which depends only on $P_{\mc{S}}$) in \eqref{eq:kappaS}.
	Using the joint distribution of $(\vX_{[K]}, \vY_K)$, we get
	\begin{align}
\boldsymbol{\tilde{\imath}}_K  &\sim \frac{1}{\sqrt{n}}  \Bigg(  \frac{(n - \norm{\vZ}^2) \PS}{2( 1+ \PS)} \notag \\
&\quad + \frac{ \sum_{\substack{i, j \in \mathcal{S} \\ i < j}} \langle \vX_i, \vX_j \rangle + \langle \vZ,  \vX_{\langle \mc{S} \rangle} \rangle}{1+ \PS} \colon \mathcal{S} \in \subemp{[K]} \Bigg). \label{eq:imathK}
\end{align}
	Define the random vector 
	\begin{align}
	\mathbf{\Ip} \triangleq (\Ip_{ij} \colon 1 \leq i < j \leq K) \in \mathbb{R}^{\binom{K}{2}},
	\end{align}
	where $\Ip_{ij} = \frac{\langle \vX_i, \vX_j \rangle}{\sqrt{nP_i P_j}}$ denotes the normalized inner product of $\vX_i$ and $\vX_j$. The inner product random vector $\mathbf{\Ip}$ replaces $\Ip$ in \eqref{eq:innerprod}. Observe that for all different $(i_1, j_1)$ and $(i_2, j_2)$ pairs, $\Ip_{i_1 j_1}$ and $\Ip_{i_2 j_2}$ are independent of each other, which follows by independence of $\vX_1, \dots, \vX_K$. However, $\mathbf{\Ip}$ does not have a product distribution due to the fact that any triplets in $\mathbf{\Ip}$ are not jointly independent.\footnote{Given that $\Ip_{12} = \Ip_{13} = \sqrt{n}$, we have that $\vX_1 = \vX_2 = \vX_3$. Therefore, $\Ip_{23}$ is necessarily equal to $\sqrt{n}$ under this condition, and $\Ip_{12}, \Ip_{13}, \Ip_{23}$ are not jointly independent.} While $P_{\mathbf{\Ip}}$ is not a product distribution, \lemref{lem:QK} implies that $P_{\mathbf{\Ip}}$ converges to the distribution of $\binom{K}{2}$ i.i.d. standard Gaussian random variables in total variation, allowing us to use the Berry-Esseen theorem just as we did for the two-transmitter MAC.
	
	As for the two-transmitter MAC, the distribution in \eqref{eq:imathK} depends on $\vX_{[K]}$ only through the inner product random vector $\mathbf{\Ip}$. The conditional distribution of $\tilde{\boldsymbol{\imath}}_K $ given $\mathbf{\Ip} = \mathbf{\ip}$ is the same as the conditional distribution of
\begin{align}
\boldsymbol{\mu}(\mathbf{\ip}) +  \frac{1}{\sqrt{n}} \sum_{i = 1}^n {\mathbf{J}_i(\mathbf{\ip})} \label{eq:itildegivenQK},
\end{align}
where
\begin{align}
\boldsymbol{\mu}(\mathbf{\ip}) &\triangleq \E{\boldsymbol{\imath}_K | \mathbf{\Ip} = \mathbf{\ip}} \notag \\
&= \sum_{\substack{i, j \in [K] \\ i < j}} \ip_{ij} \Big( \frac{\sqrt{P_i P_j}}{1 + \PS}1 \cB{i, j \in \mathcal{S}} \colon \mathcal{S} \in \subemp{[K]} \Big) \label{eq:muqK}\\
\mathbf{J}_i(\mathbf{\ip}) &\triangleq \Big( \frac{(1 - Z_i^2) \PS + 2 \sum_{s \in {\mathcal{S}}} x_{si} Z_i}{2(1 + \PS)} \colon \mathcal{S} \in \subemp{[K]}  \Big) \label{eq:JiK}
\end{align}
for $i \in [n]$,
and $\vx_{[K]}$ are vectors on the $n$-dimensional power spheres, satisfying $\frac{\langle \vx_i, \vx_j \rangle}{\sqrt{n P_i P_j}} = \ip_{ij}$ for all $i < j \in [K]$.
The conditional covariance matrix given in \eqref{eq:sigmaq} is extended to $K$ transmitters as
\begin{align}
\mathsf{\Sigma}(\mathbf{\ip}) = \textrm{Cov}\sB{\iK | \mathbf{\Ip} = \mathbf{\ip}} = \mathsf{\Sigma}_K + \sum_{i, j \in [K], i < j} \frac{\ip_{ij}}{\sqrt{n}} \mathsf{B}_{ij}   \label{eq:sigmaqK},
\end{align}
where the $\nB{\mathbb{R}^{2^K - 1}} \times \nB{\mathbb{R}^{2^K -1 }}$ matrices $\mathsf{\Sigma}_K$ and $\mathsf{B}_{ij}$ have elements 
\begin{IEEEeqnarray}{rCl}
\Sigma_{\mathcal{S}_1 \mathcal{S}_2} &=& \frac{P_{\mathcal{S}_1} P_{\mathcal{S}_2}  + 2 P_{\mathcal{S}_1 \cap \mathcal{S}_2}}{2(1 + P_{\mathcal{S}_1})(1 + P_{\mathcal{S}_2)}} \\
b_{\mathcal{S}_1 \mathcal{S}_2} &=& \frac{\sqrt{P_i P_j}}{(1 + P_{\mathcal{S}_1})(1 + P_{\mathcal{S}_2})} \notag \\
&& \cdot  1\cB{\cB{i \in \mathcal{S}_1, j \in \mathcal{S}_2} \cup \cB{i \in \mathcal{S}_2, j \in \mathcal{S}_1} } \IEEEeqnarraynumspace
\end{IEEEeqnarray}
for $\mathcal{S}_1, \mathcal{S}_2 \in \subemp{[K]}$. These formulas generalize the formulas for the two-transmitter MAC given in \eqref{eq:Sigma} and \eqref{eq:B}.
By \eqref{eq:muqK}, \eqref{eq:sigmaqK}, and the pairwise independence of $\Ip_{i_1 j_1}$, $\Ip_{i_2 j_2}$ for all different $(i_1, j_1)$ and $(i_2, j_2)$ pairs, using the law of total expectation and variance, we find that
\begin{align}
\E{\iK} &= \bs{0} \\
\Cov{\iK} &= \ms{V}(P_{[K]}),
\end{align}
where the covariance matrix $\ms{V}(P_{[K]})$ is defined in \eqref{eq:VPvec}. 

The rest of the proof follows the proof in \secref{sec:be}, where we replace $\Ip$ by $\mathbf{\Ip}$, $\tilde{\Ip}$ by the $\binom{K}{2}$-dimensional standard Gaussian random vector $\tilde{\mathbf{\Ip}}$, $P_{\bi | \Ip}$ by $P_{\tilde{\boldsymbol{\imath}}_K | \mathbf{\Ip}}$, $P_{\mathbf{V}|\Ip}$ by $P_{\mathbf{V} | \mathbf{\Ip}}$, and $P_{\mathbf{W}|\Ip}$ by $P_{\mathbf{W} | \mathbf{\Ip}}$. For the probability transition kernels $P_{\mathbf{V} | \mathbf{\Ip}}$ and $P_{\mathbf{W} | \mathbf{\Ip}}$, we replace
$\bm{\mu}(\ip)$ by $\bm{\mu}(\mathbf{\ip})$, $\ms{\Sigma}$ by $\ms{\Sigma}_K$, and $\mathsf{\Sigma}(\ip)$ by $\mathsf{\Sigma}(\mathbf{\ip})$. We replace all conditions in the form $|\ip| \leq t$ by $|\mathbf{\ip}| \leq t \bs{1}$.

The only critical modification is that the bound on the total variation distance $\mathrm{TV}(P_{\Ip}, P_{\tilde{\Ip}})$ in \eqref{eq:gotzeused} is replaced by the bound on the total variation distance $\mathrm{TV}(P_{\mathbf{\Ip}}, P_{\tilde{\mathbf{\Ip}}})$, which is $\bigo{\frac{1}{\sqrt{n}}}$ by \lemref{lem:QK}. We conclude that
\begin{align}
		\left \lvert \Prob{\iK  \in \mathcal{D}} - \mathbb{P}\left[ \mathbf{W} \in \mathcal{D} \right] \right \rvert 
		& \leq \frac{C_K }{\sqrt{n}}  \label{eq:BEfinalK}
\end{align}
for some constant $C_K > 0$, where $\mathbf{W} \sim \mathcal{N}(\bs{0}, \ms{V}(P_{[K]}))$. 

By combining \eqref{eq:RCUboundK} and \eqref{eq:BEfinalK} as in \secref{sec:completion}, we complete the proof of \thmref{thm:KMAC}.
\end{enumerate}
		
		\section{Proof of \thmref{thm:nonasymRAC}} \label{sec:proof:nonasymRAC}
 The main difference between the coding strategies for the Gaussian MAC and RAC is that for the Gaussian RAC, an output typicality condition is added to the decoding function in order to reliably detect the number of active transmitters.
\subsection{Encoding and Decoding} \label{sec:RACencoding}
\textbf{Encoding}: Recall that $n_K$ is the largest decoding time. In our encoding strategy, rather than adapting the codebook to the estimate of the number of active transmitters at the receiver, we generate codewords with length $n_K$. Each active transmitter transmits one symbol of its message codeword at each time step until the decoder signals at time $n_k \in \{n_0, \dots, n_K\}$ that it is able to decode. If decoding happens at time $n_k$, only the initial sub-codeword of length $n_k$ is used.

The common randomness random variable $U \in \mathbb{R}^{M n_K}$ has the distribution
	\begin{align}
	    P_U = \underbrace{P_{U(1)} \times P_{U(2)} \times \cdots \times P_{U(M)}}_{M \text{times}},
	\end{align}
where $P_{U(m)} = P_{\vX}$ for $m \in [M]$.
The realization of $U$ defines $M$ length-$n_K$ i.i.d. codewords.
In other words, the encoding function is given by
\begin{align}
\mathsf{f}(U, m) = U(m), \quad m \in [M].
\label{eq:fPX}
\end{align}

The need for using common randomness in encoding is due to the requirement that a single code must satisfy multiple constraints, i.e., the error probability constraints in \eqref{eq:errorprobcons}. Showing that a random code satisfies multiple constraints does not imply the existence of a deterministic code. This issue arises in \cite{polyanskiy2011feedback}, where a variable-length feedback code must satisfy both an average decoding time and an average-error constraint, and in \cite{yavas2020Random}, where a RAC code must satisfy $K + 1$ average error probability constraints. See \cite[Sec.~II.C]{yavas2020Random} for details.
	
	\textbf{Decoding}: Unlike the MAC, for the Gaussian RAC, we require the decoder to determine the time $n_k \in \{n_0, \dots, n_K\}$ at which to decode. Therefore, we couple the maximum likelihood decoder given in \eqref{eq:ML} with a threshold rule, used to estimate the number of transmitters and a single bit of feedback at each time $n_i$ up to and including the time $n_k$ at which the decoder decides to decode. The maximum likelihood decoder is applied only if the threshold test is satisfied. Here, the role of the threshold rule is to reliably determine the true channel in the communication epoch. We use a threshold rule to determine the number of active transmitters because for any $P > 0$, under an input distribution $P_{\vX}$ such that the expected input power meets the power constraint in \eqref{eq:powerRAC} with equality (i.e., $\frac{1}{n_k} \E{\norm{\vX^{[n_k]}}^2} = P $), for each $k$, the normalized squared norm of the output $\vY_k^{[n_k]}$ concentrates around its mean. That mean is different for each $k \in \{0, 1, \dots, K\}$; specifically
	\begin{align}
	\frac 1 {n_k} \, \E{ \norm{\vY_k^{[n_k]}}^2} = 1 + kP, \quad \forall k \in \{0\} \cup [K].
	\end{align}

	Upon receiving the first $n_0$ symbols of the
output, $\vy^{[n_0]}$, the decoder computes the following function
\begin{align}
\mathsf{g}_0(U, \vy^{[n_0]}) = \begin{cases} 
	0 &\mbox {if } \left\lvert \frac 1 {n_0} \norm{\vy^{[n_0]}}^2 - 1 \right \rvert \leq \lambda_0 \\ 
	\mathsf{e} &\mbox{otherwise}
	\end{cases}
\end{align}
to decide whether there are any active transmitters;
here $\lambda_0$ is a parameter that is determined by the error criterion $\epsilon_0$. At time $n_0$, if $\mathsf{g}_0(U, \vy^{[n_0]}) = 0$, the receiver broadcasts a bit value 1 to all transmitters, signaling that the receiver estimates ``no active transmitters" and the epoch ends. Otherwise the receiver broadcasts a bit value 0 and the epoch continues.

For $k \geq 1$, the decoder applies the following function to make a decision at each subsequent time $n_k \leq n_K$
\begin{align}
\mathsf{g}_k(U, \vy^{[n_k]}) = \begin{cases} 
	m_{[k]} &\mbox {if } 
	\imath_{[k]}(\mathsf{f}(U, m_{[k]})^{[n_k]}; \vy^{[n_k]}) \\
	& \,\,\, > \imath_{[k]}(\mathsf{f}(U, m'_{[k]})^{[n_k]}; \vy^{[n_k]}) \,\, \\
	& \,\,\,\textnormal{for all }  m'_{[k]} \stackrel{\pi}{\neq}  m_{[k]}, \\
	& m_1 \leq \dots \leq m_k, \\
	& \left\lvert \frac 1 {n_k} \norm{\vy^{[n_k]}}^2 - (1 + kP) \right \rvert \leq \lambda_k \\
	\mathsf{e} &\mbox{otherwise}, \label{eq:MLRAC}
	\end{cases}
\end{align}
where $\lambda_k$ is a parameter chosen to satisfy the error criterion $\epsilon_k$. At time $n_k$, if $\mathsf{g}_k(U, \vy^{[n_k]}) \neq \ms{e}$ or $k = K$, then the receiver broadcasts the bit value 1 to all transmitters, signaling the end of epoch and the start of next one. Otherwise, the receiver sends feedback 0 and the epoch continues.

By the permutation-invariance of the channel in terms of the inputs $\vX_{[k]}$ and the identical encoding in \eqref{eq:fPX}, all permutations of the messages $m_{[k]}$ give the same information density. Therefore, without loss of generality, the output of our decoder is always the ordered message vector in \eqref{eq:MLRAC}. The condition $\left\lvert \frac 1 {n_k} \norm{\vy^{[n_k]}}^2 - (1 + kP) \right \rvert \leq \lambda_k$, which does not depend on the randomly generated codebook, allows us with high probability to decode at time $n_k$ when the number of active transmitters is $k$, rather than decoding earlier or failing to decode at the time $n_k$ intended for the $k$-transmitter scenario.
		
\subsection{Error Analysis} \label{sec:RACerroranalysis}
In this section, we bound the probability of error for the random access code in Definition \ref{def:RAC}. 

\emph{No active transmitters}: For $k = 0$, the only error event is that the squared norm of the output $\vY_0^{[n_0]}$ is away from its mean:
\begin{align}
\epsilon_0 \leq \Prob{\left\lvert \frac 1 {n_0}\norm{ \vY_0^{[n_0]}}^2 - 1 \right\rvert > \lambda_0}. \label{eq:noactive}
\end{align}

\emph{$k \geq 1$ active transmitters}: When there is at least one active transmitter, the encoding function \eqref{eq:fPX} and decoding rule \eqref{eq:MLRAC} yield an error if and only if at least one of the following events occurs:

\begin{itemize}
\item $\mc{E}_{\textnormal{codeword}}$: At least one of the $k$ codewords associated with the sent messages $m_{[k]}$ violates the power constraint in \eqref{eq:powerRAC} in the first $n_k$ symbols. In this case, an error occurs since it is forbidden to transmit those codewords. We do not need to include the power constraint violation beyond the $n_k$-th symbol since that event is captured by the event of decoding time error, stated next.
\item $\mc{E}_{\textnormal{time}}$: A list of messages is decoded at a wrong decoding time $n_t \neq n_k$, or no messages is decoded during the entire epoch.
\item $\mc{E}_{\textnormal{message}}$: A list of messages $m_{[k]}' \neq m_{[k]}$ is decoded at time $n_k$.
\end{itemize}
In the following discussion, we bound the probability of these events separately, and apply the union bound to combine them. 

Since we are employing identical encoders at all encoders, we simplify the analysis by treating the event $\mc{E}_{\textnormal{rep}} = \{W_i = W_j \text{ for some } i \neq j \}$ that at least one message among transmitted messages is repeated as an error. While this case is actually advantageous to decoding, it requires special treatment since it violates the assumption of codeword independence employed in our analysis.

By the union bound, 
\begin{align}
\Prob{\mc{E}_{\textnormal{rep}}} \leq \frac{k (k-1)}{2 M}. \label{eq:erep}
\end{align}
Applying the union bound, we bound the error probability as
\begin{IEEEeqnarray}{rCl}
\epsilon_k &=& \frac1{M^k}\sum_{m_{[k]}\in [M]^k} 
        \bbP \biggl[ \bigcup_{t \colon n_t \leq n_k, t \neq k} \cB{\mathsf{g}_t(U, \vY_k^{[n_t]})\neq \mathsf{e}} \notag \\
&& \bigcup \cB{\mathsf{g}_k(U, \vY_k^{[n_k]}) \stackrel{\pi}{\neq} m_{[k]}} \Bigm| W_{[k]} = m_{[k]} \biggr]\label{eq:epsk1} \\
&\leq& \Prob{\mc{E}_{\textnormal{rep}}}  + \Prob{\mc{E}^c_{\textnormal{rep}}} \Big(\Prob{\mc{E}_{\textnormal{codeword}} \middle| \mc{E}^c_{\textnormal{rep}} } \\
&& + \Prob{\mc{E}_{\textnormal{time}} \middle| \mc{E}^c_{\textnormal{rep}}} + \Prob{\mc{E}_{\textnormal{message}} \middle| \mc{E}^c_{\textnormal{rep}}} \Big) \\
&\leq&  \Prob{\mc{E}_{\textnormal{rep}}}  +  \Prob{\mc{E}_{\textnormal{codeword}} \middle| \mc{E}^c_{\textnormal{rep}} } \notag \\
&& + \Prob{\mc{E}_{\textnormal{time}} \middle| \mc{E}^c_{\textnormal{rep}}} + \Prob{\mc{E}_{\textnormal{message}} \middle| \mc{E}^c_{\textnormal{rep}}}. \label{eq:GRACRCU}
\end{IEEEeqnarray}
\emph{Power constraint violation}: The probability that a power constraint violation occurs in the first $n_k$ symbols for at least one of the $k$ distinct messages is
\begin{align}
\Prob{\mc{E}_{\textnormal{codeword}} \middle| \mc{E}^c_{\textnormal{rep}} } = \Prob{\bigcup_{i = 1}^k \bigcup_{\substack{j: n_j \leq n_k \\ j \geq 1}} \cB{\frac 1 {n_j} \norm{\vX_i^{[n_j]}}^2 > P}}. \label{eq:powerviolation}
\end{align}
\emph{Wrong decoding time:} According to the decoding rule in \eqref{eq:MLRAC}, decoding occurs at time $n_k$ if and only if the output typicality criterion is not satisfied for any $t$ with $n_t \leq n_k$ and $t \neq k$ (that is $\left\lvert \frac 1 {n_t} \norm{\vy^{[n_t]}}^2 - (1 + tP) \right \rvert > \lambda_t$), and is satisfied for $k$ (that is $\left\lvert \frac 1 {n_k} \norm{\vy^{[n_k]}}^2 - (1 + kP) \right \rvert \leq \lambda_k$). Note that it is possible that no message set is decoded during an entire epoch. This would happen if $\left\lvert \frac 1 {n_t} \norm{\vy^{[n_t]}}^2 - (1 + tP) \right \rvert > \lambda_t$ for $t \in \{0, \dots, K\}$. The probability $\Prob{\mc{E}_{\textnormal{time}} \middle| \mc{E}^c_{\textnormal{rep}}}$ is computed as
\begin{align}
\Prob{\mc{E}_{\textnormal{time}} \middle| \mc{E}^c_{\textnormal{rep}}} &= \bbP \Biggl[ \bigcup_{\substack{t: n_t \leq n_k \\ t \neq k}} \cB{\left\lvert \frac 1 {n_t} \norm{\vY_k^{[n_t]}}^2 - (1 + tP) \right \rvert \leq \lambda_t} \notag \\
&\quad \bigcup \cB{ \left\lvert \frac 1 {n_k} \norm{\vY_k^{[n_k]}}^2 - (1 + kP) \right \rvert > \lambda_k } \Biggr]. \label{eq:wrongtime_nonasym}
\end{align}
\emph{Wrong message:} By using the RCU bound in Remark \ref{rem:RCU} and the permutation-invariance of the information density, we bound $\Prob{\mc{E}_{\textnormal{message}} \middle| \mc{E}^c_{\textnormal{rep}}}$ as
\begin{IEEEeqnarray}{rCl}
&&\Prob{\mc{E}_{\textnormal{message}} \middle| \mc{E}^c_{\textnormal{rep}}} \leq \bbE \Biggl[ \min\Bigg\{ 1, \sum_{s = 1}^k \binom{k}{s}  \binom{M-k}{s} \notag \\
&&\quad \bbP \Bigl[\imath_{[s]}(\bar{\vX}^{[n_k]}_{[s]}; \vY_{k}^{[n_k]}| \vX_{[s+1:k]}^{[n_k]}) \notag \\
&& \quad \quad \geq \imath_{[s]}(\vX_{[s]}^{[n_k]}; \vY_k^{[n_k]}| \vX_{[s+1:k]}^{[n_k]}) \Bigm| \vX_{[k]}^{[n_k]}, \vY_k^{[n_k]} \Bigr] \Bigg\} \Biggr]. \IEEEeqnarraynumspace \label{eq:wrongmessage}
\end{IEEEeqnarray}
Combining \eqref{eq:noactive}, \eqref{eq:erep} and \eqref{eq:GRACRCU}--\eqref{eq:wrongmessage} completes the proof. Note that compared to the achievability proof of the Gaussian MAC in \eqref{eq:RCUKnonasymp}, the multiplicative constant in \eqref{eq:wrongmessage} is $\binom{M-k}{s}$ instead of $(M-1)^s$ since we are given that the transmitted messages are distinct.

		\section{Proof of \thmref{thm:GRAC}} \label{sec:proofRAC}
		
In this section, we analyze the achievability bound in \thmref{thm:nonasymRAC} by particularizing the input distribution, $P_{\vX}$ in \thmref{thm:nonasymRAC}, choosing the free parameters $\lambda_k$, decoding times $n_0, n_1, \dots, n_K$, and bounding the probability and expectation terms in \eqref{eq:ekbound37}. In the rest of the proof, we assume that the decoding times satisfy $n_0 < n_1 < \dots < n_K$, which we make explicit in \eqref{eq:klogM}. 

\subsection{Particularizing $P_{\vX}$}
We modify the input distribution used in \thmref{thm:MAC} for the Gaussian MAC so that the randomly generated codewords meet the power constraints with probability 1.

A random codeword distributed according to $P_{\vX}$ has length $n_K$ and consists of $K$ independent sub-codewords. The $j$-th sub-codeword has length $|\mc{N}(j)|$, where 
		\begin{align}
		\mc{N}(j) \triangleq \begin{cases}
		[n_1] &\mbox{ if } j = 1 \\
		\{ n_{j-1} + 1, n_{j-1} + 2, \dots, n_j \} &\mbox{ if } 2 \leq j \leq K
		\end{cases}
		\end{align}
for $j \in [K]$ is the index set for the $j$-th block in our code design.
		Thus, the input distribution $P_{\vX}$ in \thmref{thm:nonasymRAC} is
		\begin{align}
		P_{\vX}(\vx) = \prod_{j = 1}^K P_{\vX^{\mc{N}(j)}}\left(\vx^{\mc{N}(j)} \right) \label{eq:inputRAC},
		\end{align}
		where
	\begin{align}
		P_{\vX^{\mc{N}(j)}}\left(\vx^{\mc{N}(j)} \right) = \frac{\delta \left(\norm{\vx^{\mc{N}(j)}}^2 - |\mc{N}(j)| P \right)}{S_{|\mc{N}(j)| }(\sqrt{|\mc{N}(j)|  P})} \label{eq:inputRACpiece},
	\end{align}
	that is, $\vX^{\mc{N}(j)} \sim \textrm{Uniform}\left(\mathbb{S}^{|\mc{N}(j)|}(\sqrt{ |\mc{N}(j)| P})\right)$, and $\vX^{\mc{N}(1)}, \dots, \vX^{\mc{N}(K)}$ are independent.

Codewords chosen according to \eqref{eq:inputRAC} satisfy the power constraints in \eqref{eq:powerRAC} with equality, giving
\begin{align}
\Prob{\bigcup_{i = 1}^k \bigcup_{j = 1}^k \cB{\frac 1 {n_j} \norm{\vX_i^{[n_j]}}^2 > P}} = 0. \label{eq:powerviol0}
\end{align}

\subsection{Error Analysis}
We separate the analysis into 3 steps: deriving an output typicality bound, evaluation of the RCU bound, and evaluation of a Berry-Esseen type inequality.

\textbf{Step 1}: In this step, we bound the probability that the output $\vY_k^{[n_k]}$ does not satisfy the condition \newline $\left\lvert \frac 1 {n_k} \norm{\vY_k^{[n_k]}}^2 - (1 + kP) \right \rvert \leq \lambda_k$ given in the decoding rule \eqref{eq:MLRAC}. Since for $k \geq 1$, $\vY_k^{\mc{N}(1)}, \vY_k^{\mc{N}(2)}, \dots, \vY_k^{\mc{N}(K)}$ are independent due to the input distribution in \eqref{eq:inputRAC}, \lemref{lem:kappa} and \lemref{lem:chi2} imply
\begin{IEEEeqnarray}{rCl}
\IEEEeqnarraymulticol{3}{l}{\Prob{\left\lvert \norm{\vY_k^{[n_k]}}^2 - n_k(1 + k P) \right\rvert > n_k \lambda_k}} \notag \\
&\leq& 2 \nB{\kappa_k(P \bs{1})}^k \exp\left\{-\frac{n_k \lambda_k^2}{8(1 + kP)^2}\right\} \label{eq:outputtypicalityk}
\end{IEEEeqnarray}
for $\lambda_k \in (0, 1 + kP)$,
where $\kappa_j(P \bs{1})$ is the constant defined in \lemref{lem:kappa}.
For $k = 0$, we have
\begin{align}
\Prob{\left\lvert \norm{\vY_0^{[n_0]}}^2 - n_0 \right\rvert > n_0 \lambda_0} &\leq 2 \exp\left\{-\frac{n_0 \lambda_0^2}{8}\right\} \label{eq:outputtypicality0}
\end{align}
for $\lambda_0 \in (0, 1)$.
We pick 
\begin{align}
\lambda_0 = \sqrt{\frac{-8 \log \frac{\epsilon_0}{2 }}{n_0}} 
\label{eq:lambda0}
\end{align}
to ensure that the right-hand side of \eqref{eq:outputtypicality0} is bounded above by $\epsilon_0$.
By setting $\lambda_t = \frac{P}{2}$ for $t \geq 1$, using \eqref{eq:outputtypicalityk} and \eqref{eq:outputtypicality0}, and applying the union bound, we bound the probability of decoding time error in \eqref{eq:decodingtimeerror} by 
\begin{align}
B &\triangleq 2 \kappa_1(P) \exp\cB{-\frac{ n_0 ((k - \frac {\lambda_0}{P}) P)^2 }{8(1 + kP)^2}} \notag \\
&\quad + 2 \sum_{t = 1}^{k}  \nB{\kappa_k(P \bs{1})}^t  \exp\cB{-\frac{ n_t ((k - t - \frac 1 2) P)^2 }{8(1 + kP)^2}}.
 \label{eq:wrongtimebound}
\end{align}

\textbf{Step 2}: To bound the expectation in \eqref{eq:ekbound}, we first modify the definition of the typical output set $\mathcal{F}(\mc{S})$ in \eqref{eq:FS} as
	\begin{align}
		\mc{F}(\mc{S})_{\textnormal{RAC}} &\triangleq \Big\{\vy^{[n_k]} \in \mathbb{R}^{n_k} \colon  \notag \\
		&\quad  \frac{1}{|\mc{N}(j)|} \norm{\vy^{\mc{N}(j)}}^2 \in \mc{I}(j, \mc{S}) \mbox{ for } j \in [k]\Big\}. \label{eq:FSRAC} \\
		\mc{I}(j, \mc{S}) &\triangleq [1 + |\mc{S}|P - |\mc{N}(j)|^{-1/3}, \notag \\
		&\quad \quad 1 + |\mc{S}|P + |\mc{N}(j)|^{-1/3}].
	\end{align}
We then show that \lemref{lem:gbound} holds under input distribution \eqref{eq:inputRAC} with typical output set \eqref{eq:FSRAC}. That is,
for every $0 < s \leq k$, and $\vy^{[n_k]}$ and $\vx_{[k]\setminus [s]}^{[n_k]}$ such that $\vy^{[n_k]} - \vx_{\langle [k]\setminus [s] \rangle }^{[n_k] } \in \mc{F}([s])_{\textnormal{RAC}}$, we prove that
\begin{IEEEeqnarray}{rCl}
	\IEEEeqnarraymulticol{3}{l}{g_{[s]}(t; \vy^{[n_k]}, \vx_{[k]\setminus [s]}^{[n_k]})} \notag \\
	&\triangleq&  \bbP \Bigl[\imath_{[s]}(\bar{\vX}_{[s]}^{[n_k]}; \vY_k^{[n_k]}|\vX_{[k]\setminus [s]}^{[n_k]}) \geq t \notag \\
	&& \Bigm | \vX_{[k]\setminus [s]}^{[n_k]} = \vx_{[k]\setminus [s]}^{[n_k]}, \vY_k^{[n_k]} = \vy^{[n_k]} \Bigr] \\
	&\leq& \frac{G_{s, k}'  \exp\cB{-t}}{\sqrt{n_k}}, \label{eq:Gsprime}
\end{IEEEeqnarray}
where $G_{s, k}'$ is a positive constant depending on $s, k$ and $P$. 

The derivation of the bound in $\eqref{eq:Gsprime}$ follows the analysis in \secref{sec:boundingmutualinfo}. The critical goal is to verify steps \eqref{eq:hy2tanlemma0}--\eqref{eq:hy2tanlemma2} for the modified input distribution in \eqref{eq:inputRAC}. This requires showing that
	\begin{align}
	&\Prob{\langle \vX_{\langle[s]\rangle}^{[n_k]}, \vX_{\langle[s]\rangle}^{[n_k]} + \vZ^{[n_k]} \rangle - \sum_{j = 1}^k\frac{|\mc{N}(j)| u_j}{2}\in [a, a + \mu] \,\middle | \mc{E}} \notag \\
	&\quad \leq  \bigo{\frac{1}{\sqrt{n_k}}} \label{eq:ProbOnk},
	\end{align}
where
	\begin{align}
	\mc{E} &= \Big\{\norm{\vX_{\langle[s]\rangle}^{\mc{N}(j)} + \vZ^{\mc{N}(j)}}^2 = |\mc{N}(j)| s_j, \notag \\
	&\quad \quad \norm{\vX_{\langle[s]\rangle}^{\mc{N}(j)}}^2 = |\mc{N}(j)| u_j \mbox{ for } j\in [k] \Big\},
	\end{align}
$s_j \in \mc{I}(j, [s])$, and $u_j > 0$. The proof of \eqref{eq:ProbOnk} is similar to the one in \cite[Appendix A]{tan2015Third} for parallel Gaussian channels since we can consider $K$ independent sub-codewords with lengths $|\mc{N}(j)|$, $j \in [K]$, as $K$ parallel channels, each having blocklength $|\mc{N}(j)|$, $j \in [K]$.
	 
	 Taking an arbitrary $t \in [k]$, we get
		\begin{IEEEeqnarray}{rCl}
		\IEEEeqnarraymulticol{3}{l}{\Prob{\langle \vX_{\langle[s]\rangle}^{[n_k]}, \vX_{\langle[s]\rangle}^{[n_k]} + \vZ^{[n_k]} \rangle - \sum_{j = 1}^k\frac{|\mc{N}(j)| u_j}{2}\in [a, a + \mu] \,\middle | \mc{E}}}  \notag \\
		&=& \int_{\mathbb{R}^{k-1}} \bbP \biggl[ Z_{n_{t-1}+1} + \frac{\sqrt{|\mc{N}(j)|}}{2} \in  \biggl[ \frac{a'}{\sqrt{|\mc{N}(j)|}}, \frac{a' + \mu}{\sqrt{|\mc{N}(j)|}} \biggr] \notag \\
		&&\quad \quad \biggm|  \mc{E}, \cB{Z_{n_{j-1}+1} = z_j, j \in [k]\setminus \cB{t}} \biggr] \notag \\
		&&\quad \quad \cdot \bigg(\prod_{\substack{j \in [k] \\ j \neq t}} f_{Z_{n_{j-1}+1}|\mc{E}}(z_j) dz_j  \bigg) \label{eq:tcondk} \\
		&\leq& \frac{L(u_t, s_t) \mu}{\sqrt{|\mc{N}(t)|}}  \\
		&\leq& \frac 3  2\frac{ L(u_t, 1 + sP)  \mu}{\sqrt{|\mc{N}(t)|}} \\
		&\leq&  \frac 3  2\frac{ \max_{j \in [k]} L(u_j, 1 + sP)  \mu}{\sqrt{|\mc{N}(t)|}}, 
		\end{IEEEeqnarray}
where $a'$ is related to $a$ by a constant shift, and \eqref{eq:tcondk} follows by setting $\vX_{\langle [s] \rangle}^{\mc{N}(j)} = (\sqrt{|\mc{N}(j)| u_j}, 0, \dots, 0)$, and conditioning on the event that $\{Z_{n_{j-1}+1} = z_j \mbox{ for } j \neq t\}$.
Since $t$ is arbitrary in \eqref{eq:tcondk}, we have
\begin{IEEEeqnarray}{rCl}
\IEEEeqnarraymulticol{3}{l}{\Prob{\langle \vX_{\langle[s]\rangle}^{[n_k]}, \vX_{\langle[s]\rangle}^{[n_k]} + \vZ^{[n_k]} \rangle - \sum_{j = 1}^k\frac{|\mc{N}(j)| u_j}{2}\in [a, a + \mu] \,\middle | \mc{E}}} \notag\\
&\leq& \frac 3 2 \frac{ \max_{j \in [k]} L(u_j, 1 + sP)  \mu}{\sqrt{\max_{t \in [k]}{|\mc{N}(t)|}}} \\
&\leq& \frac 3 2 \frac{ \sqrt{k} \max_{j \in [k]} L(u_j, 1 + sP)  \mu}{\sqrt{n_k}},
\end{IEEEeqnarray}
which implies \eqref{eq:ProbOnk}, and \eqref{eq:Gsprime} follows.

In the following discussion, we modify the analysis in 
\secref{sec:evalRCU} according to the input distribution in \eqref{eq:inputRAC}.
Define the information density random vector $\bs{\imath}_k$ and the typical events analogous to \eqref{eq:setE1}--\eqref{eq:setA} as
\begin{align}
	\bs{\imath}_k &\triangleq (\imath_{\mathcal{S}}(\vX_{\mc{S}}^{[n_k]}; \vY_k^{[n_k]} | \vX_{\mc{S}^c}^{[n_k]}) \colon \mc{S} \in \subemp{[k]}) \\
		\mc{E}(\mc{S})_{\textnormal{RAC}} &\triangleq \cB{\vX_{\langle \mc{S} \rangle}^{[n_k]} + \vZ^{[n_k]} \in \mc{F}(\mc{S})_{\textnormal{RAC}}} \\
		\mc{E}_{\textnormal{RAC}} &\triangleq \bigcap_{\substack{\mc{S} \in \subemp{[k]}}}\mc{E}(\mc{S})_{\textnormal{RAC}} \\
		\mc{A}_k &\triangleq \bigg \{ \boldsymbol{\imath}_k \geq \bigg(\log \left(\binom{M-k}{|\mc{S}|}  (G_{|\mc{S}|, k}')^2 \alpha_{|\mc{S}|, k} \right) \notag \\
		&\quad \colon \mc{S} \in \subemp{[k]} \bigg)  - \frac 1 2 \log n_k \bs{1} \bigg\},
		\end{align}
		where $\alpha_{s, k}$ is given in \eqref{eq:alphas}.
By \lemref{lem:chi2} and the union bound, we have
		\begin{align}
		\Prob{\mc{E}_{\textnormal{RAC}}^c} \leq \sum_{j = 1}^k \exp\cB{-c_k |\mc{N}(j)|^{1/3}}, \label{eq:EcRAC}
		\end{align}
		where $c_k$ is a positive constant. Combining \eqref{eq:Gsprime} and \eqref{eq:EcRAC} and following the analysis in \secref{sec:evalRCU}, we bound the expectation in \eqref{eq:ekbound} by
        \begin{align}
        \Prob{\mc{A}_k^c} + \sum_{j = 1}^k \exp\cB{-c_k |\mc{N}(j)|^{1/3}} + \frac{1}{\sqrt{n_k}}. \label{eq:Pwrongmes}
		\end{align}
		
		\textbf{Step 3}: Given $M$ and $\{\epsilon_k\}_{k = 0}^K$, we set the decoding times $n_1, \dots, n_K$ according to the equalities
		\begin{align}
		k \log M &= n_k C(kP) \notag \\
		&\quad - \sqrt{n_k (V(kP) + V_{\textnormal{cr}}(k, P))} Q^{-1}\nB{\epsilon_k - \frac{D_k}{\sqrt{n_k}}} \notag \\
		&\quad + \frac 1 2 \log n_k + \eta_k -  k\log \kappa_k(P \bs{1}) \label{eq:klogM}
		\end{align}
		for all $k \in [K]$, where $D_k$ is a positive constant to be chosen later in \eqref{eq:Dkchoice}, and
		    $\eta_k \triangleq - 2 \log G_{k, k}' + (k-1) \log k - k$.
		Since $\frac 1 s C(sP) > \frac 1 k C(kP)$ for $s < k$ and \eqref{eq:klogM}, we reach a sequence of conclusions.
		\begin{enumerate}
		\item There exists a constant $c_0 > 0$ such that $\min_{j \in [k]} |\mc{N}(j)| \geq c_0 n_k$ for large enough $M$. In other words, $|\mc{N}(j)|$ is of the same order as $n_k$ for all $j \in [k]$.
		\item The bound on the probability of message repetition, $\frac{k(k-1)}{2M}$, decays exponentially with $n_k$.
		\item In order to bound the expression in \eqref{eq:wrongtimebound} as $B \leq\bigo{\frac 1 {\sqrt{n_k}}}$, we choose $n_0 \geq \frac{4 (1 + P^2)}{P^2} \log n_1 + o(\log n_1)$.
		\item By the union bound, Chebyshev's inequality, $\alpha_{k, k} = k$ in \eqref{eq:alphas}, and the fact that
		\begin{align}
		    \binom{M}{k} \leq \left(\frac{e M}{k} \right)^k,
		\end{align}
		we get
		\begin{align}
		\Prob{\mc{A}_k^c} &\leq \frac{E_k}{n_k} + \mathbb{P} \bigg[\imath_{[k]}(\vX_{[k]}^{[n_k]}; \vY_k^{[n_k]}) < k\log M \notag \\
		&\quad  - \frac 1 2 \log n_k - \eta_k \bigg] \label{eq:Akc}
		\end{align}
for some positive constant $E_k$.
		\end{enumerate}
		Therefore, it remains only to evaluate the probability term in \eqref{eq:Akc}.
Define the modified and centered information density random variable
	\begin{align}
	\tilde{\imath}_{k} &\triangleq  \frac 1 {\sqrt{n_k}} \nB{\sum_{i = 1}^{n_k} \log \frac{P_{Y_k|{X}_{[k]}}(Y_i | X_{[k], i})}{P_{\tilde{Y}_k}(Y_i)}  - n_k C(kP) },
	\end{align}
where $\tilde{Y}_k \sim \mathcal{N}(0, 1 + kP)$. By \lemref{lem:kappa} and \eqref{eq:klogM}, we get
\begin{align}
&\mathbb{P} \bigg[\imath_{[k]}(\vX_{[k]}^{[n_k]}; \vY_k^{[n_k]}) < k\log M   - \frac 1 2 \log n_k - \eta_k \bigg] \notag \\
&\quad \leq \Prob{\tilde{\imath}_k < -\sqrt{V(kP) + V_{\textnormal{cr}}(k, P)}Q^{-1}\nB{\epsilon_k - \frac{D_k}{\sqrt{n_k}}}}. \label{eq:iktildeRAC}
\end{align}
The conditional distribution of $\tilde{\imath}_k$ given $\vX_{[k]}^{[n_k]} = \vx_{[k]}^{[n_k]}$ is the same as the conditional distribution of $\tilde{\imath}_k$ given $\mathbf{\Ip} = \mathbf{\ip} $, where
\begin{align}
	\mathbf{\Ip} = (\Ip_{ij}\colon i, j \in [k], i < j) \in \mathbb{R}^{\binom{k}{2}},
\end{align}
	and $\Ip_{ij} = \frac{\langle \vX_i^{[n_k]}, \vX_j^{[n_k]} \rangle}{\sqrt{n_k P^2}}$. To bound the right-hand side of \eqref{eq:iktildeRAC}, in a manner similar to the arguments in \secref{sec:proofMACK}, we only need to verify that
	\begin{align}
	\mathrm{TV}(P_{\mathbf{\Ip}}, P_{\tilde{\mathbf{\Ip}}}) \leq \frac{\psi_k}{\sqrt{n_k}} \label{eq:TVRAC}
	\end{align}
for some constant $\psi_k$,
where $\tilde{\mathbf{\Ip}} \sim \mc{N}\nB{\bs{0}, \ms{I}_{\binom{k}{2}}}$. To show \eqref{eq:TVRAC}, we define
\begin{align}
\mathbf{\Ip}^{(t)} \triangleq (\Ip_{ij}^{(t)}\colon i, j \in [k], i < j) \in \mathbb{R}^{\binom{k}{2}},
\end{align}
where $\Ip_{ij}^{(t)} = \frac{\langle \vX_i^{\mc{N}(t)}, \vX_j^{\mc{N}(t)} \rangle}{\sqrt{|\mc{N}(t)|P^2}}$, then write
\begin{align}
\mathbf{\Ip} = \sum_{t = 1}^k \frac{\sqrt{|\mc{N}(t)|}}{\sqrt{n_k}} \mathbf{\Ip}^{(t)}.
\end{align}
By the data processing inequality of the total variation distance and the independence of $\mathbf{\Ip}^{(t)}$, $t \in [k]$, we get 
\begin{align}
\mathrm{TV}(P_{\mathbf{\Ip}}, P_{\tilde{\mathbf{\Ip}}}) 
&\leq \mathrm{TV}\left(\prod_{t = 1}^k P_{\mathbf{\Ip}^{(t)}}, P_{\tilde{\mathbf{\Ip}}}^k \right) \\
&\leq \sum_{t = 1}^k \mathrm{TV}(P_{\mathbf{\Ip}^{(t)}}, P_{\tilde{\mathbf{\Ip}}}) \label{eq:TVtriangular}\\
&\leq \sum_{t = 1}^k \frac{F_k}{\sqrt{|\mc{N}(t)|}}  \label{eq:Qlemmaused}\\
&\leq \frac{k F_k}{\sqrt{c_0 n_k}}, \label{eq:klogMused}
\end{align}
where \eqref{eq:TVtriangular} applies \cite[eq.~(4.5)]{hoeffding1958}, which bounds the total variation distance between two product measures $P^k$ and $Q^k$ by $k$ times the total variation distances between $P$ and $Q$. The bound in \cite[eq.~(4.5)]{hoeffding1958} is extended to arbitrary product measures $\prod_{i = 1}^k P_i$ and $\prod_{i = 1}^k Q_i$ in \cite[Lemma~2.1]{sendler1975}.
Inequality \eqref{eq:Qlemmaused} follows from \lemref{lem:QK}, $F_k$ is the constant from \lemref{lem:QK}, and \eqref{eq:klogMused} follows from \eqref{eq:klogM}, which proves \eqref{eq:TVRAC}. 

By \eqref{eq:klogMused}, and following arguments similar to those in \secref{sec:proofMACK}, we conclude that
\begin{IEEEeqnarray}{rCl}
\IEEEeqnarraymulticol{3}{l}{\Prob{\tilde{\imath}_k < -\sqrt{V(kP) + V_{\textnormal{cr}}(k, P)}Q^{-1}\nB{\epsilon_k - \frac{D_k}{\sqrt{n_k}}}}} \notag \\
&\leq& \epsilon_k - \frac{D_k}{\sqrt{n_k}} + \frac{C_k}{\sqrt{n_k}}, \label{eq:BErac}
\end{IEEEeqnarray} 
where $C_k$ is a Berry-Esseen constant. We choose the constant $D_k$ such that
\begin{IEEEeqnarray}{rCl}
\frac{D_k}{\sqrt{n_k}} &\leq& \frac{k(k-1)}{2M} + B + \frac{C_k}{\sqrt{n_k}} +\frac{E_k}{{n_k}}\notag \\
&&+ k \exp\cB{-c_k (c_0 n_k)^{1/3}} + \frac{1}{\sqrt{n_k}}, \label{eq:Dkchoice}
\end{IEEEeqnarray}
where $B$ is in \eqref{eq:wrongtimebound}.
For large enough $n_k$, such a constant exists by the enumerated consequences of \eqref{eq:klogM}, above. From \thmref{thm:nonasymRAC} and the inequalities \eqref{eq:powerviol0}, \eqref{eq:Pwrongmes}--\eqref{eq:Akc}, \eqref{eq:iktildeRAC}, \eqref{eq:BErac} and \eqref{eq:Dkchoice}, we conclude that the probability of error is bounded by $\epsilon_k$. By the Taylor series expansion of the function $Q^{-1}(\cdot)$ in \eqref{eq:klogM}, we complete the proof.

\section{Concluding Remarks} \label{sec:conclusion}
This paper studies the Gaussian multi-access channels in the finite-blocklength regime for two communication scenarios. In the first scenario, called the Gaussian MAC, $K$ active transmitters are fixed and known to the transmitters and the receiver; in the second scenario, called the Gaussian RAC, an unknown subset of $K$ transmitters is active, and neither the transmitters nor the receiver knows the set of active transmitter.

For the Gaussian MAC problem, we build on the RCU bound (\thmref{thm:RCUMac}) for general MACs to prove a third-order achievability result (\thmref{thm:MAC}). Our random encoder design chooses codewords distributed independently and uniformly on the $n$-dimensional sphere. At the receiver, we employ a maximum likelihood decoder. Compared to the result of MolavianJazi and Laneman \cite{molavianjazi2015second}, our coding scheme improves the achievable third-order term to $\frac 1 2 \log n \bs{1} + O(1) \bs{1}$. \thmref{thm:KMAC} extends our result for the Gaussian MAC with two transmitters to the $K$-transmitter Gaussian MAC. 

We generalize the rateless coding strategy in \cite{yavas2020Random} for the permutation-invariant random access channels by allowing non-i.i.d. input distributions at the random encoding function. For the Gaussian RAC, our strategy uses concatenated codewords such that each sub-codeword is uniformly distributed on a power sphere and independent of the other sub-codewords. In our proposed coding strategy, the decoding occurs at finitely many time instants $n_0, \dots, n_K$, with the choice of $n_k$ indicating that the decoder's estimate of the number of active transmitters is $k$. The receiver broadcasts a single bit to all transmitters at each decoding time, indicating whether or not it is ready to decode. The decoding rule combines a threshold rule based on the total received power and a maximum likelihood decoder. Building upon our result on the Gaussian MAC, we show in \thmref{thm:GRAC} that our rateless Gaussian RAC code achieves the same performance up to the third-order term as the best known code for the Gaussian MAC in operation (\corref{cor:MACsym}).
Furthermore, by forcing decoding at time $n_K$ our feedback RAC code in \thmref{thm:GRAC} can be used with a $K$-transmitter MAC without feedback. While this can only reduce the error probability determined in \thmref{thm:nonasymRAC} by eliminating the error events that result from deciding upon an incorrect number of active transmitters, that reduction is negligible in our asymptotic regime (see the proof of \thmref{thm:GRAC}). Thus, \thmref{thm:GRAC} also describes the performance of the length-$n_K$ codebook of the RAC code when used with a $K$-transmitter MAC without feedback.  That means that although the length-$n_K$ codebook of the RAC code is supported on only a subset of the power sphere (see \figref{fig:sphere}), it achieves the same first three order terms on a $K$-transmitter MAC as the more traditional code in \corref{cor:MACsym} that uses the entire power sphere.

		\appendices
		\section{Proof of Corollary \ref{cor:MACsym}} \label{app:proofcorMAC}
In order to prove Corollary~\ref{cor:MACsym}, we show that for any $M$ that satisfies the inequality \eqref{eq:corollaryMAC}, it holds that
\begin{IEEEeqnarray}{rCl}
\nB{|\mc{S}| \log M \colon \mathcal{S}} \in \subemp{[K]} &\in& n \mathbf{C}(P \bs{1})
-  {\sqrt{n}}  Q_{\mathrm{inv}}(\mathsf{V}(P \bs{1}), \epsilon) \notag \\
&& + \frac 1 {2} \log n \boldsymbol{1} + \bigo{1} \boldsymbol{1}.
\end{IEEEeqnarray}
Let $\vZ = (Z(\mc{S}): \mc{S} \in \subemp{[K]})\sim \mc{N}(\bs{0}, \mathsf{V}(P \bs{1}), \epsilon))$. Take $M$ such that the asymptotic expansion in \eqref{eq:corollaryMAC} holds, implying that
\begin{align}
\bbP \Bigl[ Z([K]) &> \sqrt{n} \nB{C(KP) - \frac{K \log M}{n}} \notag \\
&+ \frac{1}{2} \frac{\log n}{\sqrt{n}} + \bigo{\frac{1}{\sqrt{n}}} \Bigr]  \leq \epsilon. \label{eq:logMKineq}
\end{align}
Consider any $\mc{S} \in \subemp{[K]}$ with $|\mc{S}| < K$. Then
\begin{align}
\bbP \Bigl[Z(\mc{S}) &> \sqrt{n} \nB{C(|\mc{S}|P) - \frac{|\mc{S}| \log M}{n}} + \frac{1}{2} \frac{\log n}{\sqrt{n}} \notag \\
&+ \bigo{\frac{1}{\sqrt{n}}} \Bigr] \leq \bigo{\frac{1}{n}}, \label{eq:chebused}
\end{align}
which follows from Chebyshev's inequality since $C(sP) - \frac{s}{K}C(KP) > 0$ for $s < K$.

By the union bound, \eqref{eq:logMKineq} and \eqref{eq:chebused}, we get
\begin{align}
\bbP \Biggl[ &\bigcup_{\mc{S} \in \subemp{[K]}} \Big \{ Z(\mc{S}) > \sqrt{n} \nB{C(|\mc{S}|P) - \frac{|\mc{S}| \log M}{n}} \notag \\
&+ \frac{1}{2} \frac{\log n}{\sqrt{n}} + \bigo{\frac{1}{\sqrt{n}}} \Big\} \Biggr] \leq \epsilon + \bigo{\frac{1}{n}},
\end{align}
which, by the definition \eqref{eq:defqinv}, is equivalent to
\begin{align}
&\nB{|\mc{S}| \log M \colon \mathcal{S} \in \subemp{[K]}} \in n \mathbf{C}(P \bs{1}) \notag \\
&\quad -  {\sqrt{n}}  Q_{\mathrm{inv}}\left(\mathsf{V}(P \bs{1}), \epsilon + \bigo{\frac 1 n} \right) + \frac 1 {2} \log n \boldsymbol{1} + \bigo{1} \boldsymbol{1}.
\end{align}
Applying the Taylor series expansion to $Q_{\mathrm{inv}}(\mathsf{V}(P \bs{1}), \cdot)$ completes the proof.

\section{Code Design Variations} \label{app:modifications}
\subsection{Adopting the Codebooks Based on the Channel Estimate at Time $n_0$}\label{app:mod1}
In our encoder and decoder design, we use the fact that the received output power concentrates around its mean value. In the proof of \thmref{thm:MAC}, we show that $n_0 = O(\log n_1)$ symbols are sufficient to ensure that the probability that the decision is made at the correct decoding time, i.e., $n_k$ when $k$ transmitters are active, decays with $O\left( \frac 1 {\sqrt{n_k}} \right)$. In our strategy, we make a binary decision at each decoding time $n_0, \dots, n_K$ of whether or not to decode. An alternative to this strategy would be to decide the number of active transmitters at time $n_0$, which is much smaller than the rest of the decoding times, and to inform the transmitters about the decoding time in the epoch at time $n_0$. This alternative allows for a code design that depends on the feedback from the receiver to the transmitters at time $n_0$. Using its knowledge of the typical interval, in which the squared norm of the output, $ \frac 1 {n_0} \norm{\vY_k^{[n_0]}}^2$, lies for each $k \leq K$, the decoder estimates the number of active transmitters. We denote this value by $t$. The decoder could then transmit $t$ to all transmitters, so that all parties understand that the communication epoch is going to end at time $n_t$. This strategy requires $\lceil \log (K + 1) \rceil$ bits of feedback from the receiver to transmitters at time $n_0$; in contrast, the strategy in the proof of \thmref{thm:GRAC} requires a number of bits of feedback that varies with the decoder's estimate of the number of active transmitters with a maximum of $K + 1$ bits. Let the decoder choose $t$ as the nearest integer to $\frac 1 P \left(\frac{1}{n_0}{\norm{\vy^{[n_0]}}^2} - 1 \right)$. Then, the bound in \eqref{eq:wrongtimebound} on the probability that the decoder errs in determining the number of active transmitters can be bounded~as
\begin{align}
\Prob{\mc{E}_{\textnormal{time}} \middle| \mc{E}^c_{\textnormal{rep}}} \leq 2 \nB{\prod_{j = 1}^k \kappa_j(P \bs{1})}  \exp\cB{-\frac{ n_0 (\frac P 2)^2 }{8(1 + kP)^2}}
\end{align}
in the case when the decision is made at time $n_0$. Like \eqref{eq:wrongtimebound}, this bound
decays exponentially with $n_0$. Here, however, the exponential rate is smaller than \eqref{eq:wrongtimebound}. Hence, this modification in the strategy increases the constant $c$ in \eqref{eq:n0C}, and affects the achievable $O(1)$ term in \eqref{eq:RACmainresult}.

As the encoders learn the estimate of the number of active transmitters at an earlier time, an encoding function that depends on the feedback from the receiver could be employed as follows. Recall from \eqref{eq:powerRAC} that the maximal-power constraints apply to the decoding times $n_1, \dots, n_K$, but not to $n_0$. Given the estimate $t$ of the number of active transmitters $k$, length-$n_t$ codewords are drawn such that the first $n_1$ symbols are uniformly distributed on $n_1$-dimensional sphere with radius $\sqrt{n_1 P}$, and the symbols indexed from $n_1 + 1$ to $n_t$ are distributed on $(n_t - n_1)$-dimensional sphere with radius $\sqrt{(n_t - n_1)P}$, i.e., instead of $K$ independent spherical sub-codewords, we use two independent sub-codewords. The length of the second sub-codeword depends on the estimate $t$. The effect of this modification on the error analysis is that under this input distribution,
the total variation bound in \eqref{eq:klogMused} can be improved to \begin{align}
\mathrm{TV}(P_{\mathbf{\Ip}}, P_{\tilde{\mathbf{\Ip}}}) &\leq \frac{F_k}{\sqrt{n_1}} +  \frac{F_k}{\sqrt{n_k - n_1}},
\end{align}
which decays with the same asymptotic rate as \eqref{eq:klogMused}. Therefore, this modification affects only the $O(1)$ term in \eqref{eq:RACmainresult}, meaning that the same expansion as \thmref{thm:GRAC} is achieved.

\subsection{Decoding Transmitter Identity}\label{app:mod2}
Another scenario of possible interest is the case, where the decoder must decode transmitter identities as well as messages. For this scenario, as we observed in the context of RACs without codeword cost constraints \cite[Section V.D]{yavas2020Random}, one can again use the encoding and decoding rules employed in \secref{sec:RACencoding} with a message set size of $K M$ such that the messages indexed from $(k-1) M + 1$ to $k M$ are associated with transmitter $k$. If the decoder decides to decode $k$ messages at time $n_k$, a list of $k$ out of $K M$ messages is decoded, which automatically reveals the identities of the transmitters. By our RAC code design, it is possible that the decoder decodes multiple messages belonging to the same transmitter; this would not have been possible with the MAC decoder. Given that there are $k$ active transmitters, in the proposed coding scheme, the random message $W_{[k]}$ is not uniformly distributed over a set of size $\binom{KM}{k}$, as it would be if any combination of $k$ of the $KM$ codewords could be transmitted. Instead, it is uniformly distributed over a set of size $M^k$ since each of the $k$ active transmitters sends exactly one of its $M$ codewords. Nonetheless, replacing $M$ by $KM$ in \thmref{thm:GRAC} gives a valid upper bound on the resulting error probability since the codewords for each transmitter and each message are generated i.i.d. While the inclusion of all $\binom{KM}{k}$ codeword combinations in \eqref{eq:epsk1}--\eqref{eq:GRACRCU} may yield looser than necessary bounds on the error probability, their implication is that to decode transmitter identities the RAC code pays a penalty of $- k \log K$ on the right-hand side of \eqref{eq:RACmainresult}. Since $K$ does not grow with $n_1$, decoding transmitter identities affects only the $O(1)$ term in \eqref{eq:RACmainresult}. 

		\section{Proof of \lemref{lem:gaussiancdf}}\label{app:gaussian}
		 
			Pinsker's inequality (e.g., \cite[Th. 6.5]{polyanskiyLectureNotes}) states that for any distributions $P$ and $Q$,
			\begin{align}
			\mathrm{TV}(P, Q) \leq \sqrt{\frac 1 2 D(P \|Q)}. \label{eq:pinsker}
			\end{align}
			Let $\textrm{tr}(\cdot)$ denote trace of its matrix argument. The relative entropy between two $d$-dimensional Gaussian distributions with positive covariance matrices is given (e.g., \cite[eq. (1.18)]{polyanskiyLectureNotes}) by 
			\begin{IEEEeqnarray}{rCl} 
			\IEEEeqnarraymulticol{3}{l}{D(\mathcal{N}(\bm{\mu}_1, \mathsf{\Sigma}_1 ) \| \mathcal{N}(\bm{\mu}_2, \mathsf{\Sigma}_2 ))} \notag \\
			&=& \frac 1 2 \Big( \mathrm{tr}(\mathsf{\Sigma}_1^{-1/2} \mathsf{\Sigma}_2 \mathsf{\Sigma}_1^{-1/2} - \mathsf{I}_d) + (\bm{\mu}_1 - \bm{\mu}_2)^T \mathsf{\Sigma}_1^{-1} (\bm{\mu}_1 - \bm{\mu}_2) \notag \\
			&&- \log \det(\mathsf{\Sigma}_1^{-1/2} \mathsf{\Sigma}_2 \mathsf{\Sigma}_1^{-1/2} ) \Big).  \label{eq:gaussrel}
			\end{IEEEeqnarray} 
			Define 
			\begin{align}
			    \ms{G} &\triangleq \mathsf{\Sigma}_1^{-1/2} \mathsf{\Sigma}_2 \mathsf{\Sigma}_1^{-1/2} - \mathsf{I}_d \\
			    a &\triangleq \frac 1 2 \sqrt{(\bm{\mu}_1 - \bm{\mu}_2)^T \mathsf{\Sigma}_1^{-1} (\bm{\mu}_1 - \bm{\mu}_2)}.
			\end{align}
			Combining \eqref{eq:pinsker} and \eqref{eq:gaussrel} and using the inequality $\sqrt{x + y} \leq \sqrt{x} + \sqrt{y}$, we get
			\begin{IEEEeqnarray}{rCl}
			\IEEEeqnarraymulticol{3}{l}{\mathrm{TV}(\mathcal{N}(\bm{\mu}_1, \mathsf{\Sigma}_1 ) , \mathcal{N}(\bm{\mu}_2, \mathsf{\Sigma}_2 ))}  \notag \\
			&\leq& a + \frac 1 2 \sqrt{\mathrm{tr}(\ms{G}) - \log \det(\ms{I}_d + \ms{G})}.  \label{eq:tvmain}
			\end{IEEEeqnarray}
			To bound the logdeterminant term in \eqref{eq:tvmain} from below, we use the following result from \cite[Th. 1.1]{rump2018estimates}. Let $\rho(\mathsf{\cdot})$ denote the spectral radius, i.e., the maximum absolute eigenvalue, and let $\norm{\cdot}_F$ denote the Frobenius norm. 
				If $\rho(\mathsf{G}) < 1$, then
				\begin{align}
				\exp\cB{\mathrm{tr}(\mathsf{G}) - \frac{\norm{\mathsf{G}}_F^2}{2(1-\rho(\mathsf{G}))}}\leq  \mathrm{det}(\mathsf{I}_d + \mathsf{G}). \label{eq:rump}
				\end{align}
		    For $ \rho(\ms{G}) < 1$, we apply \eqref{eq:rump} to \eqref{eq:tvmain} and get
		    \begin{align}
		   \mathrm{TV}(\mathcal{N}(\bm{\mu}_1, \mathsf{\Sigma}_1 ), \mathcal{N}(\bm{\mu}_2, \mathsf{\Sigma}_2 )) \leq \frac 1 {2\sqrt{2}} \frac{\norm{\ms{G}}_F}{\sqrt{1 - \rho(\ms{G})}} + a. \label{eq:case1}
		    \end{align}
		    In addition, trivially, we have that
		    \begin{align}
		    \mathrm{TV}(\mathcal{N}(\bm{\mu}_1, \mathsf{\Sigma}_1 ), \mathcal{N}(\bm{\mu}_2, \mathsf{\Sigma}_2 )) &\leq 1 \\ 
		    &\leq \frac{\norm{\ms{G}}_F}{\rho(\ms{G})} + a, \label{eq:case2}
		    \end{align}
		    where in \eqref{eq:case2}, we use the fact that Frobenius norm is an upper bound to the spectral radius.
		    Taking the tighter bound among \eqref{eq:case1} and \eqref{eq:case2}, we conclude that for $ \rho(\ms{G}) < 1$,
		    \begin{IEEEeqnarray}{rCl}
		    \IEEEeqnarraymulticol{3}{l}{\mathrm{TV}(\mathcal{N}(\bm{\mu}_1, \mathsf{\Sigma}_1 ), \mathcal{N}(\bm{\mu}_2, \mathsf{\Sigma}_2 ))} \notag \\ &\leq& \min\cB{\frac{1}{2 \sqrt{2}} \frac{1}{\sqrt{1 - \rho(\ms{G})}}, \frac{1}{\rho(\ms{G})}} \norm{\ms{G}}_F  + a \label{eq:tvboundmin1}\\
		    &\leq& \frac{2 + \sqrt{6}}{4} \norm{\ms{G}}_F  + a .  \label{eq:tvboundmin}
		    \end{IEEEeqnarray}
		    Inequality \eqref{eq:tvboundmin} follows since the maximum of the minimum term in \eqref{eq:tvboundmin1} is achieved by $\rho(\mathsf{G}) = 2\sqrt{6}-4 \approx 0.899$ and that maximum value is $\frac{2 + \sqrt{6}}{4}$. 
		    Since the coefficient $\frac{2 + \sqrt{6}}{4} > 1 \geq \frac{1}{\rho(\ms{G})}$ for $\rho(\ms{G}) \geq 1$, we conclude that \eqref{eq:tvboundmin} holds for any $\rho(\ms{G})$.
		
		\section{Proof of \eqref{eq:concentU}} \label{app:Ubound}
	We show a more general result. Fix any constant $u < \Ponetwo$. We prove below that for $n$ large enough,
			\begin{IEEEeqnarray}{rCl}
			g(y) &\triangleq&  \Prob{\norm{\vXltwo}^2 \leq n u \,\middle|  \norm{\vXltwo + \vZ}^2 = y} \IEEEeqnarraynumspace \label{eq:gyeqP} \\
			&\leq& \exp\cB{-nC} \label{eq:gyprove}
			\end{IEEEeqnarray}
			for all $y \in \mc{I}$, where
		\begin{align}
		\mc{I} &\triangleq [n(1 + \Ponetwo - \epsilon), n(1 + \Ponetwo + \epsilon)] \\
		 \epsilon &\triangleq n^{-1/3},
		\end{align}	
		and $C$ is a positive constant depending on $u$. Taking $u = P_1 + P_2 - \sqrt{P_1 P_2}$ in \eqref{eq:gyeqP} then proves the desired inequality \eqref{eq:concentU}. 
			
			We proceed to prove \eqref{eq:gyprove}.
			Since the support of $\norm{\vXltwo}^2$ is \begin{align}
			    \mc{S} = [n (\sqrt{P_1} - \sqrt{P_2})^2, n(\sqrt{P_1} + \sqrt{P_2})^2],
			\end{align}
			inequality \eqref{eq:gyprove}
            is trivially satisfied for $u < (\sqrt{P_1} - \sqrt{P_2})^2$. 
			To show \eqref{eq:gyprove} for $(\sqrt{P_1} - \sqrt{P_2})^2 \leq u < \Ponetwo$, we show two concentration results. First, we show that
			\begin{align}
			g(y) = g(n(1 + \Ponetwo)) \exp\{O(n \epsilon)\} \label{eq:consgy}
			\end{align} 
			for all $y \in \mc{I}$; second, we show that for $n$ large enough,
			\begin{align}
			p &\triangleq \Prob{\norm{\vXltwo}^2 \leq n u \,\middle| \, \mc{A}} \label{eq:probp} \\
			&\leq \exp\{-nC'\} \label{eq:consp}
			\end{align}
			for some $C' > 0$, where the event $\mc{A}$ is defined as
			\begin{align}
\mc{A} \triangleq \left\{\norm{\vXltwo + \vZ}^2 \in \mc{I} \right\}.
			\end{align}
			Using \eqref{eq:consgy} and \eqref{eq:consp}, we show \eqref{eq:gyprove} as follows. 
			By conditioning the probability in \eqref{eq:probp} on each value of $\norm{\vXltwo + \vZ}^2$, we express $p$ as
			\begin{IEEEeqnarray}{rCl}
			p &=& \int_{\mc{I}} g(y) f_{\norm{\vXltwo + \vZ}^2 \mid \mathcal{A}}(y)dy \\
			&=& g(n(1 + \Ponetwo)) \exp\{O(n \epsilon)\} \label{eq:gyused} \\
			&\leq& \exp\{-nC'\}, \label{eq:pgycons}
			\end{IEEEeqnarray}
			where \eqref{eq:gyused} follows from \eqref{eq:consgy} and 
			\begin{align}
			    \min_{y \in \mc{I}} g(y) \leq \int_{\mc{I}} g(y) f_{\norm{\vXltwo + \vZ}^2 \mid \mathcal{A}}(y) dy \leq \max_{y \in \mc{I}} g(y).
			\end{align}
			Inequality \eqref{eq:pgycons} follows from \eqref{eq:consp}.
			Inequalities \eqref{eq:consgy} and \eqref{eq:pgycons} imply that since $O(n \epsilon) = o(n)$, there exists a constant $C > 0$ such that for $n$ large enough, \eqref{eq:gyprove} holds for all $y \in \mc{I}$. 
			
			We proceed to show \eqref{eq:consp}. By Bayes' rule, we have
			\begin{align}
			p &= \frac{\Prob{\norm{\vXltwo}^2 \leq n u} \Prob{ \mathcal{A} \,\middle| \, \norm{\vXltwo}^2 \leq n u }}{\Prob{\mathcal{A}}}. \label{eq:pbayes}
			\end{align}
			Changing measure from $P_{\vXltwo}P_{\vZ}$ to $P_{\tilde{\mathbf{U}}}P_{\vZ}$, where $\tilde{\mathbf{U}} \sim \mathcal{N}(0, (\Ponetwo) \mathsf{I}_n)$, and then applying \lemref{lem:kappa}, we get
			\begin{align}
			p&\leq  \frac{\kappa_2(P_1, P_2) \Prob{ \norm{\tilde{\mathbf{U}}}^2 \leq n u} \cdot 1}{1 - \kappa_2(P_1, P_2) \Prob{ \left\lvert \norm{\tilde{\mathbf{U}} + \vZ}^2 - n(1 + \Ponetwo) \right \rvert > n \epsilon}} \label{eq:bayes1} \\
			&\leq  \kappa_2(P_1, P_2) \frac{\exp\left\{\frac{-n(\Ponetwo -u)^2}{4 (\Ponetwo)^2}\right\}}
			{1 - 2 \kappa_2(P_1, P_2) \exp\{\frac{-n\epsilon^2}{8 (1 + \Ponetwo)^2}\}} \label{eq:bayes2}\\
			&\leq 2 \kappa_2(P_1, P_2) \exp\left\{\frac{-n(\Ponetwo -u)^2}{4 (\Ponetwo)^2}\right\} \label{eq:bayes3}\\
			&\leq \exp\{-nC'\} \label{eq:nCprime}
			\end{align}
			for all $n$ large enough,
			where $\kappa_2(P_1, P_2)$ is the constant defined in \eqref{eq:kappa2}, and $C'$ is a positive constant. Inequality \eqref{eq:bayes2} follows from the tail bounds on the chi-squared distribution in \lemref{lem:chi2}, and \eqref{eq:bayes3} follows since the denominator on the right-hand side of \eqref{eq:bayes2} is greater than $\frac 1 2$ for $n$ large enough. Inequality \eqref{eq:nCprime} holds since $u < \Ponetwo$.
			
			We proceed to prove \eqref{eq:consgy}. Define the events $\mathcal{B} \triangleq \{\norm{\vXltwo}^2 \leq nu\}$ and $\mathcal{B}(\lambda) \triangleq \{\norm{\vXltwo}^2 = \lambda\}$ for any $ \lambda  \in \mc{S}$. By Bayes' rule, we can express $g(y)$ as
			\begin{align}
			g(y) =  \frac{\Prob{\mathcal{B}} f_{\norm{\vXltwo + \vZ}^2 | \mathcal{B}}(y)}{f_{\norm{\vXltwo + \vZ}^2}(y)}.  \label{eq:gyfrac}
			\end{align}
			By the spherical symmetry of the distribution of $\vXltwo$, the conditional distribution of
			$\norm{\vXltwo + \vZ}^2$ given $\mathcal{B}(\lambda)$ does not depend on $\mathbf{u}$ when we fix $\vXltwo$ to any $\mathbf{u}$ such that $ \norm{\mathbf{u}}^2 = \lambda  \in \mc{S} $. Therefore, the conditional distribution of 	$\norm{\vXltwo + \vZ}^2$ given $\mathcal{B}(\lambda)$ equals the distribution of
			\begin{align}
			\sum_{i = 1}^n \norm{Z_i + \frac{\sqrt{\lambda}}{\sqrt{n}}}^2,
			\end{align} 
			which has non-central chi-squared distribution with $n$ degrees of freedom and non-centrality parameter $\lambda$. That is, the probability density function is 
			\begin{align}
			f(x; n, \lambda) = \frac 1 2 \exp\cB{-\frac{(x + \lambda)}{2}} \nB{\frac{x}{\lambda}}^{\frac{n}{4} - \frac 1 2} I_{\frac{n}{2}- 1}(\sqrt{\lambda x}), \label{eq:noncentralchi}
			\end{align}
			where $I_{\nu}(x)$ denotes the modified Bessel function of the first kind with order $\nu$. Fix some $\lambda > 0$, $x_1 = nb$, and $x_2 = n(b + \delta)$, where $0 < \delta \leq \epsilon$ and $b > 0$. Consider the ratio
			\begin{align}
			\frac{f(x_1; n, \lambda)}{f(x_2; n, \lambda)} = \exp\{x_2 - x_1\} \nB{\frac{x_1}{x_2}}^{\frac{n}{4} - \frac 1 2} \frac{I_{\frac{n}{2}- 1}(\sqrt{\lambda x_1})}{I_{\frac{n}{2}- 1}(\sqrt{\lambda x_2})}. \label{eq:fratio}
			\end{align}
			Paris \cite{paris1984inequality} bounds $I_{\nu}(x) / I_{\nu}(y)$ as
			\begin{align}
			\exp\cB{x-y} \nB{\frac x y}^{\nu} < \frac{I_{\nu}(x)}{I_{\nu}(y)} < \nB{\frac x y}^{\nu} \label{eq:bessel}
			\end{align}
			for any $0 < x < y$ and $\nu > -1/2$.
			Using \eqref{eq:bessel}, we bound \eqref{eq:fratio} as
			\begin{IEEEeqnarray}{rCl}
			\IEEEeqnarraymulticol{3}{l}{\exp\{n \delta\} \nB{1 - \frac {\delta}{b + \delta}}^{\frac{n}{2}- 1} \exp\left\{- \sqrt{n \lambda}\nB{\sqrt{b + \delta} - \sqrt{b}}\right\}} \notag \\
			&\leq& \frac{f(x_1; n, \lambda)}{f(x_2; n, \lambda)} \label{eq:flower} \\
			&\leq& \exp\{n \delta\} \nB{1 - \frac {\delta}{b + \delta}}^{\frac{n}{2}- 1}. \label{eq:fupper}
			\end{IEEEeqnarray}
			Applying the Taylor series expansion at $\delta = 0$ gives
			\begin{align}
			\log \nB{1 - \frac {\delta}{b + \delta}} &= -\frac{\delta}{b} + O(\delta^2)  \label{eq:fbound1} \\
			- \sqrt{n \lambda}\nB{\sqrt{b + \delta} - \sqrt{b}} &=  - \sqrt{n \lambda}\nB{\frac {\delta}{2 \sqrt{b}} + O(\delta^2)}. \label{eq:fbound2}
			\end{align}
			Substituting \eqref{eq:fbound1} and \eqref{eq:fbound2} in \eqref{eq:flower} and \eqref{eq:fupper}, we get
			\begin{align}
			\frac{f(x_1; n, \lambda)}{f(x_2; n, \lambda)} = \exp\{O(n \delta)\}. \label{eq:fOndelta}
			\end{align}
			We can also verify the validity of \eqref{eq:fOndelta} for $\lambda = 0$ by using the probability density function of chi-squared distribution with $n$ degrees of freedom instead of \eqref{eq:noncentralchi}.
			Particularizing \eqref{eq:fOndelta} to $b = 1 + \Ponetwo$, we get for all $\lambda \in \mc{S}$ that 
			\begin{IEEEeqnarray}{rCl}
			\IEEEeqnarraymulticol{3}{l}{f_{\norm{\vXltwo + \vZ}^2 | \mathcal{B}(\lambda)}(y)} \notag \\
			&=& 	f_{\norm{\vXltwo + \vZ}^2 | \mathcal{B}(\lambda)}(n(1 + \Ponetwo))  \exp\{O(n \epsilon)\},
			\end{IEEEeqnarray}	
			which together with \eqref{eq:gyfrac} implies \eqref{eq:consgy}.

\section{Proof of \lemref{lem:QK}}\label{app:proofQK}
		 We use the induction technique from \cite[Th. 4]{stam1982} to prove this lemma, showing that the total variation distance in \eqref{eq:tvK} diminishes as $n$ goes to infinity. We here prove that the convergence rate is $\bigo{\frac 1 {\sqrt{n}}}$. Since the distribution of $\mathbf{\Ip}$ is invariant to rotation, we fix 
		\begin{align}
		\vX_1 = (1, 0, 0, \dots, 0).
		\end{align}		
Then $\Ip_{1j} = \sqrt{n} X_{j1}$ for $2 \leq j \leq K$. Define the vectors
\begin{align}
\mathbf{\Ip}_1 &\triangleq (\Ip_{1j}: 2 \leq j \leq K)\\
\mathbf{\Ip}_{2} &\triangleq (\Ip_{ij}: 2 \leq i < j \leq K),
\end{align}
which consist of all the inner product random variables including $\vX_1$, and not including $\vX_1$, respectively.
Hence $\mathbf{\Ip} = (\mathbf{\Ip}_1, \mathbf{\Ip}_2)$. Notice that $\mathbf{\Ip}_1$ is a product distribution since $X_{j1}$'s are independent. 

Note that we have for $2 \leq i < j \leq K$
        \begin{align}
		\Ip_{ij} &= \sqrt{n} X_{i1} X_{j1} + \frac{\sqrt{n}}{\sqrt{n-1}} (1-X_{i1}^2)^{\frac 1 2} (1-X_{j1}^2)^{\frac 1 2} V_{ij}  \label{eq:qij} \\
		V_{ij} &= \sqrt{n-1} \langle \vY_i, \vY_j \rangle,
		\end{align}
		where $\vY_i = (1-X_{i1}^2)^{-\frac 1 2} (X_{i2}, \dots, X_{in}) \in \mathbb{R}^{n-1}$ for $i = 2, \dots, K$. Denote by $p_K^{(n)}$ the distribution of the $\binom{K}{2}$-dimensional random vector $(\sqrt{n} \langle \vZ_i, \vZ_j \rangle\colon 1 \leq i < j \leq K)$, where 
		the $\vZ_i$, $i \in [K]$, are distributed independently and uniformly on $\mathbb{S}^{n}(1)$.
		
		Since $\vY_i$, $i \in \{2, \dots, K\}$, are distributed independently and uniformly on $\mathbb{S}^{n-1}(1)$, the joint distribution of $\mathbf{V} = (V_{ij}: 2 \leq i < j \leq K)$ is $p_{K-1}^{(n-1)}$. By \eqref{eq:qij}, the conditional distribution of $\Ip_{ij}$ given $\mathbf{\Ip}_1 = \mathbf{\ip}_1$ is the same as the distribution of
		\begin{IEEEeqnarray}{rCl}
		\frac{\ip_{1i} \ip_{1j}}{\sqrt{n}} + \frac{\sqrt{n}}{\sqrt{n-1}} \nB{1-\frac {\ip_{1i}^2}{n}}^{\frac 1 2 } \nB{1-\frac {\ip_{1j}^2}{n}}^{\frac 1 2 } V_{ij} \IEEEeqnarraynumspace \label{eq:probtranskern}
		\end{IEEEeqnarray}
		 for $2 \leq i < j \leq K$.
		 We define the random vector $\mathbf{\Ip}^*_2 = (\Ip_{ij}^*\colon 2 \leq i < j \leq K)$ through $\mathbf{\Ip}_1$ as follows. The conditional distribution of $\Ip_{ij}^*$ given $\mathbf{\Ip}_1 = \mb{q}_1$ is the same as the distribution of
\begin{align}
\frac{\ip_{1i} \ip_{1j}}{\sqrt{n}} + \frac{\sqrt{n}}{\sqrt{n-1}} \nB{1-\frac {\ip_{1i}^2}{n}}^{\frac 1 2 } \nB{1-\frac {\ip_{1j}^2}{n}}^{\frac 1 2 } Z_{ij} \label{eq:qstar}
\end{align}
for $2 \leq i < j \leq K$,
where $Z_{ij} \sim \mathcal{N}(0, 1)$, and $\Ip_{ij}^*$, $2 \leq i < j \leq K$, are conditionally independent given $\mathbf{\Ip}_1$. Now, we are ready to apply the mathematical induction.

\textbf{Base case:} For $ K = 2$, we have
\begin{align}
\mathrm{TV}(p_2^{(n)}, \mathcal{N}(0, 1)) \leq \frac 4 n
\end{align}
by \lemref{lem:gotze} with $k = 1$.

\textbf{Inductive step:} For $K > 2$, assume that for any $n$, 
\begin{align}
\mathrm{TV}\left(p_{K-1}^{(n)}, \mathcal{N}\left(\boldsymbol{0}, \mathsf{I}_{\frac{1}{2} (K-1)(K-2)}\right)\right) \leq \frac{C_{K-1}}{\sqrt{n}} \label{eq:inductiveassump}
\end{align}
for some constant $C_{K-1}$. Let $P_{\tilde{\mathbf{\Ip}}_1} = \mathcal{N}(\boldsymbol{0}, \mathsf{I}_{K-1})$ and $P_{\tilde{\mathbf{\Ip}}_2} = \mathcal{N}\left(\boldsymbol{0}, \mathsf{I}_{\binom{K-1}{2}}\right)$. Since the total variation distance is $\ell_1$ norm, applying the triangle inequality gives
\begin{IEEEeqnarray}{rCl}
\IEEEeqnarraymulticol{3}{l}{\mathrm{TV}\left(p_K^{(n)}, \mathcal{N}\left(\boldsymbol{0}, \mathsf{I}_{\binom{K}{2}}\right) \right)} \notag \\
&=& \mathrm{TV}\left(P_{\mathbf{\Ip}_1} P_{\mathbf{\Ip}_2 | \mathbf{\Ip}_1}, P_{\tilde{\mathbf{\Ip}}_1}P_{\tilde{\mathbf{\Ip}}_2} \right) \\
&\leq& \mathrm{TV}\left(P_{\mathbf{\Ip}_1} P_{\mathbf{\Ip}_2 | \mathbf{\Ip}_1}, P_{\tilde{\mathbf{\Ip}}_1}P_{\mathbf{\Ip}_2 | \mathbf{\Ip}_1} \right)  \label{eq:tv11}\\
&&+ \mathrm{TV}\left(P_{\tilde{\mathbf{\Ip}}_1}P_{\mathbf{\Ip}_2 | \mathbf{\Ip}_1}, P_{\tilde{\mathbf{\Ip}}_1}P_{\mathbf{\Ip}_2^{*}| \mathbf{\Ip}_1} \right) \label{eq:tv12} \\
&&+  \mathrm{TV}\left(P_{\tilde{\mathbf{\Ip}}_1}P_{\mathbf{\Ip}_2^{*}| \mathbf{\Ip}_1} , P_{\tilde{\mathbf{\Ip}}_1}P_{\tilde{\mathbf{\Ip}}_2} \right). \label{eq:tv13}
\end{IEEEeqnarray}
Here, \eqref{eq:tv11} approximates the input measure $P_{\mathbf{\Ip}_1}$ with the corresponding i.i.d. Gaussian measure $P_{\tilde{\mathbf{\Ip}}_1}$, \eqref{eq:tv12} approximates the inner product random variables $V_{ij}$ in the definition of the probability transition kernel given in \eqref{eq:probtranskern} with i.i.d. standard Gaussian random variables, and \eqref{eq:tv13} approximates the mean in \eqref{eq:qstar} by 0 and the variance by 1. We next bound the right-hand sides of \eqref{eq:tv11}--\eqref{eq:tv13} in that order. We have
\begin{IEEEeqnarray}{rCl}
\IEEEeqnarraymulticol{3}{l}{\mathrm{TV}\left(P_{\mathbf{\Ip}_1} P_{\mathbf{\Ip}_2 | \mathbf{\Ip}_1}, P_{\tilde{\mathbf{\Ip}}_1}P_{\mathbf{\Ip}_2 | \mathbf{\Ip}_1} \right)} \notag \\
&=&  \mathrm{TV}\left(P_{\mathbf{\Ip}_1}, P_{\tilde{\mathbf{\Ip}}_1} \right) \\
&\leq& (K-1) \mathrm{TV}\left(P_{\Ip_{12}}, \mathcal{N}(0, 1) \right) \label{eq:tvprod} \\
&\leq& \frac{4 (K-1)}{n}, \label{eq:tvlemma}
\end{IEEEeqnarray}
where \eqref{eq:tvprod} follows from \cite[Lemma~2.1]{sendler1975} since $P_{\mathbf{\Ip}_1} = (P_{\Ip_{12}})^{K-1}$ and $P_{\tilde{\mathbf{\Ip}}_1} = (\mc{N}(0, 1))^{K-1}$ are both product distributions, and \eqref{eq:tvlemma} follows from \lemref{lem:gotze}.
The total variation distance in \eqref{eq:tv12} is bounded as
\begin{IEEEeqnarray}{rCl}
 \IEEEeqnarraymulticol{3}{l}{\mathrm{TV}\left(P_{\tilde{\mathbf{\Ip}}_1}P_{\mathbf{\Ip}_2 | \mathbf{\Ip}_1}, P_{\tilde{\mathbf{\Ip}}_1}P_{\mathbf{\Ip}_2^{*}| \mathbf{\Ip}_1} \right)} \notag \\
 &=& \mathbb{E}\sB{\mathrm{TV}\nB{P_{\mathbf{\Ip}_2| {\mathbf{\Ip}}_1 = \tilde{\mathbf{\Ip}}_1}, P_{\mathbf{\Ip}_2^{*}| {\mathbf{\Ip}}_1 = \tilde{\mathbf{\Ip}}_1}} \middle| \tilde{\mathbf{\Ip}}_1 } \label{eq:triangle1}\\ 
 &=&  \mathrm{TV}\left(p_{K-1}^{(n-1)},  \mathcal{N}\left(\boldsymbol{0}, \mathsf{I}_{\binom{K-1}{2}}\right) \right) \label{eq:shiftscale}\\
 &\leq& \frac{C_{K-1}}{\sqrt{n-1}}, \label{eq:qassump}
\end{IEEEeqnarray}
where \eqref{eq:shiftscale} follows from the definitions \eqref{eq:probtranskern} and \eqref{eq:qstar} since the total variation distance is shift and scale invariant and $\eqref{eq:qassump}$ follows from the inductive assumption \eqref{eq:inductiveassump}.
The total variation distance in \eqref{eq:tv13} is bounded as
\begin{IEEEeqnarray}{rCl}
\IEEEeqnarraymulticol{3}{l}{\mathrm{TV}\left(P_{\tilde{\mathbf{\Ip}}_1}P_{\mathbf{\Ip}_2^{*}| \mathbf{\Ip}_1} , P_{\tilde{\mathbf{\Ip}}_1}P_{\tilde{\mathbf{\Ip}}_2} \right)} \notag \\
&=& \mathbb{E}\sB{\mathrm{TV}\nB{P_{\mathbf{\Ip}_2^{*}| \mathbf{\Ip}_1 = \tilde{\mathbf{\Ip}}_1} , P_{\tilde{\mathbf{\Ip}}_2}} \middle| \tilde{\mathbf{\Ip}}_1}\\
&\leq& \mathbb{E}\sB{\sum_{2 \leq i < j \leq K} \mathrm{TV}\nB{P_{\Ip_{ij}^{*}| \mathbf{\Ip}_1 = \tilde{\mathbf{\Ip}}_1} , \mathcal{N}(0, 1)} \middle | \tilde{\mathbf{\Ip}}_1} \label{eq:qiid}\\
&=& \binom{K-1}{2} \bbE \biggl[ \mathrm{TV} \bigg( \mathcal{N} \bigg(\frac{\tilde{\Ip}_{12} \tilde{\Ip}_{13}}{\sqrt{n}}, \frac n {n-1} \bigg(1 - \frac{\tilde{\Ip}_{12}^2}{n} \bigg) \notag \\
&& \quad  \bigg(1 - \frac{\tilde{\Ip}_{13}^2}{n} \bigg) \bigg) ,  \mathcal{N}(0, 1) \bigg) \biggr] \label{eq:tviid} \\
&\leq&  \binom{K-1}{2}  \Bigg\{ \frac 1 2 \frac{\E{ \left\lvert \tilde{\Ip}_{12} \right \rvert }^2 }{\sqrt{n}} \notag \\
&&+ \frac{2 + \sqrt{6}}{4} \left \lvert \frac n {n-1} \left(\E{1 - \frac{\tilde{\Ip}_{12}^2}{n}}\right)^2 - 1\right \rvert  \Bigg\} \label{eq:tvgauss} \\
&=&  \binom{K-1}{2} \left( \frac{1}{\pi \sqrt{n}}  +  \frac{2 + \sqrt{6}}{4n }  \right) \label{eq:expreplaced},
\end{IEEEeqnarray}
where \eqref{eq:qiid} follows from \cite[Lemma~2.1]{sendler1975} since the conditional distribution of $\mathbf{\Ip}_2^*$ given $\mathbf{\Ip}_1 = \mathbf{\ip}_1$ is a product distribution and $P_{\tilde{\mathbf{\Ip}}_2}$ is an i.i.d. standard Gaussian. Equality \eqref{eq:tviid} follows since the conditional distribution of $\Ip_{ij}^*$ given $\mathbf{\Ip}_1 = \mathbf{\ip}_1$ is identically distributed for $2 \leq i < j \leq K$. Inequality \eqref{eq:tvgauss} follows from \lemref{lem:gaussiancdf} with $d = 1$ using the i.i.d. distribution of $\tilde{\Ip}_{12}$ and $\tilde{\Ip}_{13}$. Combining \eqref{eq:tvlemma}, \eqref{eq:qassump}, \eqref{eq:expreplaced}, and the inequality in \eqref{eq:tv11} completes the proof by induction.

We note that the convergence rate of the total variation distance of interest is $\bigo{\frac{1}{\sqrt{n}}}$ for $K > 2$, while it is faster $\nB{\bigo{\frac 1 n}}$ for $K = 2$.

\section*{acknowledgment}
    We are grateful to Peida Tian for pointing out the paper \cite{devroye2018total} that led to an improvement in \lemref{lem:gaussiancdf}.

\bibliographystyle{IEEEtran}
\bibliography{mac} 	

\begin{IEEEbiographynophoto}
{Recep Can Yavas}(S'19) is currently a Ph.D. candidate in electrical engineering at the California Institute of Technology (Caltech). He received the B.S. degree from Bilkent University in Ankara, Turkey, in 2016 and the M.S. degree from Caltech in 2017, both in electrical engineering. His research interests include information theory and probability theory. 
\end{IEEEbiographynophoto}

\begin{IEEEbiographynophoto}{Victoria Kostina}(S'12--M'14)
is a Professor of Electrical Engineering and of Computing and Mathematical Sciences at Caltech. She received a bachelor's degree from Moscow Institute of Physics and Technology (2004), where she was affiliated with the Institute for Information Transmission Problems of the Russian Academy of Sciences, a master's degree from University of Ottawa (2006), and a PhD from Princeton University (2013). She received the Natural Sciences and Engineering Research Council of Canada postgraduate scholarship (2009--2012), the Princeton Electrical Engineering Best Dissertation Award (2013), the Simons-Berkeley research fellowship (2015) and the NSF CAREER award (2017). Kostina's research spans information theory, coding, control, learning, and communications. 
\end{IEEEbiographynophoto}

\begin{IEEEbiographynophoto}{Michelle Effros}(S'93-M'95-SM'03-F'09) 
received the B.S. degree with distinction in 1989, the M.S. degree in 
1990, and the Ph.D. degree in 1994, all in electrical engineering from 
Stanford University. She joined the faculty at the California Institute of 
Technology in 1994, where she is currently the George Van Osdol Professor 
of Electrical Engineering. Her research interests include information 
theory, network coding, data compression, and communications.

Prof. Effros received Stanford's Frederick Emmons Terman Engineering 
Scholastic Award (for excellence in engineering) in 1989, the Hughes 
Masters Full-Study Fellowship in 1989, the National Science Foundation 
Graduate Fellowship in 1990, the AT\&T Ph.D. Scholarship in 1993, the NSF 
CAREER Award in 1995, the Charles Lee Powell Foundation Award in 1997, the 
Richard Feynman-Hughes Fellowship in 1997, and an Okawa Research Grant in 
2000. She was cited by Technology Review as one of the world's top young 
innovators in 2002. She and her co-authors received the Communications 
Society and Information Theory Society Joint Paper Award in 2009. She 
became a fellow of the IEEE in 2009. She is a member of Tau Beta Pi, Phi 
Beta Kappa, and Sigma Xi. She served as the Editor of the IEEE Information 
Theory Society Newsletter from 1995 to 1998 and as a Member of the Board 
of Governors of the IEEE Information Theory Society from 1998 to 2003 and 
from 2008 to 2017, serving in the role of President of the Information 
Theory Society in 2015. She was a member of the Advisory Committee and the 
Committee of Visitors for the Computer and Information Science and 
Engineering (CISE) Directorate at the National Science Foundation from 
2009 to 2012 and in 2014, respectively. She served on the IEEE Signal 
Processing Society Image and Multi-Dimensional Signal Processing (IMDSP) 
Technical Committee from 2001 to 2007 and on ISAT from 2006 to 2009. She 
served as Associate Editor for the 2006 joint special issue on Networking 
and Information Theory in the {\em IEEE Transactions on Information 
Theory} and the {\em IEEE Transactions on Networking/ACM Transactions on 
Networking} and as Associate Editor for Source Coding for the {\em IEEE 
Transactions on Information Theory} from 2004 to 2007. She has served on 
numerous technical program committees and review boards, including serving 
as general co-chair for the 2009 Network Coding Workshop and technical 
program committee co-chair for the 2012 IEEE International Symposium on 
Information Theory.
\end{IEEEbiographynophoto}

	\end{document}